\documentclass[sigconf,nonacm]{acmart}
\usepackage{placeins}
\usepackage{afterpage}
\usepackage[ruled, vlined, linesnumbered]{algorithm2e}
\usepackage{graphicx}
\usepackage{textcomp}
\usepackage{xcolor}
\usepackage{comment}
\usepackage{tikz}
\usepackage[normalem]{ulem}
\usepackage{soul}
\usepackage{cancel}
\usepackage{xspace}
\usepackage{threeparttable}
\usepackage{booktabs}
\usepackage[utf8]{inputenc}
\usepackage{multirow}
\usepackage{caption}
\usepackage{enumitem}
\usepackage{pgfplots}
\pgfplotsset{compat=1.16}
\usepackage[maxfloats=36,morefloats=18]{morefloats}
\usepackage{listings}
\definecolor{codegreen}{rgb}{0,0.6,0}

\definecolor{codegray}{rgb}{0.5,0.5,0.5}

\definecolor{codepurple}{rgb}{0.58,0,0.82}

\definecolor{backcolour}{rgb}{0.95,0.95,0.92}

\definecolor{white}{rgb}{1.0, 1.0, 1.0}

\colorlet{mygray}{black!20}
\colorlet{mygreen}{green!60!blue}
\colorlet{mymauve}{red!60!blue}

\usepackage{tcolorbox}
\usepackage{lipsum}
\newtcolorbox{mybox}{colback=gray!30,
boxrule=0pt,arc=0pt,boxsep=2pt,left=2pt,right=2pt,leftrule=1pt}

\theoremstyle{definition}

\usepackage{tabularx}

\usepackage{tabularray}
\usepackage{amsthm}
\newtheorem{definition}{Definition}
\usepackage{makecell}
\usepackage{boldline}
\usepackage{url}

\usepackage{mathtools}

\usepackage{stackengine}
\newcommand\xrowht[2][0]{\addstackgap[.5\dimexpr#2\relax]{\vphantom{#1}}}

\makeatletter
\DeclareFontFamily{U}{FdSymbolA}{}
\DeclareFontShape{U}{FdSymbolA}{m}{n}{<-> s * FdSymbolA-Book}{}
\DeclareSymbolFont{fdsymbols}{U}{FdSymbolA}{m}{n}
\DeclareMathSymbol{\newcheckmark}{\mathord}{fdsymbols}{"B3}
\makeatother
\usepackage{utfsym}
\newcommand{\newcrossmark}{\scalebox{0.75}{\usym{2613}}}

\def\BibTeX{{\rm B\kern-.05em{\sc i\kern-.025em b}\kern-.08emT\kern-.1667em\lower.7ex\hbox{E}\kern-.125emX}}

\newcommand{\tool}{{\sc Uriah}\xspace}

\newif\ifprintcomments

\newcommand{\new}[1]{\textcolor{black}{#1}}

\newcommand{\mat}[1]{
    \ifprintcomments
        \textcolor{red}{{\em\bf Mathias:} #1}
    \fi}
\newcommand{\kaiming}[1]{
    \ifprintcomments
        \textcolor{mygreen}{{\em\bf Kaiming:} #1}
    \fi}
\newcommand{\trent}[1]{
    \ifprintcomments
        \textcolor{blue}{{\em\bf Trent:} #1}
    \fi}
\newcommand{\gang}[1]{
    \ifprintcomments
        \textcolor{blue}{{\em\bf Gang:} #1}
    \fi}
\newcommand{\JS}[1]{
    \ifprintcomments
        \textcolor{orange}{{\em\bf Jack:} #1}
    \fi}
\newcommand{\zhiyun}[1]{
    \ifprintcomments
        \textcolor{cyan}{{\em\bf Zhiyun:} #1}
    \fi}

\newcommand{\rt}[1]{\textcolor{red}{#1}}

\newif\ifsoutversion
\newcommand{\redout}[1]{\ifsoutversion\rt{\setstcolor{red}\sout{#1}}\else\fi}

\soutversionfalse
\begin{document}

\printcommentstrue
%\printcommentsfalse

%\title{Selective and Comprehensive?\\Efficient Heap Data Protection Against Memory Errors\\}
\title{Top of the Heap: Efficient Memory Error Protection \\ of Safe Heap Objects}
%\title{0 to 75 (Percent) Heap Memory Protection: Comprehensive Memory Safety Enforcement for Many Heap Objects without Runtime Checks}

%\author{{\rm Anonymous Submission}}
\if 0
\makeatletter
\newcommand{\linebreakand}{%
  \end{@IEEEauthorhalign}
  \hfill\mbox{}\par
  \mbox{}\hfill\begin{@IEEEauthorhalign}
}
\makeatother

\author{\IEEEauthorblockN{Kaiming Huang}
\IEEEauthorblockA{\textit{Penn State University} \\
kzh529@psu.edu}
\and
\IEEEauthorblockN{Mathias Payer}
\IEEEauthorblockA{\textit{EPFL} \\
mathias.payer@nebelwelt.net}
\and
\IEEEauthorblockN{Zhiyun Qian}
\IEEEauthorblockA{\textit{UC Riverside} \\
zhiyunq@cs.ucr.edu}
\linebreakand
\IEEEauthorblockN{Jack Sampson}
\IEEEauthorblockA{\textit{Penn State University} \\
jms1257@psu.edu}
\and
\IEEEauthorblockN{\ \ \ \ Gang Tan}
\IEEEauthorblockA{\textit{\ \ \ \ Penn State University} \\
\ \ \ \ gxt29@psu.edu}
\and
\IEEEauthorblockN{Trent Jaeger}
\IEEEauthorblockA{\textit{Penn State University} \\
trj1@psu.edu}

}
\fi

\author{Kaiming Huang}
\affiliation{%
  \institution{Penn State University}
  \country{}}
\email{kzh529@psu.edu}

\author{Mathias Payer}
\affiliation{%
  \institution{EPFL}
  \country{}}
\email{mathias.payer@nebelwelt.net}

\author{Zhiyun Qian}
\affiliation{%
  \institution{UC Riverside}
  \country{}}
\email{zhiyunq@cs.ucr.edu}

\author{Jack Sampson}
\affiliation{%
  \institution{Penn State University}
  \country{}}
\email{jms1257@psu.edu}

\author{Gang Tan}
\affiliation{%
  \institution{Penn State University}
  \country{}}
\email{gxt29@psu.edu}

\author{Trent Jaeger}
\affiliation{%
  \institution{UC Riverside}
  \country{}}
\email{trentj@ucr.edu}

%\author{Anonymous Submission}

\if 0



\fi

\begin{abstract}

% area
Heap memory errors remain a major source of software vulnerabilities.  
% problem
% related 
% Proposed generic memory safety defenses are often incomplete and expensive, hindering adoption in production software. 
Existing memory safety defenses aim at protecting all objects, resulting in high performance cost and incomplete protection.
% insight 
Instead, we propose an approach that accurately identifies objects that are inexpensive to protect, 
%i.e., to maximize the number of objects that can be protected cheaply, 
and design a method to protect such objects comprehensively from all classes of memory errors.
%Our goal is to find heap objects that are inexpensive to protect from memory errors (e.g., not requiring runtime checks) and isolate these objects from accesses that may violate memory safety.  
Towards this goal, 
we introduce the \tool system that (1) statically identifies the heap objects whose accesses satisfy spatial and type safety, 
%accounting for challenges presented by heap usage, such as reallocation and concurrent access, 
and (2) dynamically allocates such "safe" heap objects on an isolated safe heap to enforce a form of temporal safety while preserving spatial and type safety, 
%restricting reuse only for objects with the same size and type for all fields, 
called {\em temporal allocated-type safety}. \tool finds 72.0\% of heap allocation sites produce objects whose accesses always satisfy spatial and type safety in the SPEC CPU2006/2017 benchmarks, 5 server programs, and Firefox, which are then isolated on a safe heap using \tool allocator to enforce temporal allocated-type safety. \tool incurs only 2.9\% and 2.6\% runtime overhead, along with 9.3\% and 5.4\% memory overhead, on the SPEC CPU 2006 and 2017 benchmarks, while preventing exploits on all the heap memory errors in DARPA CGC binaries and 28 recent CVEs. Additionally, using existing defenses to enforce their memory safety guarantees on the unsafe heap objects significantly reduces overhead, enabling the protection of heap objects from all classes of memory errors at more practical costs. 
\end{abstract}

\settopmatter{printfolios=true}

\maketitle

%auto-ignore

\section{Introduction}
\label{sec:introduction}

Memory errors in C/C++ programs continue to cause the most significant security problems. The White House recently stated that future software should be memory safe~\cite{whitehousememsafe}. According to the NSA~\cite{nsa-report},  Microsoft~\cite{microsoft-report}, and Google~\cite{google-report}, $\approx$70-80\% of vulnerabilities are caused by memory errors. Known since the Anderson Report~\cite{anderson72refmon}, memory 
errors have led to high-impact attacks such as the Morris Worm~\cite{morris-worm-seeley}, Slammer~\cite{sql_slammer}, Heartbleed~\cite{heartbleed}, and Blastpass~\cite{Blastpass}. Memory errors 
%shed light on the failures of protection against memory errors, 
are exploited by ransomware that costs organizations billions of dollars~\cite{sophos2021state,emsisoft2020cost,cybersecurityventuresrdr}, and are even found by studies~\cite{stamatogiannakis2022asleep,chen2021evaluating,xu2022systematic} of software produced by LLM-based code generators~\cite{gpt,chatgpt,copilot}.  A wide variety of notable attack techniques have been discovered to exploit memory errors~\cite{rop,data-only-attack,heap-feng-shui,dirtypipe,dirtycred,smashing,hu2016data,bopc,kleak,syzbridge,syzscope,eloise,grebe,fuze17usenix,slake,kepler,cheng21tops} that enable attackers to gain control of process memory.

%which can lead to a multitude of risks to serious system corruptions, including denial of service, leak of sensitive data, privilege escalation, arbitrary (remote) code execution, etc. The memory errors existed in the software and system have exposed the need for robust security measures and prompted the development of comprehensive memory protections. 

Among these errors, \textit{heap} memory errors are notorious for their prevalence and severity, posing significant challenges for developers.  There are several reasons why heap memory is more prone to errors.  First, heap memory may be used to store objects of variable size, which can also change through reallocations.  Errors in tracking sizes of these objects can violate {\bf\em spatial safety}, where memory operations access locations outside expected bounds. Second, heap objects are often complex,  consisting of hierarchically structured data types that may be cast into multiple views to different formats depending on the program context.  Mistakes in interpreting the layouts of data structures can violate {\bf\em type safety}, where memory operations may access objects at invalid offsets and/or using incorrect data types. Third, unique to heap memory is the challenge of managing dynamic allocation and deallocation.  Unlike stack memory,
%which is automatically allocated and deallocated, 
heap memory management in C/C++ requires manual intervention, causing errors that violate {\bf\em temporal safety}
%\footnote{In this paper, we refer to all temporal errors that occur before a referent is assigned as {\em use-before-initialization} (UBI) and all temporal errors that occur after a referent is deallocated as {\em dangling pointer misuse}.},
where pointers may not be initialized before use (i.e., UBI) and/or may be used after deallocation (i.e., UAF).  Finally, heap objects may be accessed across multiple threads complicating the challenges above.

Despite years of efforts to develop heap memory safety defenses, heap data is not systematically protected from all three classes of memory errors comprehensively.  Researchers have produced techniques to enforce memory safety requirements to prevent spatial memory errors~\cite{softbound,baggybounds,memsafe10scam,lowfat-heap}, type memory errors~\cite{hextype,caver,typesan,vtrust}, and temporal memory errors~\cite{ubSAN, uafchecker,clangstaticanalyzer, mluafdetect, gueb, uafdetector,cred,Zhai2020UBITectAP, van2017dangsan,dangnull,freesentry,Nagarakatte2010CETSCE,zhou2022fat, cfixx} independently, as well as multiple classes of memory errors~\cite{safecode,duck2018effectivesan,ccured,asan,camp24usenix,checkedc, neug2017dimva}.  However, none of these defenses have been widely-adopted in production (although several are used in fuzz testing), as they often incur significant overheads.  Production defenses such as AutoSlab~\cite{autoslab}, kalloc\_type~\cite{kalloc_type}, and PartitionAlloc~\cite{palloc} only focus on temporal safety, Microsoft Defender~\cite{windowsheap} preserves the integrity of heap metadata for code with memory errors. However, such defenses only make exploitation harder (i.e., memory errors and exploitations are still possible). Thus, a lingering question is how to introduce heap protection 
%\zhiyun{suggest simply say ``heap protection''} \kaiming{Agreed}
that can be adopted in practice (i.e., is effective and efficient) and can serve as a foundation for enforcing memory safety comprehensively for the entire heap.

We observe that producing defenses for heap memory is very difficult because any defense must satisfy three {\bf\em mutually conflicting} properties: An ideal defense must provide memory protection against all \textbf{\em classes} of memory errors with \textbf{\em coverage} across all memory objects for reasonable \textbf{\em cost} in performance and memory overheads.
%, which 
%we 
%has been called the \textbf{\em 3-Cs Principle}~\cite{huang24ieeesp}
The past 20 years of research has demonstrated that it is very hard to find a silver bullet that achieves all three goals simultaneously~\cite{huang24ieeesp}.  Traditionally, researchers have focused on the goal of complete coverage across all objects, but some memory operations may occur frequently causing a high overhead for the defenses, resulting in trade-offs on protecting only one or some of the memory error classes~\cite{safecode, asan, baggybounds,softbound, diehard, safeinit}.
%\zhiyun{can we cite some papers here?} \kaiming{sure.} 
An alternative approach, 
%Recent static-analysis-driven defenses, 
exemplified by  DataGuard~\cite{dataguard}, targets protection against all memory error classes, but only for objects that can be protected efficiently, finding that over 90\% of stack objects can be protected for 4.3\% overhead on SPEC CPU2006 benchmarks.  DataGuard enables a significant fraction of "safe" stack objects to be protected comprehensively from memory errors, but does not address heap objects nor consider the impact on the remaining unsafe objects.

In this paper, we examine the challenge of memory safety validation to protect {\em safe heap objects}, whose memory references must comply with all classes of memory safety, from being corrupted by memory errors on other unsafe objects for \textit{low cost}.  Our broader vision is that such an approach will reduce the overhead of applying defenses to protect the remaining heap objects comprehensively as well. Similar ideas have been examined in selective symbolic execution~\cite{driller} and removing redundant type checks~\cite{tprunify}. We will demonstrate: (1) this idea \textit{\textbf{can}} make heap memory protection efficient, (2) protecting safe heap objects against all memory error classes \textit{\textbf{can}} provide promising exploit mitigation, and (3) applying existing defenses only to the remaining unsafe heap objects \textit{\textbf{can}} offer the same security guarantees at lower cost.

%that even partial coverage over memory objects can provide useful safety when the target objects needed for an exploit to succeed are among those protected. 

One idea would be to apply the techniques for stack memory safety validation~\cite{dataguard} to heap objects. Unfortunately, the original methods do not apply to the heap.  First, heap objects typically have more complex representations and a greater number of aliases, are involved in more dynamic changes and shared between threads, and have much longer lifetimes than stack objects. As a result, we must design new approaches for static safety validation, while addressing the greater number of false positives that may be produced by static analysis of heap usage.   Second, no general static analysis exists to validate temporal safety for heap objects. As a result, defenses have been proposed to enforce temporal safety at runtime, but these efforts have shown that fully eliminating temporal errors at runtime incurs high performance costs~\cite{duck2018effectivesan,camp24usenix}. Thus, we must adapt this idea strategically and effectively to enforce temporal safety at runtime, while keeping costs low.

%the first approach that fully protects
%\textbf{\textit{safe}} heap objects from attacks on memory errors

%Heap objects have more complex types, more dynamic changes (i.e., sizes and types), and longer and unpredictable lifetimes, which cause prior techniques to classify near all heap objects as unsafe.  In addition, heap  and may to  propose a novel approach that provides memory safety enforcement for a large fraction of heap objects efficiently.  Similar to prior work for the stack~\cite{dataguard}, our approach applies a static memory safety validation to identify heap objects whose pointers {\em cannot ever} violate spatial and type memory safety.  We then design a heap memory allocator to isolate those heap objects, cheaply, from accesses that may violate memory safety while preventing temporal errors.  However, the heap presents several challenges for both static memory safety validation and runtime isolation.

To address these challenges, we develop the \tool system to 
%enforce {\bf\em comprehensive memory safety} of heap object through: 
(1) validate heap objects whose accesses must always satisfy {\bf\em spatial} and {\bf\em type safety} and (2) enforce {\bf\em isolation} and a form of 
%temporal allocated-type safety 
{\bf\em temporal safety} over validated heap objects in (1) \new{to preserve their spatial and type safety at runtime}.  First, \tool provides static memory safety validation methods  \new{that ensure that all operations that access a heap object through all its aliases must satisfy both spatial and type safety, while accounting for the challenges above, including reallocations and concurrent access.}
%\new{(e.g., type confusion). However, spatial and type safety can also be violated by temporal exploits at runtime.} 
%by developing novel techniques to account for dynamic resizing for spatial validation, to identify a wider variety of safe type casts for type validation, and to
%remove positives without allowing any unsafe object to be mis-classified as a safe object. 
Second, 
given results showing that the type-safe memory reuse can be enforced efficiently to prevent temporal errors exploits~\cite{tat,tdi,autoslab},
\tool leverages this idea to enforce a more restrictive form of type-safe memory reuse \new{to maintain spatial and type safety at runtime for all the validated heap objects on a separate {\em safe heap} against temporal exploits, and isolate it from
%The validated heap objects are isolated on a separate safe heap to protect them from 
memory errors in accesses to objects on the regular (unsafe) heap.  This version of type-safe memory reuse, called {\em temporal allocated-type safety}, only allocates objects with the same fields of the same size and type for all fields in each memory region within an isolated safe heap.}
%\zhiyun{Is this something we invented? Somehow it reminds me of autoslab.} \kaiming{Thanks for bringing up this, actually enforcing temporal allocated-type safety on spatial and type safety object guarantees spatial and type safe comprehensively. However, many spatial errors (i.e., overflows) cannot be prevented by autoslab, we should probably say it explicitly.}\zhiyun{My recollection is that autoslab also puts objects of the same type in the same memory region (i.e., slab). I understand it doesn't deal with spatial errors. But the discussion here seems to be focused narrowly on type-safety, and we seem to be applying the same idea as autoslab (perhaps I am missing something but it would be good to clarify).}\kaiming{Thanks, yes, I agree that the above text create a wrong impression to the readers that diminish our contribution of ensuring memory safety comprehensively, not limited to the type safety. I will update the text above.}
\redout{This method ensures all accesses, even using a dangling pointer, must satisfy spatial and type safety for a memory region since the memory layout must be the same for all allocations to that region.  Objects on the safe heap are isolated from memory errors in malicious/erroneous accesses to objects on the unsafe heap.}\new{The result is the following \textit{\textbf{security guarantee}}: 
%Uriah guarantees spatial and type safety for all objects on the safe heap. 
%by: (1) validating that object accesses cannot violate spatial and type safety and (2) preventing temporal exploits that corrupt spatial and/or type safety using temporal-allocated type safety. 
For objects whose own accesses all satisfy spatial and type safety statically, which we call {\em safe objects} in this paper, \tool ensures that no access to any object can violate the spatial and type safety of any safe object at runtime.
%by enforcing temporal allocated-type safety on and isolating the safe heap (through SFI). 
\tool protects safe objects from any memory errors in accesses to unsafe objects by separating safe objects onto a safe heap isolated using SFI. Further, \tool protects the spatial and type safety of safe objects at runtime from temporal exploits that leverage dangling pointers to the safe heap
by enforcing temporal allocated-type safety. }

By employing \tool
%\footnote{\new{\bf We have open-sourced \tool's codebase in \href{http://www.google.com}{Place Holder}.}}
, we find major benefits in protecting memory safety for heap objects\footnote{The term "heap objects" refers to all objects allocated to an allocation site, which are equivalent to the static analysis, considering concurrency.}.  First, \tool finds that 72.0\% of heap allocation sites can be validated to produce objects that satisfy spatial and type safety statically for a variety of programs, including Firefox, servers (nginx and httpd), and the SPEC CPU2006/2017 benchmarks. This represents a substantial increase relative to the 39.9\% found equivalently safe by employing prior techniques (see Table~\ref{tab:OverviewEval}). \tool protects all objects produced at these allocation sites by isolation on the safe heap. 
%\zhiyun{The notion of safe heap seems to be connected to the isolated memory region which we mention above. It may be a good idea to move it up there to highlight the unique property of safe heap --- objects we put on safe heap are not only statically verified to be spatial}
This accounts for 73.6\% of the memory object allocations made by the SPEC CPU2006 benchmarks at runtime.  Second, \tool exhibits only 2.9\% and 2.6\% runtime and 9.3\% and 5.4\% memory overhead for the SPEC CPU2006 benchmarks and the SPEC CPU2017 benchmarks, respectively.  We compare \tool's memory and runtime overheads to seven  state-of-the-art defenses (see Table~\ref{tab:Overhead}), finding that \tool has lower overheads for allocating all objects
%for allocating only safe objects and has a lower overhead than 
for all but one system, but stronger security guarantees. 
%\trent{Lower overhead per object/site protected?}  
%Runtime and memory overhead of SPEC CPU2017 benchmarks is 2.6\% and 5.4\% respectively.
%As \tool manages a greater number of per-type pools than Type-After-Type, we find that its overhead is greater when applied to all allocations, but only by a modest amount despite the increased precision of \tool. 
As a defense, \tool prevents exploitation of all 102 heap vulnerabilities in Cyber Grand Challenge binaries and 28 recent heap CVEs. %\zhiyun{these numbers don't match those in abstract}. 
\new{More importantly, while \tool does not protect objects on the unsafe heap, it reduces the number of unsafe objects, lowering the effort to enforce memory safety overall. %We measure a reduction of \trent{range} for applying the TDI~\cite{tdi} and CAMP~\cite{camp24usenix}memory safety defense to the remaining objects on the unsafe heap. 
To examine this, } 
%\trent{Removed the second sentence - try to avoid making qualitative statements when you have concrete ones.}
\redout{Moreover}we combine \tool with existing defenses, TDI~\cite{tdi} and CAMP~\cite{camp24usenix}, saving $\approx$70\% of their original overhead to efficiently enforce their security guarantee over the unsafe heap.

The contributions of \tool{} include:
\begin{itemize}[leftmargin=*]
\vspace{-0.1in}
    \item We build the \tool system for heap memory safety enforcement, which provides memory safety validation to determine which heap objects satisfy spatial and type safety and a heap allocator that protects these objects by enforcing temporal allocated-safety over an isolated safe heap. Heap objects in the safe heap are isolated from memory errors efficiently.
    %leveraging the spatial and type safety of its objects to improve security guarantees efficiently.
    \item \tool includes new analyses for spatial and type safety validation that address the challenges of heap objects, such as complex representations, dynamic resizing, multi-threading, and up/downcasts.  In addition, \tool applies new symbolic execution analyses to validate unsafe cases, converting them to safe cases when all unsafe aliases lack feasible execution paths.
    %, which has a significant impact in proving given over-approximate aliasing of heap objects. 
    %This validation enables the \tool allocator to implement type-safe reuse within a single heap to enforce memory safety with low memory overheads.
    \item Our evaluation shows that \tool protects 72.0\% of the allocation sites of CPU2006/2017 benchmarks, 5 server programs, and Firefox, with only 2.9\% and 2.6\% runtime and 9.3\% and 5.4\% memory overheads, on the SPEC CPU2006/2017 benchmarks, respectively, while preventing exploits of 28 known CVEs and CGC programs.
    %The evaluation of CVEs and CGC programs demonstrates that \tool is effective in preventing real-world exploits. The evaluation of combining \tool with existing works demonstrates its adaptability with defenses to reduce the cost while ensuring same security guarantee.
\end{itemize}

%auto-ignore

\section{Motivation}
\label{sec:motivation}

%\trent{Please start with a brief summary of the problem of CVEs due to heap memory safety.  Perhaps simply a count of recent CVEs.  Something to set the scale of the problem - which should not require much justification.} 

In this section, we motivate our research by examining how heap memory errors may be exploited and past research on mitigating such errors, showing the limitations, and proposing our ideas.

\subsection{Exploiting Heap Memory Errors}
\label{subsec:exploitheap}
Despite decades of work on mitigation, memory errors remain an active source of vulnerabilities, particularly for the heap.
Over the past 10 years, there have been more than 10,000 CVEs of heap memory errors, including hundreds in the current calendar year. 

A program contains a memory error when a program's memory reference may violate a memory safety property.  Researchers have identified three classes of memory safety properties defined below.

\begin{itemize}[leftmargin=*]
\vspace{-0.05in}
\item {\bf {\em Spatial Safety}}:
%and only if 
Every pointer that may reference the object must only access memory within the object's allocated region.    
\item {\bf {\em Type Safety}}: 
% and only if 
Every pointer that may reference the object must only access the same data types for each offset and each field.  
%every memory access to the object (or its fields for a structured type) must only be to valid memory locations and based on the object's data type and 
%interprets the  the same value and data type.  
%if all pointers that may reference that object must interpret the object's value (i.e., if a singleton) or the values of all of its fields (i.e., if a structured type) consistently (i.e., same data type and value).
\item {\bf {\em Temporal Safety}}: 
%and only if 
Every pointer that may reference the object must not be used to access the object's allocated region before being assigned to the object nor after the object's deallocation.
%must be assigned to the object before any access to its allocated region
%of that object must have been assigned to that object use that may reference that object must have been assigned that object as a referent 
%and not be used after deallocation.
\end{itemize}

\begin{lstlisting}[language=C,float,floatplacement=t,abovecaptionskip=-15pt,belowcaptionskip=0pt,caption={Spatial memory error for vulnerability CVE-2023-1579, which allows attackers to exploit a buffer over-read through {\tt unit} to access objects aliased by {\tt block}\vspace{-0.25in}},captionpos=b,label={lst:CVE1}]
static uint64_t read_indexed_address (uint64_t idx, 
                        struct comp_unit *unit){
  ...
  bfd_byte *info_ptr; // used for memory access
  size_t offset;
  offset += unit->dwarf_addr_offset;
  if (offset < unit->dwarf_addr_offset
      || offset > file->dwarf_addr_size
      || file->dwarf_addr_size - offset < unit->offset_size) 
      // wrong check always true
    return 0;
  info_ptr = file->dwarf_addr_buffer + offset;
  return bfd_get_64 (unit->abfd, info_ptr);
  //unsafe memory access
}

static struct dwarf_block * read_blk (bfd *abfd,
        bfd_byte **ptr, bfd_byte *end, size_t size){
  ...
  struct dwarf_block *block;
  block = (struct dwarf_block*) bfd_alloc (abfd, ...)
}
\end{lstlisting}

Listing~\ref{lst:CVE1} shows an example of a heap vulnerability (CVE-2023-1579) in {\tt binutils}. A size check against the {\tt unit->offset\_size} field at line 9 always returns true, because this is the incorrect field for this size check; the check should be made against the {\tt unit->addr\_size} field. This error enables the {\tt info\_ptr} pointer assigned in line 12 to reference out-of-bounds memory, eventually resulting in heap buffer over-reads in the function called at line 13.  In this case, the {\tt block} allocated at line 21 is often exploited, and, as it may alias any data on the heap allocated by bfd, an attacker can read any heap data by illicitly accessing memory through {\tt block}.

To exploit heap memory errors, attackers utilize a memory error in accessing one object to exploit other objects.  We refer to an object whose accesses have memory errors as a {\bf\em vulnerable object}. In this example, the vulnerable object aliased by {\tt unit} has a spatial memory error that permits attackers to read outside its allocation.  We call the objects that can be accessed illicitly due to a memory error, {\bf\em target objects}.  In this example,  heap objects aliased by {\tt block} are target objects. Existing, commonly-adopted defenses cannot prevent the exploitation effectively.  For example, ASLR~\cite{aslr} can be bypassed by {\em disclosure attack}~\cite{snow2013just,readactor} and cannot prevent illicit reads, as the target object is often allocated at a known or configurable~\cite{heap_spray,heap-feng-shui, auto_fengshui} offset. Spatial defenses, such as ASan~\cite{asan}, are known to present high costs. Other defenses~\cite{autoslab,kalloc_type, palloc} only focus on temporal safety, neglecting spatial and type errors.

\vspace{-0.1in}
\subsection{Limitations of Heap Memory Defenses}
\label{subsec:limit}

\begin{table}[t!]
    \aboverulesep=0.25ex 
    \belowbottomsep=0.5ex
    \belowrulesep=0.25ex
    \abovetopsep=0.5ex
    \centering
    \scriptsize

	\resizebox{\columnwidth}{!}{
	\begin{tabular}{lccccc}
	    \toprule
     
		 &\multicolumn{1}{l}{\bf\em Defense}&\multicolumn{1}{l}{\bf\em Spatial}&\multicolumn{1}{c}{\bf\em Type }&\multicolumn{1}{c}{\bf\em Temporal}&\multicolumn{1}{c}{\bf\em Scope}\\
   
		\midrule
		{\bf\em \tool} ({\em this work}) &A&$\newcheckmark$&$\newcheckmark$&$\newcheckmark$\textsuperscript{\textdagger}&H$^\ast$\\
  
	    \midrule
     
		{\bf\em CCured}~\cite{ccured,ccured-realworld,ccured-toplas}&{R}&{$\newcheckmark$}&{$\newcheckmark$}&{$\newcrossmark$}&{S \& H}\\

            \midrule
            
            {\bf\em Checked-C}~\cite{checkedc,zhou2022fat}&\multirow{2}{*}{R}&\multirow{2}{*}{$\newcheckmark$}&\multirow{2}{*}{$\newcheckmark$}&\multirow{2}{*}{UAF}&\multirow{2}{*}{S \& H}\vspace{0.04in}\\

            {\bf\em EffectiveSan}~\cite{duck2018effectivesan}&&&&&\\
		\midrule
  
            {\bf\em ASan}~\cite{asan}&{R}&{$\newcheckmark$}&{$\newcrossmark$}&{UAF}&{S \& H}\\

            \midrule
  
		{\bf\em Baggy-Bounds}~\cite{baggybounds}&\multirow{4}{*}{R}&\multirow{4}{*}{$\newcheckmark$}&\multirow{4}{*}{$\newcrossmark$}&\multirow{4}{*}{$\newcrossmark$}&\multirow{4}{*}{S \& H}\vspace{0.04in}\\

            {\bf\em SoftBound}~\cite{softbound}&&&&\vspace{0.04in}\\
		
            {\bf\em Low-Fat}~\cite{lowfat,lowfat-heap}&&&&\\
            
		\midrule
  
            {\bf\em HexType}~\cite{hextype}&\multirow{4}{*}{R}&\multirow{4}{*}{$\newcrossmark$}&\multirow{4}{*}{$\newcheckmark$}&\multirow{4}{*}{$\newcrossmark$}&\multirow{4}{*}{S \& H}\vspace{0.04in}\\
		
            {\bf\em TypeSan}~\cite{typesan}&&&&\vspace{0.04in}\\
		
            {\bf\em CaVer}~\cite{caver}&&&&\\		
            
		\midrule
  
            {\bf\em DangSan}~\cite{van2017dangsan}&\multirow{4}{*}{R}&\multirow{4}{*}{$\newcrossmark$}&\multirow{4}{*}{$\newcrossmark$}&\multirow{4}{*}{UAF}&\multirow{4}{*}{H}\vspace{0.04in}\\
		
            {\bf\em DangNull}~\cite{dangnull}&&&&\vspace{0.04in}\\

            {\bf\em FreeSentry}~\cite{freesentry}&&&&\\   

            \midrule

            {\bf\em SAFECode}~\cite{safecode}&A&$\newcrossmark$&$\newcheckmark$&{UAF\textsuperscript{\textdagger}}&S \& H\\

            \midrule
  
            {\bf\em FFMalloc}~\cite{ffmalloc21usenix}&\multirow{2}{*}{A}&\multirow{2}{*}{$\newcrossmark$}&\multirow{2}{*}{$\newcrossmark$}&\multirow{2}{*}{UAF}&\multirow{2}{*}{H}\vspace{0.04in}\\
		
            {\bf\em MarkUs}~\cite{markus20oakland}&&&&\\

            \midrule

            {\bf\em Cling}~\cite{cling}&A&$\newcrossmark$&$\newcrossmark$&{UAF\textsuperscript{\textdagger}}&H\\
            
		\midrule

            {\bf\em Type-After-Type}~\cite{tat}&A&$\newcrossmark$&$\newcrossmark$&{UAF\textsuperscript{\textdagger}}&S \& H\\
  
		\midrule
  
            {\bf\em DieHard}~\cite{diehard}&\multirow{5}{*}{A}&\multirow{5}{*}{$\newcheckmark$}&\multirow{5}{*}{$\newcrossmark$}&\multirow{5}{*}{$\newcheckmark$\textsuperscript{\textdagger}}&\multirow{5}{*}{H}\vspace{0.04in}\\

            {\bf\em DieHarder}~\cite{dieharder}&&&&\vspace{0.04in}\\

            {\bf\em FreeGuard}~\cite{freeguard17ccs}&&&&\vspace{0.04in}\\ 
            
            {\bf\em TDI}~\cite{tdi}&&&&\\

            \midrule

            {\bf\em CAMP}~\cite{camp24usenix}&A\&R&$\newcheckmark$&$\newcrossmark$&{UAF}&S \& H\\

		\midrule

            {\bf\em DataGuard}~\cite{dataguard}&A&$\newcheckmark$&$\newcheckmark$ &$\newcheckmark$&S$^\ast$\\
            
            \midrule

            {\bf\em Safe Stack}~\cite{kuznetsov14osdi}&A&$\newcheckmark$&{$\newcrossmark$} &{$\newcrossmark$}&S$^\ast$\\
            
		\bottomrule
	\end{tabular}
        }
        \captionsetup{font=footnotesize}
	\caption{{\bf Comparison of \tool with Previous Works}. {\em Defense} includes custom allocator (A) or runtime checks (R). ``$\protect\newcheckmark$" indicates an approach protects that class of memory errors and ``$\protect\newcrossmark$" when it does not. \new{UAF in {\em Temporal} column indicates the protection includes use-after-free, double-free, and invalid-free, but not use-before-initialization. \textdagger indicates the approach enforces a well-defined subset temporal safety, such as temporal type safety~\cite{tat}.
 %does not fully mitigates the attack towards the declared protection, e.g., type-based allocation, probablistic memory safety.
 } The {\em Scope} column indicates the protection covers stack (S) and/or heap (H), possibly for a subset of objects that adhere to a property (e.g., are "safe") ``$\ast$".}  
	\label{tab:previouswork}
 \vspace{-0.4in}
\end{table}

 %SAFECode allows overflow within its allocation pool. EffectiveSan and Type-After-Type only protect temporal errors partially. DangSan, DangNull, FreeSentry, FFmalloc, MarkUs focus on mitigating UAF bugs

% t1: two defenses (alloc or check)
% t1: partial coverage by class or scope
% t1: two perspectives (all or safe only)

% hw - forward point to relwork
% dont protect safe heap stuff
% fewer runtime checks is always better
% hw: checks are too expensive

Researchers have explored a variety of techniques to enforce a subset or all classes of memory safety, but none of these techniques have yet found broad acceptance in production systems. %which we attribute to the challenge of simultaneously achieving three conflicting goals.  
Most current defenses aim for full {\bf {\em coverage}} of vulnerable objects, but 
have issues enforcing comprehensive memory safety for all {\bf {\em classes}} of memory errors for reasonable performance and memory {\bf {\em costs}}, hindering their deployment.
%\gang{"Full coverage" vs "comprehensive memory safety" are hard to distinguish. Maybe you meant full coverage for one vulnerability class?}
We characterize software-based defenses\footnote{Section~\ref{sec:relatedwork} summarizes related work, including hardware-based defenses.} in  Table~\ref{tab:previouswork}. 
%Initially, researchers focused on defenses that apply runtime checks, which aim to make all accesses to potentially vulnerable objects safe for one or more classes of memory errors. 
Initially, runtime  defenses were designed for a single class of memory errors, such as spatial error protections~\cite{softbound,lowfat-heap,ccured,baggybounds,asan} that restrict memory operations on objects within bounds metadata, type error protections~\cite{vtrust,ubSAN,typesan,hextype,caver, duck2018effectivesan} that validate type casts at runtime, and temporal error protections that often seek to invalidate pointers when memory is deallocated~\cite{van2017dangsan,dangnull,freesentry,duck2018effectivesan,zhou2022fat}. However, despite optimizations, these defenses have performance overheads that prevented their adoption for production systems.  Defenses that enforce memory safety for multiple classes of memory errors, such as ASan~\cite{asan} and EffectiveSan~\cite{duck2018effectivesan}, are applied commonly for detecting vulnerabilities in fuzzing.  ASan performance has been optimized~\cite{asan--}, but its use of red-zones can be circumvented, limiting its effectiveness in preventing exploits. \new{Checked-C~\cite{checkedc} introduces additional language semantics for spatial and type safety, but faces backward compatibility issues with legacy codebases. Checked-C was extended to enforce temporal safety with compatible use of fat pointers~\cite{zhou2022fat}, but the original compatibility issues of Checked-C persist.} 

\if 0
First, the column {\em Defense} specifies the general approach of either applying runtime checks (R) (e.g., {\em sanitizers}~\cite{sanitization_sok}) to restrict accesses to potentially vulnerable objects or supplying a security-aware memory allocator (A) to constrain the memory reuse (e.g., by type) and/or set memory barriers (e.g., guard pages) to prevent illicit accesses to target objects.  
Second, the {\em Spatial, Type, Temporal} columns identify how {\bf {\em comprehensive}} a defense is.
While some defenses enforce multiple classes of memory safety, comprehensive memory safety defense is only the goal of very few systems.  Third, the {\em Scope} column identifies the {\bf {\em coverage}} of a defense by the memory regions  protected (i.e., heap (H) and/or stack (S)) and fraction of objects, where defenses that do not aim for full coverage of a memory region are 
%.  In some cases, defenses only protect a subset of objects in each region (e.g., only ones that satisfy some property) 
signified by an $\ast$.  In most cases, defenses aim for full coverage of all objects in the memory regions they protect.
\fi

More recently, researchers have proposed defenses that leverage secure allocators~\cite{tat,tdi,diehard,dieharder,ffmalloc21usenix,freeguard17ccs,markus20oakland,safecode,camp24usenix} 
to prevent exploitation to target objects through temporal errors.  Most of these defenses aim to prevent UAF~\cite{safecode,tat,ffmalloc21usenix,markus20oakland,camp24usenix}, but some  works~\cite{diehard,dieharder,freeguard17ccs, tdi} include prevention of UBI. 
%These defenses either prevent use-before-initialization~\cite{diehard,dieharder,tdi} (i.e., prevent use of uninitialized memory) or use-after-free~\cite{ffmalloc21usenix, markus20oakland,freeguard17ccs} (i.e., prevent use of deallocated memory).  
Interestingly, researchers have found that enforcing type-safe memory reuse~\cite{tat,safecode,tdi} is an effective approach to prevent dangling pointer misuse by associating data regions with a single data type, as such exploits typically result in type confusion. Unfortunately, the pioneer SAFECode~\cite{safecode} ignores spatial safety and requires an exact alias analysis to apply its per-function, per-type allocation pool for all heap objects, so it can only be applied to small programs and is expensive. Recent work, Type-After-Type~\cite{tat} demonstrates such a technique can be enforced efficiently. However, it places objects from the same allocation site into the same pool, risking type errors if multiple types are dynamically determined and allocated at that site.  Other works focus on restricting memory reuse~\cite{markus20oakland, ffmalloc21usenix}, trading memory overhead for preventing dangling pointer misuse (See Table~\ref{tab:Overhead}). In a broader sense of memory safety, the impact for all the secure allocators is limited as they do not account for all classes of memory errors, as attacks on spatial and/or type errors remain possible on objects allocated by these systems.
\vspace{-0.1in}
\subsection{Protecting Target Objects without Checks}
\label{subsec:target}

%\trent{Probably this is the argument for the prior section - 1st couple of sentences.  Then, seems that protecting target objects from vulnerable ones is the key insight - this section should be about the challenges of protecting target heap objects.}

% goal: all classes for low overhead
% 

\new{In practice, attacks often exploit memory errors in accesses to unsafe objects to corrupt safe object~\cite{kleak,syzbridge,syzscope,eloise,grebe,bopc,hu2016data,fuze17usenix,slake,kepler}. As a result, defenses that prevent accesses to unsafe objects from reaching safe objects can block many attack options.} %\trent{I don't think that there has been an explicit focus on attacking safe objects, but that happens to be the case - especially since many objects are safe.  Tone down.} 
Recent work has proposed an \redout{alternative} approach that aims to achieve this goal in the classes-coverage-cost space of memory defenses. DataGuard~\cite{dataguard} proposes to \new{protect stack}\redout{enforce all classes of memory safety for} objects that pass a static memory safety validation for all classes, finding that protection can be achieved at low cost using multiple stacks, such as Safe Stack~\cite{kuznetsov14osdi}, without runtime checks.  Although this solution is biased toward enforcement of all classes of memory safety  at a low cost at the expense of complete coverage,  such a trade-off is worthwhile, as over 85\% of stack objects can only have memory references that satisfy all classes of memory safety in a large study of over 1,200 Linux packages~\cite{huang23secdev}.  The overhead for protecting memory-safe stack objects in SPEC CPU2006 benchmark programs is only 4.3\%~\cite{dataguard}.
\if 0
Motivated by low overhead defenses for protecting stack objects using multiple stacks, such as Safe Stack~\cite{kuznetsov14osdi}, the DataGuard system provides a suite of analyses to validate all three classes of memory safety for stack objects.  Only stack objects whose references can be validated to be safe for all three classes of memory safety will be isolated on a safe stack, preventing them from being used as targets of a vulnerable object access.  While DataGuard applies an existing isolation technique~\cite{kuznetsov14osdi}, this tool provides a combination of analyses to validate memory safety accurately and comprehensively for stack objects, finding that over 85\% of stack objects can only have memory references that satisfy memory safety in a large study of over 1,200 Linux packages~\cite{huang23secdev}.  The overhead for protecting memory-safe stack objects in SPEC CPU2006 benchmark programs is only 4.3\%~\cite{dataguard}.
\fi

Thus, a research question is whether and how memory safety validation can be applied to identify heap objects that can be protected from all classes of memory errors cheaply. Unfortunately, heap usage introduces challenges that Dataguard's stack analyses cannot handle. For spatial safety, DataGuard's use of {\em value range analysis}~\cite{value-range-book} relies on the predefined, fixed-size objects.  However, heap objects may be resized dynamically, and even reallocated, as well as being accessed across multiple threads. For type safety, 
DataGuard only validates integer type casts, considering any cast between non-identical compound types as unsafe.  Heap objects are much more likely to be involved in the latter (approximately 60\%, see Table~\ref{tab:OverviewEval}).  For temporal safety, DataGuard performs a static liveness analysis, which is not feasible for heap objects. Heap objects have much longer lifetimes than stack objects, so researchers have not yet produced a satisfactory static analysis to validate the temporal safety of heap objects.  In addition, all these static analyses suffer more false positives (i.e., falsely classifying objects as unsafe) due to the greater over-approximation of aliases to heap objects.  DataGuard utilizes symbolic execution to validate legal executions to remove a significant fraction of these false positives, but the depths of the def-use chains for aliases to heap objects often exceed the limits that DataGuard uses to prevent path explosion.

As a result, a method to protect heap objects from memory errors that retains low overhead requires a variety of different problems to be solved.  The spatial and type memory safety validation analyses must be extended to solve challenges specifically 
%that are not possible or are rare for stack objects to 
for validating heap objects effectively.  The additional complexity of these analysis problems and heap analysis in general will result in more false positives, so we need alternative methods to resolve the greater number of false positives that will occur while retaining good scalability.  Finally, we need runtime methods to enforce temporal safety while preserving spatial and type safety.
%are necessary to address the must greater number of false positives while  protects stack objects statically-validated as safe by isolating them in a separate stack at runtime (i.e., {\em safe stack} defense~\cite{kuznetsov14osdi}), preventing accesses from unsafe objects and their pointers.  However, the lack of a static temporal safety validation for  heap objects exposes them to exploitation from temporal attacks (e.g., use-after-free).  
Inspired by the fact that illicit memory accesses caused by temporal errors require (re)use memory of different types, existing temporal defenses for heap objects enforce type-safe memory reuse to ensure only objects of the same type are allocated in each memory region~\cite{safecode,tat}, but the aim for complete coverage introduces limitations (See Section~\ref{subsec:limit}).  

\begin{figure*}
    \centering
    \includegraphics[width=.9\textwidth]{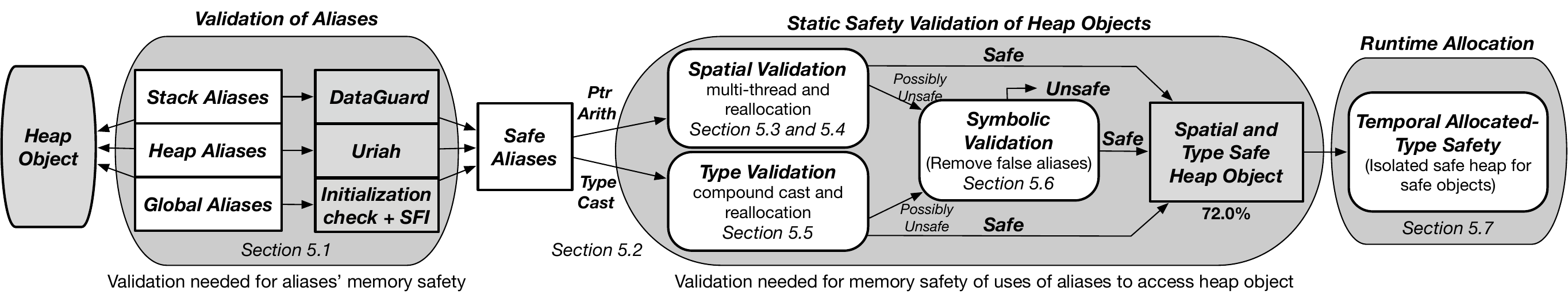}
    \vspace{-0.15in}\captionsetup{font=footnotesize}\caption{Overview of the \tool approach.  %For each {\em heap object} with only memory-safe aliases, \tool determines whether all aliases are only used in accesses of the heap object that must satisfy spatial and type safety. \tool uses constraints extracted from the heap object to validate spatial and type safety for its aliases. Infeasible unsafe cases are pruned through symbolic validation.  Safe heap allocation enforces temporal allocated-type safety for safe heap objects. 
    } 
    \Description{Overview of the \tool approach.  %For each {\em heap object} with only memory-safe aliases, \tool determines whether all aliases are only used in accesses of the heap object that must satisfy spatial and type safety. \tool uses constraints extracted from the heap object to validate spatial and type safety for its aliases. Infeasible unsafe cases are pruned through symbolic validation.  Safe heap allocation enforces temporal allocated-type safety for safe heap objects. }
    %preserving spatial and type safety for safe heap objects.
    } 
    \label{fig:overview}\vspace{-12pt}
\end{figure*}

\section{\tool Overview}
\label{sec:insight}

%A heap object is classified as safe if every operation that may access the object can be proven to satisfy spatial and type safety with respect to the concrete size and type of the object. 
%The goal of this work is to maximize the number of heap objects that can be protected from memory errors for low cost, reducing the number of remaining objects requiring more complex and expensive defenses.  Towards this end, 
\redout{In \tool}\new{To ensure the security guarantee (Section~\ref{sec:introduction}) of \tool}, we aim to: (1) statically validate the heap objects whose accesses must satisfy spatial and type safety, \new{which form its {\em allocated-type} (Definition~\ref{def:alloctype}),} and (2) prevent exploits at runtime from permitting accesses that violate the  \redout{corrupting the spatial or type safety}\new{allocated-type of validated safe heap objects in (1), through enforcing {\em temporal allocated-type safety} (Definition~\ref{def:temporalalloctypesafety}) and isolation.} 
%Many exploits of temporal errors violate spatial and/or type safety, so 
\redout{Our goal is to enforce temporal allocated-type safety while preserving spatial and type safety of the validated safe heap objects
% heap object is safe if all accesses to the heap object must satisfy spatial and type safety statically and a 
%form of temporal safety called 
%We find that protecting heap objects against all classes of memory errors can be achieved efficiently by isolating memory-safe objects (i.e., as target objects) from memory errors on accesses to unsafe heap objects. In this paper, we propose the \tool system that accomplishes the task by enforcing 
In temporal allocated-type safety, a memory region can only be 
used to access objects of one declared size and type for all fields, which we call an {\em allocated-type} (Definition~\ref{def:alloctype}).}\new{The \tool allocator 
%enforces temporal allocated-type safety \new{(Definition~\ref{def:temporalalloctypesafety})} for objects whose accesses satisfy spatial and type safety, ensuring 
ensures that only safe objects with identical memory layout of sizes and types are (re)allocated in the same memory locations.}
\redout{By enforcing temporal allocated-type safety on objects that are validated to be spatial and type safe, 
%memory safety of those objects can be preserved without any runtime checks. 
memory reuses are restricted to the same allocated-type, preventing exploits of temporal errors that rely on type confusion}\new{Memory (re)uses are restricted to the same allocated-type for validated safe heap objects, preventing temporal exploits that corrupt their allocated-type\footnote{\new{Similar to temporal type safety~\cite{tat}, temporal allocated-type safety does not prevent exploits that reuse memory of different objects of the same allocated-type.}}}. 
\vspace{-0.05in}
\begin{definition}
\label{def:alloctype}
The \textbf{allocated-type} of a heap object (consists of $n$ fields) is a tuple $(S, T, \{(f_i, s_i, \tau_i, o_i)\}_{i=1}^n)$ where:
\begin{itemize}[leftmargin=*]
  \item $S$ is the declared total size of the object in memory.
  \item $T$ represents the declared type of the object.
  \item $\{(f_i, s_i, \tau_i, o_i)\}_{i=1}^n$ is the set of quadruples for each field $f_i$ of the object, where $s_i$ is the size of field $f_i$, $\tau_i$ is the type of field $f_i$, and $o_i$ is the offset of $f_i$ in the object.
  \item No two fields overlap, i.e., for any pair of distinct fields $f_i$ and $f_j$ with offsets $o_i$ and $o_j$, sizes $s_i$ and $s_j$, respectively, the intervals $[o_i, o_i + s_i)$ and $[o_j, o_j + s_j)$ do not intersect.
\end{itemize}
\end{definition}
\vspace{-0.1in}
\begin{definition}
\label{def:temporalalloctypesafety}
\new{\textbf{Temporal allocated-type safety} is a property that requires a memory region be used to access objects of one allocated-type. Memory reuses are restricted to objects of one declared size and type for all fields, without partial overlapping.}
\end{definition}
\vspace{-0.08in}
Figure~\ref{fig:overview} shows an overview of the \tool approach. 
%to validate and enforce memory safety for heap objects.   
This approach must: (1) validate that {\bf\em accesses to all the aliases of heap objects} must satisfy memory safety 
%(i.e., using temporal allocated-type safety for heap aliases), 
(2) validate that {\bf\em all accesses to a heap object} through all of its aliases must satisfy spatial and type safety, and (3) enforce temporal allocated-type safety on heap objects found to satisfy (1) and (2) to prevent exploitation of temporal errors by preserving spatial and type safety in all allocations.  First, \tool collects all possible aliases of heap objects conservatively.  Aliases may reside in the stack, heap, or global regions, so \tool validates heap aliases and leverages the DataGuard analysis~\cite{dataguard} for stack aliases and adapts that analysis for global aliases conservatively (Section~\ref{subsec:alias}).  Second, \tool validates that all uses of aliases to access heap objects must satisfy spatial and type safety.  % Aliases that are not used in pointer arithmetic or type cast operations are trivially safe (Section~\ref{subsec:validationoverview}). Third, 
\tool validates spatial safety considering the impact of reallocations (Section~\ref{subsec:spatialvalidate}) and multi-threading (Section~\ref{subsec:multithread}), and type safety  considering the impact of compound type casts and reallocations (Section~\ref{subsec:typevalidate}).  \tool uses symbolic execution to remove false aliases and infeasible unsafe paths to validate false positives due to the over-approximations of static analysis (Section~\ref{subsec:applyse}). Third, heap objects validated to satisfy spatial and type safety are allocated on a single isolated safe heap that enforces temporal allocated-type safety (Section~\ref{subsec:temporal}). Unsafe heap objects are isolated on an unsafe heap, and existing protection can be applied to the unsafe heap.

The \tool approach for protecting heap objects draws inspiration from the \emph{Anna Karenina Principle}~\cite{AKP-wikipedia}\footnote{From the opening line in Tolstoy's \textit{Anna Karenina}~\cite{AnnaKarenina-Tolstoy}: ``All happy families are alike; each unhappy family is unhappy in its own way."}. 
%Specifically, statically validated safe heap objects and associated accesses are the same in terms of \textit{always} fulfilling \textit{spatial} and \textit{type} safety guarantees. 
%As a result, those objects can be protected in one isolated  with the secure allocator to preserve  {\em temporal type safety}~\cite{safecode,tat} with secure initialization without the need for runtime checks. 
\tool aims to maximize the number of safe target objects (i.e., the ``happy family") that may be co-located in the isolated {\em safe heap} (i.e., creating the largest ``happy family") by (1) validating spatial and type safety statically, 
%Statically validated safe heap objects and associated accesses are the same in terms of \textit{always} fulfilling \textit{spatial} and \textit{type} safety guarantees, 
and (2) enforcing temporal allocated-type safety at runtime to prevent illicit accesses caused by temporal errors.  Despite the complexity in structure (e.g., compound types) and usage (e.g., reallocation and multi-threading) of heap objects, we find that many allocation sites only produce heap objects whose accesses can be proven to satisfy spatial and type safety.  No runtime checks are necessary on safe objects, as \tool's heap allocator ensures that accesses to safe objects will satisfy temporal allocated-type safety. \tool isolates these safe objects from illicit accesses from the unsafe heap via software-fault isolation.  %allocated on the safe heap can be protected against illicit accesses caused by all classes of memory errors without runtime checks. Isolation of unsafe heap guarantees that memory errors in unsafe heap never affect anything outside.
%objects in the unsafe heap can be protected through other defenses efficiently.

\vspace{-0.08in}
\section{Threat Model}
\label{sec:threatmodel}

We assume that every program protected by \tool may have any classes of memory errors on the heap. We also assume that every heap object that \tool has deemed unsafe may have memory error.  We assume adversaries will try to attack any unsafe heap object to corrupt safe heap objects. We aim to protect the safe heap objects from being affected by such attacks by the construction of \tool. Protection of unsafe heap objects can be realized by applying existing defenses on \tool's unsafe heap objects only.

\redout{To ensure that our static analyses truly examines an \\overapproximation of the possible program executions} \new{To ensure that the CFG we analyze is an over-approximation of program executions to guarantee the soundness of \tool's static analyses}, we assume the presence of CFI~\cite{cfi}, \new{so that attackers cannot leverage the remaining unmitigated exploits by \tool to synthesize malicious control flows beyond the expected CFG.} We further assume the code memory is not writable, and the data memory is not executable~\cite{dep}. To ensure that \tool\redout{'s static safety validation and runtime allocation} operates correctly, we assume the computed safety constraints are protected from tampering, and the underlying allocator (i.e., TcMalloc) is free from flaws and maintains the integrity of its state and metadata.

%\input{5_Design}
%auto-ignore

\section{Design}
\label{sec:design}

 In this section, we detail \tool's design to perform the tasks outlined Figure~\ref{fig:overview}. Soundness\footnote{As defined by the static analysis community~\cite{yannis18ecoop}, sound analysis over-approximates the program's executions. \new{We also summarize soundness in Section~\ref{subsec:sound}}.} (ensuring that no unsafe objects are misclassified as safe) is discussed individually in each section.

\subsection{Collecting and Validating Aliases}
\label{subsec:alias}

All aliases that may alias each heap object must be identified to validate that all possible accesses to each heap object satisfy spatial and type safety and to validate the memory safety of accesses to the aliases themselves. \new{Assume program executions are restricted to the computed CFG via CFI (Section~\ref{sec:threatmodel})}, \tool computes an overapproximation of all aliases from the stack, heap, and global regions for each heap object inter-procedurally in a context and flow-sensitive way, using static value-flow analysis~\cite{svf} (SVF) on a program-dependence graph~\cite{Liu17CCS} (PDG) representation. SVF and PDG are both claimed to be sound\footnote{Computing global aliases may suffer from scalability and precision issues for large programs such as Linux Kernel~\cite{hybridglobalalias}.  We leave adoption of the proposed technique to reason about global aliases for future work.}. SVF~\cite{svf,sui2012issta,sui2014tse,sui2016fse,sui2018tse} uses {\em heap cloning}~\cite{Lattner2007MakingCP,javaheapcloningase,javaheapcloningpldi} for flow-sensitive aliases analysis~\cite{Barbar2020FlowSensitiveTH} at intra-procedural level, the result is fed into PDG~\cite{Liu17CCS}, which uses a parameter-tree model to compute context-sensitive aliases inter-procedurally.  

A challenge is to verify the safety of all aliases to each heap object, regardless of whether they are stored in the stack, heap, and global region.
The memory safety of accesses to stack aliases (67.29\% of all aliases) is validated using DataGuard~\cite{dataguard}.  Heap aliases (31.88\% of all aliases) will be validated by \tool as heap objects as described in the remaining subsections. 
% global - what's the overall story?
Global aliases make up only 0.83\% of the aliases in all the tested benchmarks\footnote{Using global aliases to reference heap objects raises both security and performance concerns~\cite{duck2018effectivesan,lowfat-heap,hybridglobalalias}, and is thus explicitly inadvisable in much production software (e.g., a stated in reference to the Linux kernel~\cite{kernelCodingStyle}, Chrome~\cite{googleCppStyleGuideGlobals}, and Nginx~\cite{nginxCommonPitfalls}).}, so we provide a conservative approach.  Global aliases are either: (a) singletons, (b) fields of global objects that only have singleton fields, or (c) fields of global objects that have fields of compound types.  
%Global aliases of (c) may be corrupted through memory errors in other compound fields.  
With \tool, we validate global aliases of classes (a) and (b), which constitute over 87.3\% of all global aliases, as follows.  First, no spatial errors are possible for accesses to singleton objects or objects with only singleton fields (i.e., no pointer arithmetic except field access).  Second, \tool detects any type casts of these global aliases, as described in Section~\ref{subsec:typevalidate}.  Third, \tool requires global aliases to be initialized immediately following their declaration. Uninitialized global aliases are deemed unsafe to avoid temporal errors. Since protecting global memory is not a contribution of this work, we assume SFI techniques may be applied to protect safe global aliases from being corrupted by memory errors on accesses to unsafe global memory similar to how we isolate the safe heap, but implementing that is future work.

\vspace{-0.2in}
\subsection{Spatial Safety Validation}
\label{subsec:spatialvalidate}

A heap object is validated to be {\em spatially safe} if all of its aliases must only access memory locations within the object's memory region, defined by its size.  \tool requires that all heap objects must be declared to have or always be bounded by a constant size, which forms the {\em spatial constraint} checked in spatial safety validation.  If an object's size is not a constant, the object is classified as unsafe.
%due to the failure to determine its spatial constraint. 
%Spatial constraint of heap object must be collected before validating complied accesses. For spatial constraint collection, \tool requires the heap objects must be declared to have a constant size (or always be bounded by a constant), otherwise the object is unsafe for spatial constraint collection failure. 

\begin{algorithm}[t]
\DontPrintSemicolon
\small
\KwIn{\textbf{object} - the heap object to be validated}
\KwOut{\textbf{classification} of the object (safe or unsafe)}
\SetKwBlock{Begin}{function}{end function}
\Begin(SpatialValidation{(}object{)})
{
    \For{each alias of object}
    {
        $\textit{alias.index $=$ def(alias)}$ \;
        \uIf{(alias.index $\geq$ object.size}
            {
                object.safety(unsafe) \textbf{   return}\;
            }
    
        \uIf{SingleThread(alias)}
        {
            \For{each {\bf use} of alias to {\bf access} object}
            {
                \uIf{IsConstant(offset)}
                {          
                    \uIf{use.offset < 0 {\bf or !}(0 $\leq$ (alias.index + use.offset) $<$ object.size}
                    {
                        object.safety(unsafe) \textbf{   return}\;
                    }
                    \uIf{Increment\_Index(use))}
                    {   
                            $\textit{alias.index $+=$ use.offset}$\;
                    }  
                }
            }
        }
        \uElseIf{MultiThread(alias)}
        {
            MultiThreadValidation(object, alias)\;
        }
    }
    object.safety(safe) \textbf{   return}\;
}
\caption{Spatial Safety Validation for Heap Objects}
\label{algo:spatial}
\end{algorithm}

\setlength{\textfloatsep}{5pt}

CCured~\cite{ccured} finds that only aliases used in pointer arithmetic operations may violate spatial safety, \tool applies sound context-sensitive~\cite{sharir-pnueli} value-range analysis~\cite{value-range-book,Douglas11SBLP,dataguard} to validate the spatial safety of each heap object using Algorithm~\ref{algo:spatial} for all pointer arithmetic operations on each of its aliases. Each alias is checked to comply with the collected constraints (i.e., {\em size}) for all accesses with an {\em index} (i.e., a position of the object between 0 and {\em size}) along the def-use chain \new{(line 3)}. When referencing an object through an alias, the reference may use {\em offset} (i.e., in pointer arithmetic) to change the index. The initial index must be less than the size of heap object (line 4). The offset is required to be constant and the index+offset must always be
%greater than 
between 0 and the size of the object for every use
%to prevent underflow 
\new{(line 5-9)}. 
\tool classifies an object as unsafe if any access uses a negative offset to prevent underflows, as the PDG overapproximates possible executions.  
%The offset can be used to either change the alias (line 7) or access the object through index+offset (line 9), where the index in the former case must be updated. 
The index may be incremented by the offset in some cases \new{(line 10-11)} and needs to be updated in such cases.  
%For all uses of aliases to access the heap object, \tool checks that index is always greater than 0 and less than the declared size, otherwise, the heap object is unsafe. 
The validation for multithreading \new{(line 13)} is discussed in Section~\ref{subsec:multithread}. %\trent{Is case at line 11 necessary?  Not mentioned.  Is there a case where we just access at the index w/o an offset?} \kaiming{Yes, line 11 captures when we access out of bounds}

The challenge is that spatial safety validation must account for reallocations since heap objects can be {\em dynamically-resized} through reallocations. Previous approaches~\cite{duck2018effectivesan,baggybounds,softbound,lowfat,tdi,camp24usenix} treat reallocations as new allocations, ignoring the impact of unsafe reallocations on accesses using the original aliases. Reallocation can compromise accesses through the original aliases by: (1) shrinking the object's size, as accesses using the original aliases to the reallocated object may exceed the new/reallocated bounds, and (2) moving the object to a different location, which can turn the original aliases into dangling pointers if not managed correctly. For case (1), \tool classifies the objects whose sizes are reduced through reallocations (which is rare) as unsafe. \tool limits the cases objects can be reallocated safely only to those that extend the size of the last field or append fields to the object at the end of its original data type (which are common cases for reallocation).  Aliases to the reallocated object will be analyzed (i.e., for their def-use chains) using the newly reallocated size 
%Compliant reallocations will assign the object a new allocated-type with a new size {\em newsize}, 
as the {\em size} in Algorithm~\ref{algo:spatial}.  
%Upon reallocation, {\em newsize} is collected and updated as the size constraint for the def-use chain from that point forward for the new aliases only. As a result, accesses using the original aliases of the heap object are naturally spatially safe on the new size constraints after reallocation. \trent{but the original aliases reference the old memory location, so shouldn't they be restricted to the old size.  Are they?} \kaiming{Yes, the intention of the above sentence is to show that no extra validation is needed for the original alias after we update the size constraints. Kind of redundant I agree if that is obvious.}\trent{I read this as the "newsize" is applied to the original aliases, which would not be correct.}  
For case (2), accesses using the original (possibly dangling) aliases will be evaluated using the original size for their def-use chains to detect an spatial errors relative to the original object memory rather.  The temporal safety concerns for case (2) are addressed in Section~\ref{subsec:temporal}.  
%\trent{put the temporal discussion closer to the discussion of the old aliases.}

%Heap objects after reallocations are assessed separately from the original heap objects. Different from previous approaches~\cite{duck2018effectivesan,baggybounds,softbound,lowfat,tdi,camp24usenix}, 
%Uriah propagates the new spatial constraints and safety classifications of the reallocated object to all aliases of the original object, enabling spatial validation using the updated constraints. 

%We first use a real-world example to illustrate the challenges in spatial safety validation for heap objects, then introduce how \tool deals with those challenges.

\begin{algorithm}[t]
\DontPrintSemicolon
\small
\KwIn{\textbf{object} - the heap object used in concurrent context}
\KwIn{\textbf{alias} - the alias to the heap object}
\KwOut{\textbf{classification} of the object (safe or unsafe)}
\SetKwBlock{Begin}{function}{end function}
\Begin(MultiThreadValidation{(}object, alias{)})
{
    \uIf{$\textit{IsShared(object)}$}
    {
        \uIf{{\bf !}IsConstant(Threads\_Set)}
        {
            object.safety(unsafe) \textbf{   return}\;
        }
        \For{each {\bf use} of alias used in {\bf multiple threads}}
        %\uIf{$\textit{Used\_In\_Multiple\_Threads(alias)}$}
        {

            \For{each {\bf thread} in Threads\_Set}
            {
                \uIf{use.offset < 0 {\bf or !}(0 $\leq$ (alias.index + use.offset) $<$ object.size}
                {
                    object.safety(unsafe) \textbf{   return}\;
                }
                \uIf{Increment\_Index(use))}
                {   
                        $\textit{alias.index $+=$ use.offset}$\;
                }   
            }
        }
    }
    classify(object, SAFE) \textbf{   return}\;
}
\caption{Spatial Safety Validation for Shared Objects}
\label{algo:multithread}
\end{algorithm}
\vspace{-0.15in}
\subsection{Spatial Validation with Concurrency}
\label{subsec:multithread}

As heap objects may be {\em shared} among threads (i.e., can be accessed concurrently in multiple threads), spatial safety validation of \tool must account for concurrent accesses to heap objects through aliases while indexes may be modified in different threads\footnote{Race conditions, as not memory errors, are not discussed in this paper.}. The problem is to identify the heap objects that may be accessed concurrently and then reason about the safety of such concurrent accesses conservatively.
%such that no unsafe concurrent accesses will be misclassified as safe.  
%The lack of Rust {\em ownership}~\cite{rustownership} which simplifies such identification makes the shared heap objects and their aliases in C/C++ hard to track.  
While SVF provides intra-thread CFGs~\cite{svfmta} that identify the aliases accessed 
%and their aliases used 
in functions that may be executed concurrently in multiple threads, such a method introduces a large number of false positives, as many heap objects 
%and aliases 
accessed in the intra-thread CFG may not be used concurrently (i.e., are thread-local). The challenge is to remove false positives while still maintaining an overapproximated set of objects that may be accessed concurrently.

\tool identifies the shared heap objects from the heap objects accessed in an intra-thread CFG by determining whether they are assigned to any alias that may be used concurrently in multiple threads.  Such aliases should be used in intra-thread CFG and either: (1) are on the heap, (2) are in global memory, or (3) are on the stack and passed as parameters of thread creation API. Other aliases used in the intra-thread CFG are thread-local. Since the construction of inter-procedural, intra-thread CFGs and SVF aliasing are claimed sound~\cite{svfmta, threadcfg}, \tool over-approximates the heap objects that may be shared and the aliases that may access them concurrently.

%Accesses of thread-local objects/aliases do not affect other concurrent contexts, 
To reason about spatial safety conservatively, updates to the aliases of shared heap objects must consider whether the aliases may be used concurrently in multiple threads or not.  
Algorithm~\ref{algo:multithread} shows the spatial validation of shared heap objects among threads. Accessing shared heap objects consists of two cases, depending on whether the alias is: (1) thread-local 
%\zhiyun{I didn't see this term used in the algorithm (better use consistent naming to make it easy for readers} \kaiming{Thanks, this part was not included in Algorithm 2 but instead using the method in algorithm 1. However, we cannot use thread-local to describe aliases in algorithm 1 since there are many alias are not involved in multithreading at all. I added a sentence below to say that (1) is omitted from algorithm 2.} 
or (2) used in multiple threads.
%\zhiyun{for some reason, I was mistakenly thinking that we should reason about the objects as opposed to aliases. But then I realize we are not trying to reason about race condition bugs where the same object can be read/write by different threads.}\kaiming{Agreed, maybe we need to say it explicitly as concurrency bugs (i.e., race conditions) are not memory errors.} \zhiyun{good idea. I also think it may be helpful to tell people what kind of concurrency problems we are looking at, e.g., index may be modified by a different thread when used to access an array? This was not quite clear to me and I don't have a very good intuition even after reading the algorithms.}
%(i.e., shared). 
We observe that case (1) is the common method for accessing shared heap objects in production software. For example, Nginx and httpd create thread-local copies of aliases to access shared heap data. For case (1), since the alias is thread-local, aliases in different threads are used independently and can be examined 
%by \zhiyun{by can be removed} 
using Algorithm~\ref{algo:spatial} discussed in Section~\ref{subsec:spatialvalidate} (thus excluded in Algorithm~\ref{algo:multithread}).
%The {\em index} will not be synchronized (i.e., remains per alias) since the alias is local data.  
For case (2), where the number of threads can be concretized statically \redout{(line 5)}, we extend the value-range analysis to calculate the index by accumulating the access range \new{(line 5-10)} across memory operations in all threads.  
%\trent{Does this buy us many safe cases?  Or are we simply declaring that all shared aliases that update their indices are unsafe?}  \kaiming{Yes, since in many cases the number of threads can be determined, and there are only limited number (i.e., 1 or 2) indexes updating among such cases}. 
For example, if 2 threads access the heap object through incrementing the alias \redout{(line 5)}by 3 and 5, then the index calculated by \tool will be increased by 8. Accumulation only applies to accesses that increment the index of the alias using constant offsets. Other accesses (e.g., access to fields without changing the index of the alias) do not need to be accumulated. Scenarios where the index is incremented while the number of threads cannot be determined statically are unsafe \new{(line 3-4)}. Thread-local objects are validated as described above in Section~\ref{subsec:spatialvalidate}. If the calculated index is greater than the size of the heap object, then the object is unsafe \new{(line 7-8)}.

For reallocations on shared heap objects, \tool relies on the common convention that the reallocation operation is atomic\footnote{POSIX realloc() employs mutexes internally to protect the memory management metadata to avoid data corruption in races~\cite{posixrealloc}. Intel suggested programmers adopt thread-safe memory management API~\cite{intelmultithread}, e.g., Intel oneTBB~\cite{intelonetbb}. Allocators~\cite{tcmalloc,ptmalloc,jemalloc} have been designed to leverage thread-local storage/cache and per-thread freelists. Applications such as Nginx and Apache Httpd utilize atomic reallocation operations and tend to only use realloc on thread-local data.} (i.e., reallocation and memory access to the same heap object cannot be executed concurrently in multiple threads). Spatial constraints are updated for all threads upon reallocation.  \new{Shared heap objects are unsafe if reallocated to variable size in concurrent contexts.}\redout{Reallocation of a shared heap object to a variable size in concurrent contexts results in the object being classified as unsafe. }

\subsection{Type Safety Validation}
\label{subsec:typevalidate} 

%\zhiyun{I haven't thought carefully about this, but is it possible to apply further optimizations like the ones we did in Yizhuo's usenix security 24 paper to further identify spatially-safe heap objects? If so, it may be something worth mentioning in the discussion section (and in general it would be nice to discuss how one can further improve the precision of static analysis).}\kaiming{Agreed, plus the discussion is kind of behind the current version. I assume you mean to identify the type safe heap object. If it is the case, then definitely optimizations like Yizhuo's work that identifies programmer-implemented checks and ensures their correctness would make it possible to identify more type safe heap objects further.} \zhiyun{Yes that is what I meant. It would also be nice to look at how one can improve other kinds of analysis (beyond type-safety)}\kaiming{Sure}
As defined in Section~\ref{subsec:exploitheap}, A heap object is considered {\em type safe} only if all of its aliases only access objects of the same data types at each offset per the definition of its allocated-type, see Definition~\ref{def:alloctype}.  The task is to identify the types used by aliases to access the object and verify if type casts among these types ensure type safety of all accesses. %Prior work only considered type safety in limited ways. 
CCured~\cite{ccured} finds that only aliases used in type cast operations may violate type safety, and identified that {\em upcast} must be safe~\cite{ccured-toplas}.  They define an {\em upcast} is a type cast from type $TN$ to type $T$ when the layout of $T$ in memory is a prefix of the layout of $TN$.  
%\trent{Is this CCured def or ours?  They only consider the prefix embedded within the type, so that is different than us.}  \kaiming{This is exactly what in the CCured.} 
At the time of the CCured work (around 2005), they claimed that 63\% of type casts are between identical types, and of the remaining casts, 93\% are safe upcasts.  Thus, a significant benefit may be seen by validating safe type casts statically.  However, SAFECode conservatively classifies all objects in type casts as unsafe, CCured only applied its approach to casts where the type $T$ is an explicit subtype in $TN$, ignoring cases where the subtype $T$ does not physically exist, but instead of its all its fields are present as a prefix of type $TN$, as such cases are prefixes using allocated-types.
%\trent{only consider safe upcasts for types contained within the object. NOTE: how to say this clearly?} \kaiming{How about CCured }
Moreover, in recent systems, only C++ upcasts are classified ~\cite{hextype,caver,typesan, duck2018effectivesan,tprunify} by leveraging class hierarchy and RTTI.  Thus, redundant runtime checks are still applied to casts that can be validated as type safe.

\begin{algorithm}[t]
\DontPrintSemicolon
\small
\KwIn{\textbf{object} - the memory object to be validated}
\KwOut{\textbf{classification} of the object (SAFE or UNSAFE)}
\SetKwBlock{Begin}{function}{end function}
\Begin(TypeValidation{(}object{)})
{
    \For{each type cast cast(TN, T) for each alias of object}
    {
        \uIf{IsIntegerCast(cast)}
        {
            DataGuardTypeValidation(TN, T)\;
        }
        \Else
        {
            \uIf{$T!= TN$}
            {
                ${layout_{TN}} \gets {\textit {GetTypeLayout}(TN)}$\;
                ${layout_{T}} \gets {\textit{GetTypeLayout}(T)}$\;
                ${prefix_{TN}} \gets {\textit{GetPrefix}(layout_{TN}}, {\textit{sizeof}(T))}$\;
                \For{{\bf each field} $f_{T}$ {\bf in} {$layout_{T}$}}
                {
                    $f_{TN} \gets \textit{GetField}  (f_{T}, prefix_{TN})$\;
                    \If{$offset(f_{T}) \neq offset(f_{TN}) {\bf  or } !f_{TN}$}
                    {
                        object.safety(unsafe) \textbf{   return}\;
                    }
                }
            }
        }
    }
    object.safety(safe) \textbf{   return}\;
}
\caption{\footnotesize Type Safety Validation for Heap Objects and Aliases}
\label{algo:type}
\end{algorithm}

We find that any upcast where the type resulting type is an exact prefix of the allocated-type satisfies temporal type safety~\cite{tat} for allocated-types, providing more opportunities for type-safe casts.  To identify safe type casts\footnote{We consider all C and C++ type casting operations. C++ offers 5 types of casts: {\em static\_cast}, {\em dynamic\_cast}, {\em reinterpret\_cast}, {\em const\_cast}, and {\em C-style\_cast}. Among them, {\em dynamic\_cast} and {\em const\_cast} have no security concerns~\cite{hextype,caver,typesan}, {\em C-style\_cast} is translated to other casts. We concentrate on validating {\em static\_cast} and {\em reinterpret\_cast}.}, \tool performs type safety validation following Algorithm~\ref{algo:type} for each type cast operations for all aliases of the heap object. DataGuard~\cite{dataguard} only validates safe type casts among integers types.  \tool reuses DataGuard's type safety validation for integer casts while focusing on casts among compound types \new{(line 2-4)}. \tool validates compound type casts to determine if two data types are {\em compatible}, which we call the {\em compatible-type-cast analysis} \new{(line 6-14)}. Two types in a cast are compatible types if the types are identical or the cast is an upcast. Other type casts are considered unsafe. Initially, all heap objects and their aliases are type safe. 
%\new{(e.g., {\tt ngx\_core\_module\_t} is the physical subtype of {\tt ngx\_event\_module\_t} in Listing~\ref{lst:typeexample}). 
Different from C++ where $T$ usually serves as a field of $TN$ due to inheritance and coercion, we observed that in many cases, $T$ is not contained in $TN$ for C programs, but instead all of its fields are. 
The compatible-type-cast analysis does not require $T$ to be contained in $TN$, as long as all the fields of the $T$ are at the same offsets and of the same sizes and types in $TN$, cast from $TN$ to $T$ ensures that an access at any offset within $T$ references the same type \new{(line 10-13)}. Also, it is worth mentioning that type safety validation on shared heap objects among threads is validated similarly since the safety of type cast is regardless of whether it can be executed concurrently.  The compatible-type-cast analysis is built on the sound definition of upcasts in CCured~\cite{ccured-realworld,ccured-toplas}. Our validation is conservative, requiring the types to match concretely and exactly to retain soundness.

Type safety validation is complicated by several challenges: (1) heap objects can be dynamically typed (e.g., through polymorphism in C++); (2) interchange of {\tt void*}, {\tt char*}, and other types (e.g., {\tt memcpy} takes {\tt void*} as parameter for arbitrary types); (3) heap memory is not always assigned concrete types upon allocation (e.g., the {\tt auto} keyword may result in the pointer being assigned {\tt void*}); (4) reallocations change original allocated-type (e.g., by extending the length of a field), \new{and (5), memory safety concerns of using unions}.  These challenges may make it harder to determine the correct memory layout of an object. \tool resolves challenges in determining an object's type (i.e., cases (1-3) by delaying type assignment. \tool does not consider {\tt void*} or {\tt char*} as a concrete type. For heap objects whose allocated-type cannot be concretized or aliased by {\tt void*} or {\tt char*} pointers, \tool delays assignment of a concrete type until the object is interpreted as a specific type. If the type cannot be concretized statically, such objects are classified as unsafe.  For (4), \tool only allows reallocations that extend the size of the last field or append fields to the object at the end of its original data type (also discussed in Section~\ref{subsec:spatialvalidate}). For complied cases, a new allocated-type is assigned to the heap object accordingly. \new{For (5), LLVM represents unions as structures. Union members are declared as distinct SSA variables~\cite{llvm_union} through type casts from such structures to the member. Uriah validates such casts using type safety validation before validating all following operation.}

\vspace{-0.15in}
\subsection{Symbolic Validation}
\label{subsec:applyse}

Because \tool's static analysis over-approximates the possible executions, it may classify a heap object as unsafe that could really be safe, producing false positives. Researchers have identified two approaches to remove false positives found from static analysis using symbolic execution: (1) executing all paths to verify legal executions~\cite{dataguard}, and (2) pruning infeasible paths until only compliant paths remain~\cite{Zhai2020UBITectAP,increlux}. Using (1) for the heap, we find that many false positives are generated by infeasible aliasing (i.e., pointers cannot be defined to reference the object) and operations (i.e., pointers uses cannot access the object), due to the over-approximation of static analysis, indicating the executions that result in the objects to be classified as unsafe is not feasible (i.e., infeasible path).

%\tool uses symbolic execution to remove such false positives to increase the number of statically validated heap objects.   First, prior work on stack object safety~\cite{dataguard} uses symbolic execution in (1) to follow all the def-use chains found in static analysis to validate that only legal executions are possible. If after removing infeasible aliases and operations, all the remaining aliases of a heap object can be validated to be safe for spatial and type safety for the remaining operations, the heap object is reclassified as safe.  

%are generated by the termination caused by exceeding depth limit, since the depths of the def-use chains for many heap objects exceed these limits, meaning that only a modest fraction of cases can be validated this way. 

Instead, on top of (1), \tool applies symbolic execution in (2) to prune infeasible paths. Specifically, \tool removes infeasible paths until all unsafe aliases can be removed as false positives.  False positives may occur because: (1) an object cannot be assigned (i.e., defined) to an alias on a path with an unsafe operation (i.e., infeasible definition); (2) an object cannot be used by an alias in an unsafe operation (i.e., infeasible use); and (3) the path cannot be executed in a manner that causes the unsafe operation (i.e., infeasible path).  First, to prune infeasible definitions, \tool symbolically executes each unsafe alias from its declaration to definitions that involve later unsafe uses to determine whether the definition is reachable in this path.  Second, to prune infeasible uses, \tool symbolically executes from the pointer definition to its unsafe uses following its def-use chain.   In both of these cases, infeasibility is detected by the failure of the symbolic execution engine (S2E~\cite{s2e}) to generate path constraints.  Finally, the infeasible paths are detected by evaluating the path constraints on the def-use chains.\redout{\tool limits the depth of functions executed symbolically to avoid path explosion and classify heap objects in such cases as unsafe.} All possibly unsafe paths are considered before an object is reclassified as safe, preventing any unsafe object from being classified as safe.  If after removing infeasible aliases and operations, all the remaining aliases of a heap object can be validated to be safe for spatial and type safety for the remaining operations, the heap object is reclassified as safe.  

\new{Using symbolic execution for memory safety validation suffers from scalability issues, such as path explosion. To resolve this, \tool applies LLVM's loop canonicalization~\cite{llvm-loop-canonicalization} and loop simplification and unrolling~\cite{llvm-loop-simplify,huang19oakland} features, as well as S2E's symbolic state merging~\cite{s2e-state-merging}. \tool limits functions executed symbolically via a configurable depth to avoid path explosion and classify heap objects in such cases as unsafe.}

\vspace{-0.05in}
\subsection{Safety Validation Soundness}
\label{subsec:sound}

%\trent{collect arguments here.  Gang should check.}

In this section, we assess the soundness\footnote{We use the term {\em soundness} by the static analysis community~\cite{yannis18ecoop}, where a sound analysis over-approximates the program's executions.} of \tool's memory safety validation. Validation must be sound to ensure that no unsafe object can be misclassified as safe.  \redout{Similar to DataGuard~\cite{dataguard}, \tool relies on {\em relative soundness}, meaning its static analyses are sound if the techniques it applies and the compositions are sound.} \new{Uriah’s analysis is sound {\em relative} to baseline tools, e.g., PDG~\cite{Liu17CCS}, SVF~\cite{svf}, all claimed but not formally proven sound. Uriah uses those analyses soundly, so assuming they are sound, and assuming the analyzed CFG truly overapproximated program execution (we assumed CFI, see Section~\ref{sec:threatmodel}), Uriah does not generate false negatives but may produce false positives. There is no available ground truth to quantify false positives, but Uriah is significantly more accurate (i.e., fewer false positives) than prior works in Table 9, indicated by the increased safe objects count.}
%This allows for more robust safety validation with over-approximation.
%Soundness must be retained in \tool to ensure identified safe objects do not contain unsafe objects. %that would mistakenly be included in the set of safe objects, which could compromise security.

{\bf Soundness of Analysis Foundations}: \tool employs a context-sensitive heap alias analysis, which consists of two approaches: (1) SVF alias analysis~\cite{svf,sui2012issta,sui2014tse,sui2016fse,sui2018tse} at the intra-procedural level, which utilizes {\em heap cloning}~\cite{Lattner2007MakingCP,javaheapcloningase,javaheapcloningpldi} soundly by incorporating type information within flow-sensitive pointer analysis~\cite{Barbar2020FlowSensitiveTH} and (2) PtrSplit PDG~\cite{Liu17CCS} to represent inter-procedural may-aliases, which uses the parameter-tree mechanism to convey aliases inter-procedurally.  Both are claimed sound. \new{Implementation faults are known to harm the soundness of any analyses in practice; e.g., SVF misses aliases in flow-sensitive interprocedural analysis for large programs.}

{\bf Soundness for Static Safety Validation}:
%To establish the potential index ranges when accessing heap objects
To validate spatial safety, we employ value-range analysis~\cite{value-range-book,Douglas11SBLP,baggybounds,dataguard}. This analysis is conducted using the PDG using a sound data-flow analysis~\cite{dataguard}. \tool obtains concrete size constraints, even for resizing and multi-threading. 
%The constraints on pointer declarations and/or definitions, along with concrete size allocations through allocation functions, contribute to this analysis. 
As the PDG and the data-flow analysis are sound, it is evident that \tool's value-range analysis is also sound. \tool validates the type safety of heap objects 
% TJ: nothing new and don't have room for this...
%for two cases: (1) \tool classifies the type casts on integer objects that never change their value as safe~\cite{dataguard}, which relies on the value-range analysis that has already been proven sound~\cite{value-range-book} and (2) \tool classifies 
for type casts between compatible types.
%on objects of structured types as safe. 
As shown in Section~\ref{subsec:typevalidate}, the compatible-type-cast analysis is built on the sound definition of upcasts in CCured~\cite{ccured-realworld,ccured-toplas}. %\tool also extends CCured's definition to cover more upcast cases. 
Our validation is conservative, requiring the types to match exactly to retain soundness.
%\trent{What about the second case?}
%Though our current methods of identifying upcast may not be complete (i.e., may have false negatives), 
\tool over-approximates the number of unsafe heap objects by directly classifying all heap objects without concrete constraints as unsafe, despite many of them might be safe, then validates the remaining using sound approaches discussed above. Thus, the static safety validation of \tool is sound.

{\bf Soundness of Symbolic Execution}: 
%Path explosion in symbolic execution often means that it is impractical to execute all paths in the program, even with loop canonicalization~\cite{huang19oakland} and symbolic state merging~\cite{s2e-state-merging}. 
\tool employs a depth limit to prevent costly scenarios (e.g., path explosion) in symbolic execution, where any terminated symbolic execution implies that the related heap object is unsafe. %Additionally, when conducting symbolic analyses, there might be a compromise on soundness by utilizing concrete values for certain variables~\cite{SurveySymExec-CSUR18, s2e}. \tool adapted the approach used in~\cite{dataguard}, in the spatial safety analysis, \tool only concretizes variables that hold constant values.
%\gang{If those are already constants, why do you need to concretize them?} \kaiming{Maybe I expressed it in the wrong way, my intention is to say only constant value remains constant, others are symbolic.}
%All other variables are initialized with symbolic values. 
\tool's symbolic verification follows the def-use chains of unsafe aliases from either: (1) a declaration to the definition associated with an unsafe use or (2) a definition to an unsafe use, based on Section~\ref{subsec:applyse}. Thus, all paths that could lead to memory error would be symbolically validated before considering heap object as safe.  %\trent{Different from DataGuard all symexec analyses must be complete before we can remove an infeable case.}

\subsection{\tool Runtime Allocation}
\label{subsec:temporal}

%\trent{remind of goal}
%\trent{limitations of prior - properties we want and do not}

%\trent{Too much intro.  Temporal type safety from Sect 3.  Enforced via type-safe memory reuse approaches.  Flaws: (1) by site; (2) spatial error within pool. }
\tool enforces {\em temporal allocated-type safety} for heap objects validated to satisfy spatial and type safety in an isolated, safe heap.  \tool's temporal allocated-type safety prevents temporal errors prior to initialization and after deallocation by initializing all pointers on allocation and enforcing type-safe reuse on reallocation, respectively.
%prevent the exploits caused by reusing dangling pointers, and (2) initialize all heap objects upon allocation to prevent use-before-initialization.
%(see task (4) in Section~\ref{subsec:limits}).  
Recall from Section~\ref{sec:insight} that temporal allocated-type safety requires that objects of only one allocated-type may be allocated in each memory region to prevent a dangling pointer from being used to access data of a different allocated-type.  However, there are several issues with the current approaches that prevent them from enforcing temporal allocated-type safety and preserving spatial and type safety.  We first outline \tool's allocator operations and then examine how it resolves issues in prior approaches to enforce spatial, type, and temporal allocated-type safety.

\vspace{-0.08in}
\subsubsection{The Safe Heap}
\label{subsubsec:safeheap}
$\newline$
\tool implements {\em temporal allocated-type safety} over objects in the safe heap. Table~\ref{tab:temporalops} describes the API of \tool's allocator for the safe heap.   First, \tool supports {\em allocation}, which allocates objects using type-specific freelists. All memory regions begin untyped and are assigned an allocated-type on the first allocation to that memory region.
%\footnote{Type-After-Type~\cite{tat} proposed a {\em sizeof-based} approach to distinguish {\tt char*} allocations, such a method may result in multiple types that have the same size be allocated in a same pool.}.  
Second, \tool supports {\em deallocation}, which places the memory region on a per-type free list to preserve the {\em allocated-type} of the memory region for future allocations.  \tool maintains the allocated-type with each allocation's metadata.  Third, \tool supports {\em reallocation}, which may change the size of the object allocated (i.e., changing the allocated-type). Reallocations that result in a new allocated-type of the heap object will result in the heap object being moved out from the original pool and allocated in the pool corresponding to the new allocated-type. Original aliases, though may be dangling pointers without proper handling, can only reference objects of the original allocated-type. Thus, temporal allocated-type safety is preserved. 
%The safe reallocations that \tool supports are (1) changing the size of the last field of the original data type (i.e., using constant values to pass spatial safety validation), and (2) appending extra fields to the original data type, which are the two common ways of dynamic sizing.

\begin{table}[t]
    \centering
    \footnotesize
	\scalebox{1}{
	\begin{tabularx}{8.5cm}{c|X}
	    \toprule
		\multicolumn{2}{c}{\bf\em \tool's Heap Allocator Operations}\\
		\midrule
		{\bf\em Allocation}& Allocate object X of allocated-type T to region labeled for T. X must be validated to satisfy spatial and type safety. \\
  %Upon an allocation request of a heap object X of type T, the allocation operation must ensure X is allocated in the free space within the allocation area solely dedicated for type T. \\
	    \midrule
		{\bf\em Deallocation}& Dealloc object X of allocated-type T, returning the memory region to the free-list of allocated-type T to restrict its reuse to T.\\
		\midrule
		{\bf\em Reallocation}&  Reallocates the object of original allocated-type T to new allocated-type T'. The object is deallocated in the memory for T and allocated in the memory for T'. \\
		\bottomrule
	\end{tabularx}}
    
	\caption{\footnotesize Operational semantics for the \tool safe heap allocator}
	\label{tab:temporalops}
     \vspace{-0.2in}
\end{table}

\begin{table*}[t!]
    \aboverulesep=0ex 
    \belowbottomsep=0.5ex
    \belowrulesep=0ex
    \abovetopsep=0.5ex
    \centering   
        \resizebox{0.9\textwidth}{!}{
	\begin{tabular}{l|r|rr|rrr|r|r}
	    \toprule
		 \multirow{2}{*} &\multirow{2}{*}{\bf\em Total}&\multirow{2}{*}{\bf\em VR-Spatial }&\multirow{2}{*}{\bf\em Uriah-Spatial}&\multirow{2}{*}{\bf\em CCured-Type}&\multirow{2}{*}{\bf\em CTCA-Type}&\multirow{2}{*}{\bf\em Uriah-Type\ }&\multicolumn{1}{c|}{\bf\em VR-Spatial+}&\multicolumn{1}{c}{\bf\em Uriah-Spatial+}\\
            
            &&&&&&&\multicolumn{1}{c|}{\bf\em CCured-Type}&\multicolumn{1}{c}{\bf\em Uriah-Type}\\
            \midrule
		{\bf\em Firefox}&26,162&19,857 (75.9\%)&20,432 (78.1\%)&14,101 (53.9\%)&19,700 (75.3\%)&20,040 (76.6\%)&12,270 (46.9\%)&18,392 (70.3\%)\\
		\midrule
		{\bf\em nginx}&954&705 (73.9\%)&785 (82.3\%)&585 (61.3\%)&766 (82.3\%)&819 (85.5\%)&521 (54.6\%)&744 (78.0\%)\\
            
		{\bf\em httpd}&1,074&662 (61.6\%)&816 (76.0\%)&825 (76.8\%)&918 (85.5\%)&942 (87.7\%)&575 (53.5\%)&760 (70.8\%)\\
		
		{\bf\em proftpd}&1,707&1,275 (74.7\%)&1,380 (80.8\%)&596 (34.9\%)&1,201 (70.4\%)&1,366 (80.0\%)&458 (26.8\%)&1,174 (68.8\%)\\
		%\hline
		%{\bf\em openvpn}&&&&&&&&&&&\\
		
		{\bf\em sshd}&378&270 (71.4\%)&310 (82.0\%)&170 (45.0\%)&284 (75.1\%)&304 (80.4\%)&144 (38.1\%)&274 (72.5\%)\\
		
		{\bf\em sqlite3}&761&614 (80.7\%)& 655 (85.7\%)&382 (50.2\%)& 567 (74.5\%)&587 (77.1\%)&316 (41.5\%)&513 (67.4\%)\\
	    \midrule
            {\bf\em SPEC2006}&---&{71.1\%}&{79.6\%}&{37.7\%}&{80.3\%}&{85.0\%}&{29.8\%}&{72.2\%}\\
            \midrule
            {\bf\em SPEC2017}&---&{68.9\%}&{83.4\%}&{41.8\%}&{79.0\%}&{83.9\%}&{28.2\%}&{75.6\%}\\
            \midrule
		{\bf\em AVERAGE}&---&{\bf 72.2\%}&{\bf 81.0\%}&{\bf 50.2\%}&{\bf 77.8\%}&{\bf 82.0\%}&{\bf 39.9\%}&{\bf 72.0\%}\\
		\bottomrule
	\end{tabular}
}
        \captionsetup{font=footnotesize}
	\caption{Incremental Safety Improvement of \tool, and Comparison with CCured. {\em Total} column shows the total number of heap objects (i.e., allocation sites). We omitted {\em CCured-Spatial} column since heap objects are always involved in pointer arithmetic, resulting in the {\em CCured-Spatial} to be around 0 for all benchmarks. {\em CCured-type} column shows the number of heap objects are not aliased by any pointer used in type casts. {\em VR-Spatial} column represents the number of heap objects passed value-range analysis, {\em Uriah-Spatial} column represents the number of heap objects passed the complete Uriah's static spatial safety validation. {\em CTCA-Type} column represents the number of heap objects passed Compatible-type-cast analysis, {\em Uriah-Type} column represents the number of heap objects passed the complete Uriah's static type safety validation. The {\em VR-Spatial+CCured-Type} column shows the number of heap objects passed value-range analysis and CCured-type analysis.  {\em Uriah-Spatial+Uriah-Type} column shows the number of safe heap objects passed Uriah's complete static safety validation. {\em SPEC 2006} and {\em SPEC 2017} rows show the average percentage, individual results of SPEC benchmarks are shown in Table~\ref{tab:OverviewEval}. {\em Average} row shows the average number of all tested benchmarks. }
	\label{tab:OverviewEval}
     \vspace{-0.25in}
\end{table*}

\vspace{-0.08in}
\subsubsection{Isolation from the Unsafe Heap}
\label{subsubsec:unsafeheap}
$\newline$
The unsafe heap region is built on top of the tcmalloc’s span and page heap memory management scheme. For 64-bit system, currently only 48 bits are used for addressing, with 1 bit to distinguish kernel and user space, up to 128TB memory can be used for user space. 
%The user memory space starts from 0x000000000000, ends at 0x7FFFFFFFFFFF. With BSS segment typically ends at 0x0000 08000000, that is the address where heap starts. In order to not interfere with the memory management of the safe heap region, which will be placed near the starting of the heap by Uriah’s allocator, we picked the starting address from 0x000000FF00000000 (~ 1TB for the safe region). 
%Given that most of the server/browser use less than 16 GB of memory upon their peak memory usage, 
\tool reserves 1TB for the unsafe heap region. 
%Thus, the unsafe heap region expands from 0x000000FF00000000 to 0x000000FFFFFFFFFF. 
\tool forces all access to unsafe objects to only access memory in the unsafe heap by performing bit-masking upon memory operations. %\trent{why not just memory operations?}\kaiming{We can change to memory operations, is that too vague?}.
%to ensure the pointers that alias objects in the unsafe heap region always point to unsafe heap region, 
Thus, even if a pointer is illicitly modified to an address outside of the unsafe heap, \tool will restrict the pointed address to within unsafe heap through bit-masking.  Thus, any memory errors on operations to objects located in the unsafe heap cannot be exploited for crafting pointers that reference memory outside the unsafe heap region.

\vspace{-0.08in}
\subsubsection{Security Implications of the Safe Heap}
\label{subsubsec:secimpl}
$\newline$
{\bf Spatial Safety}: SAFECode~\cite{safecode} and Type-After-Type~\cite{tat} do not enforce spatial safety within the heap.  SAFECode uses per-type heaps, so spatial errors are limited to objects of same data type, but they recommend additional runtime checks.  Other systems~\cite{tdi} allocate objects separated by guard pages, but spatial errors can evade guard pages.  \tool's static safety validation ensures that any object added to the safe heap must satisfy spatial safety. 

{\bf Type Safety}: SAFECode~\cite{safecode} enforces type safety by only pooling objects of the same data type into each per-type heap.  However, SAFECode's method for reusing per-type heaps leads to significant overheads~\cite{intelmpxperf}.  \tool's static safety validation ensures that any object added to the safe heap must satisfy type safety.  Objects of multiple types can be added to the same safe heap (i.e., in different locations) since they are guaranteed to satisfy spatial safety. 

{\bf Use-Before-Initialization (UBI)}: UBI is possible on the stack, heap, or global aliases of heap objects. 
For stack pointers,  \tool leverages DataGuard~\cite{dataguard} to ensure that safe stack aliases are never used before initialization. For heap pointers, current techniques to enforce temporal type safety~\cite{safecode,tat,camp24usenix} do not explicitly prevent UBI attacks on heap pointers, techniques to zero memory on initialization~\cite{safeinit} is expensive.  However, \tool's heap allocator is built on top of TcMalloc, which already zeroes memory upon request from the OS~\cite{tcmalloczero1, tcmalloczero2}. \tool detects uninitialized global aliases and classifies the aliased heap object as unsafe (Section~\ref{subsec:alias}).

{\bf Dangling Pointers}: Another memory safety problem occurs due to the reuse of dangling pointers after reallocation/deallocation. Dangling pointers to the reallocated/deallocated safe heap objects are restricted to only reference memory of the object's allocated-type.  Some allocation sites may allocate objects whose type cannot be determined statically, so some systems use allocation sites instead of types to determine the memory regions that may be allocated for objects~\cite{tat,tdi}.  This has been found to create cases where memory regions may be reused for  objects of multiple types and sizes, invalidating the temporal allocated-type safety.  Because \tool preserves allocated-type for memory reuse, it does not allocate any such objects on the safe heap, avoiding this problem.

%auto-ignore

\section{Implementation}
\label{sec:implementation}

%\tool has been deployed on the x86\_64 architecture, running on an Intel CPU i9-9900K with 128 GB RAM, using LLVM 10.0 on Ubuntu 20.04 with Linux kernel version 5.8.0-44-generic. Since the C library often handles heap objects, we used the uClibc library in our \tool safety analyses.

\tool has been deployed on the x86\_64 architecture, running on an Intel CPU i9-9900K with 128 GB RAM, using LLVM 10.0 on Ubuntu 20.04 with Linux kernel version 5.8.0-44-generic. %Figure~\ref{fig:impl} shows the implementation of the \tool framework. 
The CCured framework is adapted from NesCheck~\cite{nescheck}. We expand the original value-range analysis through the call-string approach for achieving scalable context-sensitivity~\cite{sharir-pnueli} and covering dynamically-sized objects.
For type safety validation, we eliminate the type casts generated by the compiler immediately after memory allocation, since it is the common way of allocating heap objects (i.e., casting to the corresponding type from void*) and it is safe. We utilize S2E~\cite{s2e} as the guided symbolic execution engine for removing false aliases. To reduce the path explosion of the symbolic execution, \tool employs a depth limit, where any terminated symbolic execution implies that the related heap object is unsafe.
%\tool applies LLVM's loop canonicalization~\cite{llvm-loop-canonicalization} and loop simplify and unrolling~\cite{llvm-loop-simplify,huang19oakland} features, as well as S2E's symbolic state merging~\cite{s2e-state-merging}. \tool limits the scope of symbolic execution employing the configurable depth of the call stack.

For runtime allocation to enforce temporal allocated-type safety, \tool creates per-allocated-type pools on the safe heap and isolates all unsafe objects found by static validation in the unsafe heap. This is achieved by adding an additional parameter that contains the hash of the allocated-type to the allocation API in TcMalloc only for safe heap objects, while all unsafe heap objects share the same unique hash. The pools are built by leveraging the spans of TcMalloc, the unsafe heap owns a separate, isolated span. The spans and metadata are originally isolated in TcMalloc through guard pages. Once acquired, the memory will never be returned to the OS for later reuse by another span (pool), but is only reused through the pool's freelist. The metadata and freelist are also isolated. 
\vspace{-0.08in}
\section{Evaluation}
\label{sec:evaluation}

\begin{table*}[t]
\aboverulesep=0ex 
\belowbottomsep=0.5ex
\belowrulesep=0ex
\abovetopsep=0.5ex
\centering
\resizebox{0.95\textwidth}{!}{
\begin{tabular}{l|rrrrrrrcc|rrrrrrrrr}
\toprule
    %\rule{0pt}{1.1EM}
\multicolumn{1}{c|}{\multirow{2}{*}{}} & \multicolumn{9}{c|}{\bf Runtime Overhead (\%)} & \multicolumn{9}{c}{\bf Memory Overhead (\%)} \\ %\cmidrule(lr){2-17}
\cmidrule(lr){2-10} \cmidrule(lr){11-19}
\multicolumn{1}{c|}{} &
  \multicolumn{1}{c}{TAT} &
  \multicolumn{1}{c}{SC} &
  \multicolumn{1}{c}{AS} &
  \multicolumn{1}{c}{DS} &
  \multicolumn{1}{c}{ES} &
  \multicolumn{1}{c}{FF} &
  \multicolumn{1}{c}{MU} &
  \multicolumn{1}{c}{\tool-R} &
  \multicolumn{1}{c|}{\tool} &
  \multicolumn{1}{c}{TAT} &
  \multicolumn{1}{c}{SC} &
  \multicolumn{1}{c}{AS} &
  \multicolumn{1}{c}{DS} &
  \multicolumn{1}{c}{ES} &
  \multicolumn{1}{c}{FF} &
  \multicolumn{1}{c}{MU} &
  \multicolumn{1}{c}{\tool-R} &
  \multicolumn{1}{c}{\tool} \\ \midrule
{\bf perlbench}                              &8.3  &2800.8 &174.2&251.2&824.1&8.5      &22.1      &8.5&{\bf 5.1} &35.7&83.5&262.4 &371.4&{\bf 21.8}&124.8&28.1&39.2&24.5\\
{\bf bzip2}                                  &4.4  &11.1   &74.4 &5.1  &122.5&2.7      &{\bf 1.7} &4.7&2.4       &5.5 &12.3&42.4  &6.2  &11.5      &17.4 &7.2 &6.4&{\bf 3.9} \\
{\bf mcf}                                    &2.6  &42.6   &12.6 &48.3 &67.1 &{\bf 1.2}&2.2       &2.4&1.5       &2.3 &11.1&14.5  &55.3 &8.2       &5.5  &6.8 &2.4&{\bf 1.8} \\
{\bf gobmk}                                  &5.6  & -     &56.3 &11.3 &225.8&7.5      &17.2      &5.9&{\bf 3.2} &8.2 & -  &1130.5&124.5&17.8      &78.5 &91.6&8.2&{\bf 5.2} \\
{\bf hmmer}                                  &2.4  &7.6    &42.7 &2.8  &324.8&6.4      &27.2      &2.4&{\bf 2.1} &51.5&24.5&6580.4&{\bf 12.4} &44.5      &85.2 &82.8&57.5&34.4 \\
{\bf sjeng}                                  &3.6  &457.2  &67.5 &2.7  &78.2 &4.1      &4.2       &4.0&{\bf 2.2} &3.6 &8.5 &18.7  &{\bf 2.7}  &12.2      &21.6 &17.4&3.8&3.1  \\
{\bf libquantum}                             &4.6  &202.4  &27.1 &3.1  &291.2&3.8      &3.0       &5.1&{\bf 2.9} &4.2 &7.6 &289.1 &4.1  &6.7       &17.4 &12.2&4.5&{\bf 3.5}   \\
{\bf h264ref}                                &6.2  &423.3  &23.2 &2.3  &655.8&{\bf 1.7}&4.8       &6.0&3.2       &12.6&42.7&72.4  &8.5  &8.8       &39.5 &78.4&14.1&{\bf 8.5}   \\
{\bf lbm}                                    &3.6  &10.5   &34.4 &3.7  &43.7 &2.2      &2.8       &3.9&{\bf 2.0} &4.6 &6.8 &27.2  &5.7  &15.4      &8.9  &9.8 &4.9&{\bf 3.2}   \\
{\bf sphinx3}                                &6.8  &23.1   &17.5 &5.8  &235.1&7.6      &5.5       &7.0&{\bf 3.7} &38.7&94.7&1150.8&182.8&{\bf 9.1} &1182.5&34.2&41.5&24.4   \\
{\bf milc}                                   &6.2  &12.8   &14.6 &21.1 &142.0&4.4      &6.4       &6.5&{\bf 3.2} &3.4 &10.5&337.1 &34.1 &14.2      &12.2 &15.6 &3.7&{\bf 3.2}   \\
{\bf omnetpp}                                &6.6  & -     &76.5 &698.4&167.4&5.1      &36.5      &6.8&{\bf 4.5} &13.2& -  &425.3 &1125.6&12.6     &436.1&73.1 &15.1&{\bf 11.5}   \\
{\bf soplex}                                 &2.2  &2135.4 &94.7 &12.2 &212.9&7.2      &5.6       &2.6&{\bf 2.1} &5.6 &17.8&355.2 &1450.0&24.1     &52.5 &59.3 &5.8&{\bf 4.3}   \\
{\bf namd}                                   &3.4  &2020.4 &27.3 &3.8  &66.5 &8.7      &{\bf 2.3} &3.9&2.5       &4.4 &9.8 &54.2  &8.5   &7.5      &24.4 &17.6 &4.8&{\bf 3.6}   \\
{\bf astar}                                  &3.5  &312.6  &62.6 &72.3 &246.1&3.3      &{\bf 2.5} &3.6&3.1       &4.8 &35.5&468.1 &515.6 &11.3     &77.1 &64.4 &5.2&{\bf 3.7}   \\ \midrule
{\bf AVERAGE}                                &4.7  &650.8&53.7 &76.2&246.9 &4.9 &9.6 &4.9&{\bf 2.9}& 13.2 &28.1 &748.6 &260.5&15.0 &145.6&39.9 &14.5&{\bf 9.3}   \\ \bottomrule
\end{tabular}
}\captionsetup{font=footnotesize}
	\caption{Runtime and Memory Overhead of \tool Compared with Prior Works for SPEC CPU 2006. We use the following abbreviations to represent the prior  works, TAT-Type-After-Type, SC-SAFECode, AS-ASan, DS-DangSan, ES-EffectiveSan, FF-FFMalloc, MU-MarkUs. The lowest overhead on a benchmark is marked as bold. 
 %\tool introduces the lowest runtime overhead on 10 out of 15 benchmarks and the lowest memory overhead on 11 out of 15 benchmarks. For other benchmarks, \tool's overhead is close to the lowest with negligible difference. Overall, \tool has the lowest runtime and memory overhead on average. We also measured the overhead of \tool's runtime allocation applied to all heap allocations (i.e., without static safety validation), for a fair full coverage comparison (\tool-R).
 }
	\label{tab:Overhead}
    \vspace{-0.2in}
\end{table*}

This section focuses on evaluating how \tool improves the security for low overhead. We conduct this evaluation by analyzing browsers, servers, SPEC CPU2006, and SPEC CPU2017
%\trent{CPU2017?} \kaiming{Yes, pending results}
benchmarks.

\vspace{-0.08in}
\subsection{Identifying Safe Heap Objects}
\label{subsec:safeheapobj}

\if 0
\begin{table}[t]
    \centering
        %\scalebox{0.85}{
        \resizebox{0.7\columnwidth}{!}{
	\begin{tabular}{l|r|r}
	    \toprule
		\ &\multicolumn{1}{c|}{\bf\em SPEC 2006 }&\multicolumn{1}{c}{\bf\em SPEC 2017}\\ \midrule%&\multicolumn{1}{c|}{\bf\em Memory}\\
		
		{\bf\em Avg. Safe Allocations}& 73.0\%&\\
            \bottomrule
	\end{tabular}
 }
        \captionsetup{font=footnotesize}
	\caption{Percentage of memory allocations protected by \tool, individual results of SPEC benchmarks are shown in Table~\ref{tab:Protection of Heap Allocations at Runtime Abstract}.}
	\label{tab:Protection of Heap Allocations at Runtime}
     \vspace{-0.4in}
\end{table}
\fi

\noindent{\bf RQ1}: {\em How many heap objects does \tool identify that can be allocated on the safe heap?} Table~\ref{tab:OverviewEval}
shows the counts and percentages of safe heap objects using existing techniques (i.e., {\em VR-Spatial+CCured-Type}) and \tool ({\em \tool-Spatial+\tool-Type}). \tool classifies 72.0\% of
heap objects as statically safe in Firefox, server programs and SPEC CPU2006/2017 benchmarks\footnote{We examined 15 out of 19 C/C++ benchmarks in SPEC CPU2006.  The remaining benchmarks (gcc, xalancbmk, povray, and dealII) are not supported by the SVF analysis due to the version is too old. The newer version in SPEC CPU2017 (gcc\_s and xalancbmk\_s) are supported. We analyzed all 12 SPEC CPU2017 Benchmarks.} on average.  
%\trent{We should say something about our coverage of the spec benchmark programs.}  
%\trent{Can we please add an average line to the table at the bottom?} \kaiming{Yes, I will add the average when we get all data. I'm using the new pdg implementation to test the unsupported SPECs, except for gcc.}

For spatial safety, CCured classified virtually no safe heap objects, which is expected as heap objects are usually compound objects and use pointer arithmetic for field accesses.  Traditional value-range analysis ({\em VR-Spatial}) classifies 72.2\% of heap objects as satisfying spatial safety, but may misclassify unsafe heap objects as safe due to its inability to handle reallocations and concurrency. \tool's extensions that handle dynamic sizing and multi-threading (Section~\ref{subsec:spatialvalidate} and~\ref{subsec:multithread}) and remove spurious
aliasing (Section~\ref{subsec:applyse})
classify around 81.0\% of heap objects as safe. %through value-range analysis and symbolic execution, given the fact that the majority of heap pointer arithmetic accesses are for accessing the fields, which can be verified to be within bounds.
For type safety, CCured classifies around 50.2\% of the heap objects as safe. Enhanced by the compatible-type-cast analysis (Section~\ref{subsec:typevalidate}) and symbolic execution (Section~\ref{subsec:applyse}), \tool classifies 82.0\% of the heap objects as safe, an additional 31.8\% of heap objects being classified as safe. The stats for individual benchmarks are attached in Table~\ref{tab:appendspec} in Appendix~\ref{append:spec}. %\trent{I guess we are still penalizing CCured here for the contained upcasts only.} \kaiming{No, CCured are low because it considers the cast between pointers/integers, void*/char* all as unsafe.}
%In total, \tool's static safety validation approach is able to verify 71.9\% of the heap objects satisfy spatial and type safety. 
%Given that CCured aimed to classify pointers instead of objects originally, to be fair, we also collected the number of safe pointers to the heap based on the original CCured approach, as well as the number of safe heap objects after applying our value-range analysis on top of CCured's spatial and type safety analyses, i.e., VR+CCured-Type column. 
%The results show that \tool outperforms it by 38.1\%. \trent{Are we comparing pointers or objects?  Seems like objects, which renders the discussion of pointers moot.}\kaiming{Objects, but the classification of objects is through the static analyses on pointers.}

The results in Table~\ref{tab:OverviewEval} show static heap object counts, raising the question: how does \tool perform with runtime allocations?
%Table~\ref{tab:Protection of Heap Allocations at Runtime} shows the number of safe and total heap allocations at runtime for the SPEC CPU2006 benchmarks.  
We use heaptrack, perf, and Mtuner as the heap memory profiler to measure such data on SPEC 2006 benchmarks. If \tool classifies a heap object as safe statically, all the corresponding runtime allocations are classified as safe as well. Based on the result, 73.6\% of total runtime heap allocations are classified as safe by \tool.

\vspace{-0.1in}
\subsection{Performance Evaluation}
\label{subsec:perf}

% First plot: Runtime Overhead
\begin{figure}[]
\centering

% First plot: Runtime Overhead
\begin{tikzpicture}
\begin{axis}[
    xbar,  % Horizontal bars
    width=0.9\linewidth,  % Adjust width to fit within a single column
    height=4cm,  % Adjusted height for better spacing
    symbolic y coords={perlbench\_s, gcc\_s, mcf\_s, xalancbmk\_s, deepsjeng\_s, x264\_s, lbm\_s, omnetpp\_s, imagick\_s, leela\_s, nab\_s, xz\_s},
    ytick=data,
    yticklabel style={font=\scriptsize},
    xmin=0, xmax=10,
    bar width=2pt,  % Thinner bars
    nodes near coords,
    nodes near coords align={horizontal},
    every node near coord/.append style={font=\tiny},
    enlarge y limits={0.05},  % Reduced space between bars significantly
    enlarge x limits={0},  % A bit of horizontal margin for aesthetics
    xticklabel style={font=\scriptsize}
]
\addplot[fill=blue] coordinates {(6.6,perlbench\_s) (4.2,gcc\_s) (0.5,mcf\_s) (8.6,xalancbmk\_s) (0.7,deepsjeng\_s) (3.8,x264\_s) (1.7,lbm\_s) (6.4,omnetpp\_s) (1.2,imagick\_s) (1.4,leela\_s) (0.8,nab\_s) (2.7,xz\_s)};
\legend{\small Runtime}
\end{axis}
\end{tikzpicture}

% Second plot: Memory Overhead
\begin{tikzpicture}
\begin{axis}[
    xbar,  % Horizontal bars
    width=0.9\linewidth,  % Adjust width to fit within a single column
    height=4cm,  % Adjusted height for better spacing
    symbolic y coords={perlbench\_s, gcc\_s, mcf\_s, xalancbmk\_s, deepsjeng\_s, x264\_s, lbm\_s, omnetpp\_s, imagick\_s, leela\_s, nab\_s, xz\_s},
    ytick=data,
    yticklabel style={font=\scriptsize},
    xmin=0, xmax=20,
    bar width=2pt,  % Thinner bars
    nodes near coords,
    nodes near coords align={horizontal},
    every node near coord/.append style={font=\tiny},
    enlarge y limits={0.05},  % Reduced space between bars significantly
    enlarge x limits={0},  % A bit of horizontal margin for aesthetics
    xticklabel style={font=\scriptsize}
]
\addplot[fill=red] coordinates {(18.6,perlbench\_s) (12.4,gcc\_s) (1.1,mcf\_s) (14.5,xalancbmk\_s) (0.2,deepsjeng\_s) (1.2,x264\_s) (0.2,lbm\_s) (7.7,omnetpp\_s) (2.4,imagick\_s) (3.2,leela\_s) (2.9,nab\_s) (0.1,xz\_s)};
\legend{\small Memory}
\end{axis}
\end{tikzpicture}
\vspace{-0.2in}
\caption{\footnotesize Runtime and Memory Overhead of \tool on SPEC CPU2017}
\label{fig:spec2017}
\end{figure}
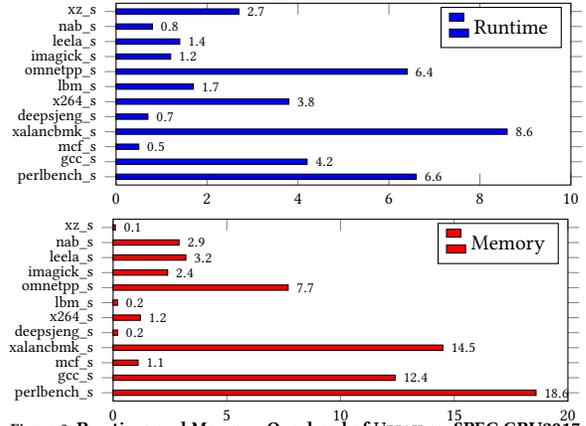

\noindent{\bf RQ2}: {\em What is the performance impact of isolating enforcing temporal allocated-type safety?} In this section, we evaluate the performance of \tool on the SPEC CPU2006 and SPEC CPU2017 benchmarks. %\tool extends Type-After-Type's typed-pool allocation for statically validated safe heap objects and isolates the unsafe heap objects in a separate allocation pool. 
We compare \tool with seven prior heap defenses on SPEC CPU2006 and provide result on SPEC CPU2017 (many compared works were not originally evaluated on SPEC CPU2017). All prior works are built from their open-sourced implementations without any updates. We note that defenses vary in their goals (see Table~\ref{tab:previouswork}).  All defenses except \tool provide coverage of all objects for the memory error classes they address, but may not cover all memory error classes, whereas \tool protects 73.6\% of heap objects (72.2\% of allocation sites) that satisfy spatial and type safety to enforce temporal allocated-type safety for the SPEC CPU2006 benchmarks.  The \tool-R column uses \tool to allocate all heap objects to emulate the complete coverage of all objects for comparison.  
%\trent{How are unsafe objects handled?  a single unsafe pool with Uriah reuse?}\kaiming{Yes}
%, and we leave them as their original build without updating, e.g., llvm versions.

\vspace{-0.1in}
\subsubsection{Performance Overhead}
\label{subsubsec:timeperf}

Table~\ref{tab:Overhead} on the left side shows that \tool has the lowest average runtime overhead on SPEC CPU2006 benchmarks of 2.9\%.  ASan, DangSan, and EffectiveSan enforce memory safety using runtime checks, including maintaining the metadata for use in such checks, resulting in high runtime overhead.  \tool does not require runtime checks, so it elides their overheads.
HexType~\cite{hextype} is not listed as it only supports C++ programs.
%ASan provides a spatial defense, so it is next most expto extract metadata of the corresponding protection and perform runtime checks based on metadata. ASan deploys redzones to facilitate the detection of illicit memory access while incurring the overhead of maintaining redzones and detecting errors. We do not compare to HexType~\cite{hextype} since it only supports C++, but its overhead is roughly similar to EffectiveSan based on our tests, and thus still higher than \tool. 
%Fortunately, 

\tool also outperforms allocator-based defenses. %\trent{Can we run Uriah all objects on SPEC?  Would be more apple-to-apples for what people expect than Firefox, although that is impressive.} 
Specifically, SAFECode exhibits much worse performance overhead than others (more than 20x on \textit{perlbench} and \textit{named}), due to SAFECode's method to reuse memory by allocating and deallocating per-type pools for each function (or call chains through escape analysis).  
%One exception is  {\em mcf}, which has the greatest heap memory usage, as shown in Table~\ref{tab:Protection of Heap Allocations at Runtime}.  However,  since {\em mcf} allocates huge heap chunks in a small number of allocations, there are few pools to maintain. SAFECode has lower performance overhead on mcf, but still more than 40\%. 
MarkUs and FFmalloc are much more efficient since they track used memory and prohibit memory reuse, although they are still more expensive than \tool in most cases. Type-After-Type improves performance by introducing a per-type pool allocation scheme by using an efficient memory allocator, {\em TcMalloc}. 
%Benefiting from the static safety validation approach, \tool is able to enforce per-type allocation only on a safe heap while enforcing a broader security guarantee.
Unfortunately, it is impractical to compare Type-After-Type and \tool on the same set of objects without breaking either system.  %\trent{Last sentence sounds wrong now.} 
For a fair comparison, we measured \tool's overhead when it applies to all heap allocations, the results show that \tool-R(untime) performs slightly worse, which is expected given the more precise identification of type by \tool. Moreover, \tool provides guarantees for spatial and type safety, and enforces the stronger temporal allocated-type safety property against temporal attacks. We also evaluated \tool on SPEC CPU2017. See Figure~\ref{fig:spec2017}, \tool exhibits a 2.6\% overhead on SPEC CPU2017 on average, which is reasonably low.

\vspace{-0.05in}
\subsubsection{Memory Consumption}
\label{subsubsec:memperf}

\tool has the lowest average memory overhead on SPEC CPU2006 benchmarks of 9.3\% (right side of Table~\ref{tab:Overhead}).  \tool-R, which provides full coverage of allocation, has a slightly higher (14.5\%) memory consumption than Type-After-Type (13.2\%), which again is expected. Previous works consume much more memory than \tool due to: (1) memory usage for checking mechanisms, e.g., red-zones of ASan; (2) memory usage for allocation metadata of EffectiveSan and DangSan; (3) memory usage by prohibiting memory reuse in FFmalloc and MarkUs.  
%DangSan has relatively low memory usage because its scope of enforcement is limited to temporal safety. \kaiming{DangSan's memory overhead is not low ~ 260\%}
%prevent memory reuse after releasing by the previous object, achieving complete protection against dangling pointer exploitation but introducing high memory overhead on heap-intensive benchmarks. 
Systems that employ type-based reuse, like \tool, sometime also have a higher memory utilization.  SAFECode also supports reusing heap pools, which enables it to be more memory-efficient than Type-After-Type and \tool in some cases, despite still being quite high for some programs, like perlbench and sjeng.  Also, pool reuse is the main cause of SAFECode's increased runtime overhead.  We have observed that the need to reuse pools is often unnecessary as allocated-types are typically reused. Again, we evaluated \tool on SPEC CPU2017 for its memory overhead. See Figure~\ref{fig:spec2017}, \tool exhibits a 5.4\% memory overhead on SPEC CPU2017 on average.

\vspace{-0.05in}
\subsubsection{Static Analysis Time}
\label{subsubsec:statictime}

Table~\ref{tab:StaticTime} shows the static safety validation time of \tool on all the benchmarks we analyzed. In general, the spatial and type safety validation approaches by \tool are efficient given the {\bf Static} column indicates the least amount of time among the three phases. The CCured analysis and constraint extraction {i.e., {\bf Foundation}}, takes more time since \tool leverages the analysis in a context-sensitive manner. We note that the time for static safety validation is a one-time cost for each program, as long as no updates are made to the program. We also provide the static analysis time for SPEC benchmarks in Table~\ref{tab:appendtime} in Appendix~\ref{append:spec}.

\begin{table}[]
\aboverulesep=0ex 
\belowbottomsep=0.5ex
\belowrulesep=0ex
\abovetopsep=0.5ex
\centering
\resizebox{.85\columnwidth}{!}{
\begin{tabular}{l|rrrrrrr}
\toprule
\multicolumn{1}{c|}{} &
  \multicolumn{1}{c}{\textbf{Foundation}} &
  %\multicolumn{1}{c}{\textbf{Porting}} &
  \multicolumn{1}{c}{\textbf{Static}} &
  \multicolumn{1}{c}{\textbf{SymExec}} &
  \multicolumn{1}{c}{\textbf{Total}} \\ \midrule
\textbf{Firefox}    &17,436.4s &2,349.2s  &10,386.8s &30,172.4s     \\
\textbf{nginx}      &2,259.8s  &132.6s &323.7s  &2,716.1s      \\
\textbf{httpd}      &249.5s   &1.6s  &64.2s  &315.3s  \\
\textbf{proftpd}    &1,328.4s  &212.7s  &564.4s  &2,105.5s    \\
\textbf{sshd}       &225.6s  &251.0s  &191.4s &668.0s    \\
\textbf{sqlite3}    &3,396.4s  &472.8s  &1,624.4s  &5493.6s     \\
%\textbf{SPEC 2006}    &1984.7 s  &757.0s  &1119.4   &3861.2s     \\
%\textbf{SPEC 2017}    &s  &s  &s   &s     \\ 
\bottomrule
\end{tabular}
}\captionsetup{font=footnotesize}

	\caption{Time Elapsed in Each Phase of \tool's Static Safety Validation. The {\bf Foundation} column represents the adapted CCured analysis time. The {\bf Static} column represents the time spent for spatial and type safety validation. {\bf SymExec} shows the time elapsed in the symbolic execution phase. {\em Total} column shows the total static safety validation time - sum of the first 3 columns.}

	\label{tab:StaticTime}
     \vspace{-0.2in}
\end{table}

\begin{comment}
\begin{figure*}[hbt!]
\centerline{\includegraphics[width=\textwidth]{Figures/PerfOverhead.pdf}}
\vspace{-0.05in}
\caption{Performance Overhead of SPEC CPU 2006 Benchmark. The overhead that are more than 100\% are marked with the actual value.}
\vspace{-0.1in}
\label{fig:perfoverhead}
\end{figure*}

\begin{figure*}[hbt!]
\centerline{\includegraphics[width=\textwidth]{Figures/MemOverhead.pdf}}
\vspace{-0.05in}
\caption{Memory Overhead of SPEC CPU 2006 Benchmark. The overhead that are more than 100\% are marked with the actual value.}
\vspace{-0.1in}
\label{fig:memoverhead}
\end{figure*}

\end{comment}

\vspace{-0.05in}
\subsubsection{Firefox}
\label{subsubsec:firefox}

We applied the \tool framework on the Firefox web browser (changeset ad179a6f) to further evaluate \tool's runtime overhead on real-world applications. We evaluated four benchmarks, namely SunSpider, Octane 2.0, Dromaeo JS, and Dromaeo DOM, which are commonly-used benchmarks for Firefox. Note that we only compared with Type-After-Type in this section for 2 reasons: (1) \tool shares a similar approach for runtime allocation (i.e., enforcing type-based temporal safety) with Type-After-Type and (2) for all the existing frameworks evaluated in Table~\ref{tab:Overhead}, only EffectiveSan was evaluated on Firefox originally. Since EffectiveSan is mostly a runtime checking mechanism (i.e., Sanitizer), it exhibits much higher overhead than the allocator-based designs.

As shown in Table~\ref{tab:firefox}, \tool incurs less overhead than Type-After-Type for all four benchmarks tested. However, a key reason is that \tool only applies its type-based pool allocation %to ensure temporal type safety for all 
for the safe heap objects (i.e., 70.3\% of allocation sites), while Type-After-Type applies a similar technique to all the heap allocations. %Thus, \tool is supposed to have less overhead since it requires less number of pools and maintain less number of freelists. 
For a fair comparison, we also include the results when \tool is applied to all heap allocations (i.e., Uriah-R). \tool generally incurs a slightly higher overhead than Type-After-Type, which is expected given that it distinguishes types more strictly and precisely (i.e., allocated-type) than Type-After-Type, creating more pools.  However, the additional overhead of \tool is modest for Firefox.

\vspace{-0.05in}
\subsection{Impact on Mitigating Exploitation}
\label{subsec:exploiteval}

\noindent{\bf RQ3}: {\em Does \tool improve the security against exploitation of heap memory errors?} While \tool may expand the range of safe objects, a legitimate concern is whether this (alone) effectively prevents exploits or only reduces the overheads of applying additional defenses. \tool places heap objects that cannot be validated as satisfying spatial or type safety on an unsafe heap, so perhaps these unsafe heap objects are both the vulnerable objects and target objects exploited in attacks. For evaluating the ability of \tool to mitigate attacks on memory errors for heap objects, we choose to examine vulnerabilities in the DARPA CGC binaries and 28 recent CVEs.
%to assess the effectiveness of \tool in preventing exploits.  

\vspace{-0.05in}
\subsubsection{Impact on Mitigating Exploits on CGC Binaries}
\label{subsubsec:cgc}

To assess the security implications of \tool, we apply it on the DARPA CGC Binaries~\cite{darpacgc}.  The CGC Binaries are crafted programs intended to encompass various types of software vulnerabilities.  
%They closely resemble real-world software with sufficient complexity to challenge both manual and automated vulnerability detection. They include comprehensive functionality tests, intentionally introduced bugs, and patches, allowing for the evaluation of patching and bug mitigation strategies. 
We use the approach implemented by Trail of Bits~\cite{trailofbitscgc} to port CGC Binaries for Ubuntu 20.04. In our experiment, we picked all 73 CGC Binaries that have heap-related memory bugs, including 65 spatial errors, 17 type errors, and 20 temporal errors, for both UAF and UBI.

%For the 65 spatial errors, 
For all the memory errors, we say that \tool successfully mitigates the attack when the vulnerable object is classified as unsafe and isolated in the unsafe heap and the target object is classified as safe and allocated in the safe heap. Additionally, temporal errors may also be mitigated by temporal allocated-type safety enforcement.  \tool successfully mitigates all 65 spatial error exploits. Of the 17 type errors, 4 of them are incompatible casts, 13 are bad casts on integers (serving as bounds or access ranges) that change signedness, objects are unsafe if index is used in bad cast. \tool successfully mitigates all 17 type error exploits. 
%Note that we also identified 6 additional bad casts on integers located on the stack that serve as the bounds or access range for heap objects.  Based on our threat model of applying DataGuard~\cite{dataguard} for stack objects, these stack integer variables would be identified as unsafe, which we would use in \tool to identify these heap objects as unsafe.  Integration of \tool with the techniques of DataGuard is future work. 
%, but with the protection of previous efforts~\cite{dataguard}, this can be mitigated effectively.
%\trent{revisit this}
12 of the temporal errors are UAF and 8 are UBI.  
%We say that the \tool successfully mitigates exploits when the target object is classified as safe and the vulnerable object involved in the temporal error is either: (1) classified as unsafe by the static validation or (2) % classified as safe for spatial or type error during the static validation and allocated using per-type pool allocation to constrain the memory reuse by exploiting the temporal error must be within the same type of objects. 
%mitigated by temporal allocated-type safety enforcement. 
%\trent{Shouldn't this be an "AND"?  We find it unsafe and prevent the exploitation.  Are any vulnerable objects "safe"?}\kaiming{No, it should be "OR", the unsafe is in terms of spatial and type safety, if a heap object is already spatial/type unsafe by the static analyses, we place it into the unsafe allocation pool and do not need to enforce per-type allocation on them. If a heap object is spatial and type safe, then there is a chance for it to be temporal unsafe, we place it into the per-type allocation pool on the safe heap to protect against temporal error exploitation.}
Of the 20 cases, 6 of them are mitigated by \tool's static safety validation since the exploits are combined with spatial or type errors. 14 of them are mitigated by \tool's runtime allocation to ensure temporal allocated-type safety. The UBI cases trigger crashes using original PoC exploit scripts, since \tool's heap allocator zeros heap memory upon requesting memory from OS, preventing exploitation.
Thus, \tool successfully mitigated all 102 heap memory errors of the 73 CGC binaries.

\begin{table}[t!]
\aboverulesep=0ex 
\belowbottomsep=0.5ex
\belowrulesep=0ex
\abovetopsep=0.5ex
    \centering
        %\scalebox{0.85}{
        \resizebox{.63\columnwidth}{!}{
	\begin{tabular}{l|r|r|r}
	    \toprule
		\ &\multicolumn{1}{c|}{\bf\em Uriah}&\multicolumn{1}{c|}{\bf\em Uriah-R} &\multicolumn{1}{c}{\bf\em TAT }\\ 
  \midrule
		
		{\bf\em SunSpider}&1.6\%&2.3\%&2.1\%\\
		
		{\bf\em Octane 2.0}&0.1\%&0.3\%&-0.2\%\\
		
		{\bf\em Dromaeo JS }&2.4\%&3.3\%&3.8\%\\
		
		{\bf\em Dromaeo DOM}&0.8\%&1.4\%&1.1\%\\
    
            \bottomrule
	\end{tabular}
 }
\captionsetup{font=footnotesize}
	\caption{Runtime Overhead of \tool vs. Type-After-Type on Firefox. 
 %{\bf\em Uriah-R} represents the runtime-only version of \tool in which allocation applies to all the heap objects,
    %we include it into this table because the full version of \tool actually applies our runtime allocation scheme to fewer heap objects, having {\bf\em Uriah-R} in this table is used for a fair comparison between Type-After-Type since it applies runtime allocation to all the heap objects. %\zhiyun{the caption is a bit too long. Suggest we move some of this to the paper body.}
    }

	\label{tab:firefox}
    \vspace{-0.2in}
\end{table}

\vspace{-0.05in}
\subsubsection{Impact on Mitigating Exploits on Recent CVEs}
\label{subsubsec:cve}

Table~\ref{eval:cve} shows the 28 recent CVEs evaluated. \new{We picked all C/C++ heap memory error CVEs in userspace programs in 2023 from CVE database reported before June 2023 when we began the evaluation. The selected CVEs must} meet the following requirements: (1)\redout{the vulnerability must be on an} open-sourced; (2)\redout{there must be a detailed description released on the vulnerability, including the class of the memory error and the vulnerable object (i.e., as defined in Section~\ref{subsec:exploitheap})} \new{at least one PoC and/or PoV is available}; and (3) \redout{a patch to fix the vulnerability must have been released} \new{a released patch}.  These requirements are necessary as we must pinpoint the vulnerable/target object and run the memory safety validation. \redout{We choose recent (2021-2023) CVEs to assess \tool's ability to mitigate recent vulnerabilities.}

%All the CVEs are from 2021 to 2023. 
%, most of the CVEs that we evaluated are from the year 2023.  while few of them are not, the reason is that at the time we conduct the evaluation, there are not enough CVEs released in the year 2023 that qualify the aforementioned 3 requirements, thus, we picked a few in the year 2022 and 2021 instead.

%To measure whether \tool successfully mitigates the vulnerability, the 
%\trent{This is not a concrete as for CGC.  Do we not know the target object in any cases?Do we just use all this information to find the vulnerable object?  So, then, the aim is to ensure that spatial and type vulnerable objects are unsafe?  Probably don't need all the discussion of UBI now.}\kaiming{No, these are newly released CVEs and no exploiting script is released so we do not know. For the CVEs, yes, as you wrote, the aim is to classify vulnerable objects as unsafe.} \trent{Please rewrite to describe the precise conditions under which \tool is said to prevent the CVE exploitation and justify why that is reasonable.  If you have to hedge, then we need to discuss.} 

\new{We evaluated the exploitation paths used in the available exploits (PoCs).  The goal is to find how often advanced exploits corrupt safe target objects to  determine whether isolating safe objects can reduce real attack targets. Of course, some vulnerabilities could be exploited by attacking unsafe target objects instead in other exploitation paths, but isolating safe objects for low cost remains worthwhile as it can increase exploit development cost.} A challenge is that recent CVEs often do not have a corresponding PoV (i.e., proof-of-vulnerability, exploit write-up/script) released.  Thus, we do not know the target objects of any example exploit.  However, the description includes information on the vulnerable objects, so we can evaluate \tool's ability to identify vulnerable objects as unsafe. For the 18 CVEs that relate to spatial/type memory errors, \tool successfully identified vulnerable objects for all 18 cases, by classifying them as unsafe. Runtime allocation of \tool ensures that those vulnerable objects are isolated on the unsafe heap to block illicit access to the safe heap objects. For the 10 CVEs that relate to temporal memory errors, since vulnerable objects are not involved in spatial/type errors, \tool's static validation classified them as safe and prevents exploits by enforcing temporal allocated-type safety on the safe heap, preventing \new{temporal exploits (e.g., UAF/DF)} that aim at reusing the memory for different allocated-type.

\vspace{-0.05in}
\subsection{Combining \tool with Existing Protections}
\label{subsec:combineuriah}

\noindent{\bf RQ4}: {\em What is the security impact and performance improvement of applying existing protections to the \tool unsafe heap?} We applied two recent frameworks of protecting heap memory safety on \tool's unsafe heap, namely TDI~\cite{tdi} and CAMP~\cite{camp24usenix}.

The results, as depicted in Table~\ref{tab:reducecost}, showcase significant reductions in both runtime and memory overhead when protections are applied to \tool's unsafe heap compared to their native implementation. 
For SPEC CPU2006, TDI reduced runtime overhead to 2.5\% from 8.4\% and memory overhead to 3.7\% from 15.5\%. CAMP decreased runtime overhead to 16.8\% from 54.9\% and cut memory overhead to 72.3\% from 237.7\%. For SPEC CPU2017, TDI achieved a runtime overhead reduction to 4.4\% from 12.5\% and a memory overhead reduction to 7.1\% from 18.6\%. CAMP showed a runtime overhead decrease to 8.2\% from 21.3\%, and a memory overhead cut to 40.6\% from 127.5\%. These findings highlight the effectiveness of integrating protections with \tool in reducing overhead. More importantly, all the CVEs presented in their paper, namely 28 CVEs from CAMP and 4 CVEs from TDI, remain prevented by only applying these defenses to \tool's unsafe heap, highlighting the effectiveness of integrating existing protections with \tool in reducing costs while keeping the their original security guarantee. 

\begin{table}[t]
    \centering
    \aboverulesep=0ex 
\belowbottomsep=0.5ex
\belowrulesep=0ex
\abovetopsep=0.5ex
    \footnotesize
	\begin{tabularx}{0.47\textwidth}{c|X}
	    \toprule
		\multicolumn{2}{c}{\bf\em CVEs Evaluated}\\
		\midrule
		{\bf\em Spatial}&2023-27781; 2023-27117; 2023-27249; 2023-27103; 2023-23456; 2023-1655; 2023-1579; 2023-0433; 2023-0288; 2023-0051\\
	    \midrule
		{\bf\em Type}&
                2023-23454; 
                2023-1078; 
                2023-1076; 
                2022-27882; 
                2021-21861; 
                2021-21860; 
                2021-3578; 
                2021-28275\\
		\midrule
		{\bf\em Temporal}&
                2023-27320; 
                2023-25136; 
                2023-22551; 
                2023-0358; 
                2023-1449; 
                2022-47093; 
                2022-45343; 
                2022-43680; 
                2022-43286; 
                2022-4292\\
		\bottomrule
	\end{tabularx}
        \captionsetup{font=footnotesize}
	\caption{Recent CVEs evaluated under the \tool defense.}
	\label{eval:cve}
    \vspace{-0.2in}
\end{table}

%auto-ignore

\vspace{-0.05in}
\section{Discussion}
\label{sec:Discussions}

While we have seen that \tool does mitigate existing exploits, as previously discussed, there remain classes of attacks that \tool does not directly protect against: (1) attacks on unsafe objects; (2) attacks where the target and vulnerable objects are the same; and (3) \new{temporal attacks on (re)using memory of objects of the same allocated-type}.  By design, \tool does not prevent attacks on heap objects that fail spatial or type safety validation \new{(i.e., unsafe objects)}, as these unsafe objects are allocated in the traditional \new{(unsafe)} heap.  For example, \tool does not prevent the exploitation of CVE-2022-23088~\cite{heapexploitdemo}, in which the attacker overflows a field of a structure and gains access to another field, achieving an arbitrary write primitive. However, \tool does greatly reduce the fraction of heap objects that are placed in the traditional  heap and are thus exposed to attacks. Fortunately, other defenses can be applied on unsafe heap with \redout{low}\new{reduced} cost, as illustrated in Section~\ref{subsec:combineuriah}.  %\trent{summary?}
%Similarly, \tool will not protect against scenarios where the target object is distinct from the (initially) vulnerable object, but both are unsafe objects of the same type. 

In some cases, attackers can perform exploits by reusing the memory of the same allocated-type of vulnerable object, but such attacks can be difficult to implement~\cite{autoslab}. \new{While Uriah does not completely prevent such temporal exploits through UAF/DF, it does prevent such reuse attacks that (1) require spatial and/or type errors (e.g., DirtyCred~\cite{dirtycred}) or (2) reuse memory of unsafe objects for safe objects of the same allocated-type (i.e., by isolation).} To assess the efficacy of attacks by reusing memory of the same type, we examine the recent DirtyCred attack. DirtyCred allows an attacker to modify authentication credentials by reusing the memory of the same type, %Such workflow of the data-only attack is not limited to kernel space but can also take place in user-level programs. 
but we found that DirtyCred also requires exploitation of other memory errors to enable such reuse, which \tool prevents successfully. Note that advanced static analysis to improve \tool's ability to statically validate more heap objects that satisfy spatial and type safety can be adopted, such as T-prunify~\cite{tprunify}.
\vspace{-0.05in}
\section{Related Work}
\label{sec:relatedwork}

To prevent spatial errors, researchers have proposed techniques to validate that memory accesses are within an object's memory bounds on each reference~\cite{softbound,memsafe10scam,lowfat-heap}.   Static analysis techniques are employed to remove checks for objects that can be proven to only be accessed safely~\cite{ccured,baggybounds}. To prevent type errors, 
%the safety in type casts is not analyzed statically to remove checks other than simple type comparisons and integer type casts~\cite{dataguard}, although it has been suggested that upcasts are safe~\cite{ccured-realworld}. 
runtime checks such as UBSan~\cite{ubSAN} and VTrust~\cite{vtrust} worked only for polymorphic classes of objects, CaVer~\cite{caver} and TypeSan~\cite{typesan} aimed to include the non-polymorphic objects and C-style type casts but introduced high overhead due to inefficient metadata tracking and inserted checks. Hextype~\cite{hextype} minimizes overhead and increases coverage for various allocation patterns but only supports C++. EffectiveSan~\cite{duck2018effectivesan} includes C programs, but it only deals with upcasts in C++.
%The incomplete coverage results in both the exploitation of vulnerable objects and the corruption of target objects remaining possible.
To prevent temporal errors, detectors of potential temporal safety violations on the heap using static analysis are proposed~\cite{uafchecker,clangstaticanalyzer, mluafdetect, gueb, uafdetector, cred, Zhai2020UBITectAP}. Unfortunately, all works present soundness limitations. Alternatively, runtime defenses have been proposed by leveraging dangling pointer elimination~\cite{van2017dangsan,dangnull,freesentry}, pointer dereference checking~\cite{duck2018effectivesan,ccured,ccured-realworld,ccured-toplas,Nagarakatte2010CETSCE,zhou2022fat}, limiting exploitability~\cite{cfixx}, modifying memory allocators to introduce new memory allocation~\cite{tat,ffmalloc21usenix,freeguard17ccs}, and garbage collection schemes~\cite{markus20oakland,safecode}. \new{However, as discussed in Section~\ref{subsec:limit}, they do not provide protection against all classes of memory errors and/or present high overhead. }
%Generally speaking, runtime temporal safety defenses incur significant runtime overheads, due to the increased number of instructions for processing metadata for tracking and validating the lifetime of objects, which limits their applicability.
%Generally speaking, incomplete coverage results in both the exploitation of vulnerable objects and the corruption of target objects due to the over-approximation of static analyses to remove runtime checks, memory accesses to many safe objects are deemed to require checks falsely. 
%Combined with inefficient metadata processing, existing software-based solutions incur overhead that prevents adoption. 

Hardware-assisted defenses ~\cite{pacmem,aos20micro,pacitup19usenix,mpxreport,infatpointer,chex86,practicalrest,arm-mte} are emerging to reduce costs introduced by software-based defenses by applying recent hardware features (e.g., ARM PA) and fat-pointer design. However, the applicability of such approaches is usually limited due to the dependency on the specific hardware platform and feature. Moreover, research has already shown that such hardware features can be compromised~\cite{pacmam22isca}. Also,  the idea of employing cryptography for memory protection and pointer integrity has been proposed ~\cite{palit21oakland,proskurin20oakland,cryptag,lemay21micro,encryptmemorysafe}. 
However, these techniques do not ensure memory safety for all objects (e.g., pointers only) and only selectively protect sensitive memory. Moreover, many approaches utilize the unused bits of pointers for encoding the metadata of checking. Given the trend of evolving the OS into 64-bit mode by Intel~\cite{intel64bit}, such designs may no longer be feasible in the near future. \new{\tool does not rely on any specific hardware design and features, avoiding compatibility issues.}

\begin{table}[t]
\aboverulesep=0ex 
\belowbottomsep=0.5ex
\belowrulesep=0ex
\abovetopsep=0.5ex
    \centering
        \resizebox{\columnwidth}{!}{
	\begin{tabular}{l|r|r|r|r}
	    \toprule
            \multicolumn{1}{c|}{\multirow{2}{*}{}} & \multicolumn{2}{c|}{\bf SPEC CPU2006} & \multicolumn{2}{c}{\bf SPEC CPU2017} \\ 
\cmidrule(lr){2-3} \cmidrule(lr){4-5}
		\ &\multicolumn{1}{c|}{\bf\em \ \ \ \ \ Native\ \ \ \ \ }&\multicolumn{1}{c|}{\bf\em w/ \tool} &\multicolumn{1}{c|}{\bf\em\ \ \ \ \ Native\ \ \ \ \ }&\multicolumn{1}{c}{\bf\em w/ \tool}\\ 
  \midrule
		
		{\bf\em TDI}&8.4\% / 15.5\%&2.5\% / 3.7\%&12.5\% / 18.6\%&4.4\% / 7.1\%\\
		
		{\bf\em CAMP}&54.9\% / 237.7\% &16.8\% / 72.3\%&21.3\% / 127.5\%&8.2\% / 40.6\%\\
  
            \bottomrule
	\end{tabular}
 }
\captionsetup{font=footnotesize}
	\caption{Overhead Reduction of Applying TDI and CAMP to \tool Unsafe Heap. Overhead is represented using the form "(runtime) / (memory)".}
	\label{tab:reducecost}
    \vspace{-0.2in}
\end{table}

%auto-ignore
\vspace{-0.05in}
\section{Conclusion}
\label{sec:conclusion}

We present the \tool system that enforces temporal allocated-type safety while preserving spatial and type safety of the validated safe heap objects. \tool leverages static safety validation approaches to validate heap objects that are
free from spatial and type memory errors, proposes an efficient memory allocator to isolate objects that passes static safety validations on a separate safe heap, and enforces temporal allocated-type safety on the safe heap, ensures no access to any object can violate the spatial and type safety of
any safe object at runtime.
%to guarantee type-safe memory reuse to block temporal error exploitations that aim to reuse memory of different types. 
Distinct from previous approaches that tried to detect and eliminate memory errors, \tool is designed to isolate the targets of exploitation from illicit accesses from vulnerable objects to target objects to thwart attacks. \tool isolates objects produced at 72.0\% of heap allocation sites from exploitations of spatial and type errors, as well as temporal attacks that violate temporal allocated-type safety for low runtime and memory overhead. Combining \tool with existing protections drastically reduces their overhead while preserving their original security guarantee, offering insights of making existing and future heap memory error defenses more targeted and efficient.

%\section*{Acknowledgment}
%\section*{Acknowledgment}
%This research was sponsored by the U.S. Army Combat Capabilities Development Command Army Research Laboratory and was accomplished under Cooperative Agreement Number W911NF-13-2-0045 (ARL Cyber Security CRA), National Science Foundation grants CNS-1801534, CNS-1801601 and CNS-1652954, the European Research Council (ERC) under the European Union's Horizon 2020 research and innovation program (grant agreement No. 850868), and DARPA grant HR0011-19-C-0106.
%Any opinions, findings, and conclusions or recommendations expressed in this paper are those of the authors and do not necessarily reflect the views of the NSF. The
%views and conclusions contained in this document are those of the authors and should not be interpreted as representing the official policies, either expressed or implied, of the Combat Capabilities Development Command Army Research Laboratory of the U.S. government. The U.S. government is authorized to reproduce and distribute reprints for government
%purposes notwithstanding any copyright notation here on.

\bibliographystyle{acm}
\bibliography{refs1,refs2}

\begin{thebibliography}{100}

\bibitem{cfi}
{\sc Abadi, M., Budiu, M., Erlingsson, U., and Ligatti, J.}
\newblock Control-flow integrity.
\newblock In {\em Proceedings of the 12th ACM Conference on Computer and Communications Security\/} (New York, NY, USA, 2005), CCS '05, ACM, p.~340–353.

\bibitem{markus20oakland}
{\sc Ainsworth, S., and Jones, T.~M.}
\newblock Markus: Drop-in use-after-free prevention for low-level languages.
\newblock In {\em 2020 {IEEE} Symposium on Security and Privacy, {SP} 2020\/} (San Francisco, CA, USA, 2020), {IEEE}, pp.~578--591.

\bibitem{cling}
{\sc Akritidis, P.}
\newblock Cling: A memory allocator to mitigate dangling pointers.
\newblock In {\em Proceedings of the 19th USENIX Conference on Security\/} (USA, 2010), USENIX Security'10, USENIX Association, p.~12.

\bibitem{baggybounds}
{\sc Akritidis, P., Costa, M., Castro, M., and Hand, S.}
\newblock Baggy bounds checking: An efficient and backwards-compatible defense against out-of-bounds errors.
\newblock In {\em Proceedings of the 18th Conference on USENIX Security Symposium\/} (USA, 2009), SSYM'09, USENIX Association, p.~51–66.

\bibitem{javaheapcloningase}
{\sc Anand, S., and Harrold, M.~J.}
\newblock Heap cloning: Enabling dynamic symbolic execution of java programs.
\newblock In {\em Proceedings of the 26th IEEE/ACM International Conference on Automated Software Engineering\/} (USA, 2011), ASE '11, IEEE Computer Society, p.~33–42.

\bibitem{anderson72refmon}
{\sc Anderson, J.~P.}
\newblock Computer security technology planning study.
\newblock Tech. Rep. ESD-TR-73-51, The Mitre Corporation, Air Force Electronic Systems Division, Hanscom {AFB}, Badford, MA, 1972.

\bibitem{heapexploitdemo}
{\sc Anonymous}.
\newblock {CVE-2022-23088 Exploiting A Heap Overflow in the Freebsd WiFi Stack}.
\newblock \url{https://www.zerodayinitiative.com/blog/2022/6/15/cve-2022-23088-exploiting-a-heap-overflow-in-the-freebsd-wi-fi-stack}, 2022.
\newblock Accessed on May 25, 2023.

\bibitem{Blastpass}
{\sc Apple}.
\newblock About the security content of ios 16.6.1 and ipados 16.6.1.
\newblock \url{https://support.apple.com/en-us/106361}.
\newblock Accessed: 2024-04-10.

\bibitem{kernelCodingStyle}
{\sc Archives, L.~K.}
\newblock Linux kernel coding style.
\newblock \url{https://www.kernel.org/doc/html/next/process/coding-style.html}.
\newblock Accessed: 2024-04-10.

\bibitem{Barbar2020FlowSensitiveTH}
{\sc Barbar, M., Sui, Y., and Chen, S.}
\newblock Flow-sensitive type-based heap cloning.
\newblock In {\em 34th European Conference on Object-Oriented Programming, {ECOOP} 2020\/} (Berlin, Germany (Virtual Conference), 2020), R.~Hirschfeld and T.~Pape, Eds., vol.~166 of {\em LIPIcs}, Schloss Dagstuhl - Leibniz-Zentrum f{\"{u}}r Informatik, pp.~24:1--24:26.

\bibitem{diehard}
{\sc Berger, E.~D., and Zorn, B.~G.}
\newblock Diehard: Probabilistic memory safety for unsafe languages.
\newblock In {\em Proceedings of the 27th ACM SIGPLAN Conference on Programming Language Design and Implementation\/} (New York, NY, USA, 2006), PLDI '06, Association for Computing Machinery, p.~158–168.

\bibitem{aslr}
{\sc Bhatkar, S., DuVarney, D.~C., and Sekar, R.}
\newblock Address obfuscation: An efficient approach to combat a board range of memory error exploits.
\newblock In {\em Proceedings of the 12th Conference on USENIX Security Symposium - Volume 12\/} (USA, 2003), SSYM'03, USENIX Association, p.~8.

\bibitem{cfixx}
{\sc Burow, N., McKee, D.~P., Carr, S.~A., and Payer, M.}
\newblock Cfixx: Object type integrity for c++.
\newblock In {\em Network and Distributed System Security Symposium\/} (San Diego, CA, USA, 2018), The Internet Society.

\bibitem{chen2021evaluating}
{\sc Chen, M., Tworek, J., Jun, H., Yuan, Q., Pinto, H. P. d.~O., Kaplan, J., Edwards, H., Burda, Y., Joseph, N., Brockman, G., et~al.}
\newblock Evaluating large language models trained on code.
\newblock {\em arXiv preprint arXiv:2107.03374\/} (2021).

\bibitem{data-only-attack}
{\sc Chen, S., Xu, J., Sezer, E.~C., Gauriar, P., and Iyer, R.~K.}
\newblock {Non-Control-Data Attacks Are Realistic Threats}.
\newblock In {\em Proceedings of the 14th Conference on USENIX Security Symposium - Volume 14\/} (USA, 2005), SSYM’05, USENIX, p.~12.

\bibitem{eloise}
{\sc Chen, Y., Lin, Z., and Xing, X.}
\newblock A systematic study of elastic objects in kernel exploitation.
\newblock In {\em Proceedings of the 2020 ACM SIGSAC Conference on Computer and Communications Security\/} (New York, NY, USA, 2020), CCS '20, Association for Computing Machinery, p.~1165–1184.

\bibitem{slake}
{\sc Chen, Y., and Xing, X.}
\newblock Slake: Facilitating slab manipulation for exploiting vulnerabilities in the linux kernel.
\newblock In {\em Proceedings of the 2019 ACM SIGSAC Conference on Computer and Communications Security\/} (New York, NY, USA, 2019), CCS '19, Association for Computing Machinery, p.~1707–1722.

\bibitem{cheng21tops}
{\sc Cheng, L., Ahmed, S., Liljestrand, H., Nyman, T., Cai, H., Jaeger, T., Asokan, N., and Yao, D.}
\newblock {Exploitation Techniques for Data-Oriented Attacks with Existing and Potential Defense Approaches}.
\newblock {\em ACM Transactions on Privacy and Security 24}, 4 (2021).

\bibitem{s2e}
{\sc Chipounov, V., Kuznetsov, V., and Candea, G.}
\newblock S2e: A platform for in-vivo multi-path analysis of software systems.
\newblock {\em SIGPLAN Not. 46}, 3 (mar 2011), 265–278.

\bibitem{googleCppStyleGuideGlobals}
{\sc Chrome, G.}
\newblock Google c++ style guide - static and global variables.
\newblock \url{https://google.github.io/styleguide/cppguide.html#Static_and_Global_Variables}.
\newblock Accessed: 2024-04-10.

\bibitem{ccured-realworld}
{\sc Condit, J., Harren, M., McPeak, S., Necula, G.~C., and Weimer, W.}
\newblock Ccured in the real world.
\newblock {\em SIGPLAN Not. 38}, 5 (may 2003), 232–244.

\bibitem{readactor}
{\sc Crane, S., Liebchen, C., Homescu, A., Davi, L., Larsen, P., Sadeghi, A.-R., Brunthaler, S., and Franz, M.}
\newblock Readactor: Practical code randomization resilient to memory disclosure.
\newblock In {\em 2015 IEEE Symposium on Security and Privacy}.

\bibitem{s2e-state-merging}
{\sc Cyberhaven}.
\newblock {Exponential Analysis Speedup with State Merging}.
\newblock \url{http://s2e.systems/docs/StateMerging.html}, 2018.

\bibitem{heap_spray}
{\sc Daniel, M., Honoroff, J., and Miller, C.}
\newblock Engineering heap overflow exploits with javascript.
\newblock In {\em Proceedings of the 2nd Conference on USENIX Workshop on Offensive Technologies\/} (USA, 2008), WOOT'08, USENIX Association.

\bibitem{darpacgc}
{\sc Darpa}.
\newblock {DARPA Cyber Grand Challenge}.
\newblock \url{https://github.com/CyberGrandChallenge/}, 2016.

\bibitem{safecode}
{\sc Dhurjati, D., Kowshik, S., and Adve, V.}
\newblock Safecode: Enforcing alias analysis for weakly typed languages.
\newblock {\em SIGPLAN Not. 41}, 6 (June 2006), 144–157.

\bibitem{lowfat}
{\sc Duck, Yap, and Cavallaro}.
\newblock {Stack Bounds Protection with Low Fat Pointers}.
\newblock In {\em Proceedings of the 2017 Network and Distributed System Security Symposium (NDSS)\/} (2017).

\bibitem{lowfat-heap}
{\sc Duck, G.~J., and Yap, R. H.~C.}
\newblock Heap bounds protection with low fat pointers.
\newblock In {\em Proceedings of the 25th International Conference on Compiler Construction\/} (New York, NY, USA, 2016), CC 2016, Association for Computing Machinery, p.~132–142.

\bibitem{duck2018effectivesan}
{\sc Duck, G.~J., and Yap, R. H.~C.}
\newblock Effectivesan: Type and memory error detection using dynamically typed c/c++.
\newblock {\em SIGPLAN Not. 53}, 4 (jun 2018), 181–195.

\bibitem{checkedc}
{\sc Elliott, A.~S., Ruef, A., Hicks, M., and Tarditi, D.}
\newblock Checked c: Making c safe by extension.
\newblock In {\em 2018 IEEE Cybersecurity Development (SecDev)\/} (Cambridge, MA, USA, 2018), IEEE, pp.~53--60.

\bibitem{emsisoft2020cost}
{\sc Emsisoft}.
\newblock The cost of ransomware in 2020: A country-by-country analysis.
\newblock \url{https://blog.emsisoft.com/en/36665/the-cost-of-ransomware-in-2020-a-country-by-country-analysis/}, 2020.
\newblock Accessed on May 13, 2023.

\bibitem{gueb}
{\sc Feist, J., Mounier, L., and Potet, M.-L.}
\newblock Statically detecting use after free on binary code.
\newblock {\em Journal of Computer Virology and Hacking Techniques 10\/} (08 2014), 211--217.

\bibitem{tcmalloc}
{\sc Ghemawat, S., and Menage, P.}
\newblock Tcmalloc: Thread-caching malloc.
\newblock \url{https://goog-perftools.sourceforge.net/doc/tcmalloc.html}, 2021.

\bibitem{copilot}
{\sc GitHub}.
\newblock {GitHub Copilot}.
\newblock \url{https://copilot.github.com/}.
\newblock Accessed on May 13, 2023.

\bibitem{tcmalloczero2}
{\sc Golick, J.}
\newblock How tcmalloc works.
\newblock \url{https://jamesgolick.com/2013/5/19/how-tcmalloc-works.html}, 2013.

\bibitem{palloc}
{\sc Google}.
\newblock Partitionalloc design.
\newblock \url{https://chromium.googlesource.com/chromium/src/+/master/base/allocator/partition_allocator/PartitionAlloc.md}, 2021.

\bibitem{tcmalloczero1}
{\sc Google}.
\newblock Restartable sequence mechanism for tcmalloc.
\newblock \url{https://github.com/google/tcmalloc/blob/master/docs/rseq.md}, 2023.

\bibitem{typesan}
{\sc Haller, I., Jeon, Y., Peng, H., Payer, M., Giuffrida, C., Bos, H., and van~der Kouwe, E.}
\newblock Typesan: Practical type confusion detection.
\newblock In {\em Proceedings of the 2016 ACM SIGSAC Conference on Computer and Communications Security\/} (New York, NY, USA, 2016), CCS '16, Association for Computing Machinery, p.~517–528.

\bibitem{hu2016data}
{\sc Hu, H., Shinde, S., Adrian, S., Chua, Z.~L., Saxena, P., and Liang, Z.}
\newblock Data-oriented programming: On the expressiveness of non-control data attacks.
\newblock In {\em 2016 IEEE Symposium on Security and Privacy (SP)\/} (2016), IEEE, pp.~969--986.

\bibitem{dataguard}
{\sc Huang, K., Huang, Y., Payer, M., Qian, Z., Sampson, J., Tan, G., and Jaeger, T.}
\newblock The taming of the stack: Isolating stack data from memory errors.
\newblock In {\em Network and Distributed System Security Symposium (NDSS)\/} (San Diego, CA, USA, 2022), The Internet Society.

\bibitem{huang24ieeesp}
{\sc Huang, K., Payer, M., Qian, Z., Sampson, J., Tan, G., and Jaeger, T.}
\newblock Comprehensive memory safety validation: An alternative approach to memory safety.
\newblock {\em IEEE Security \& Privacy}, 01 (apr 5555), 2--11.

\bibitem{huang23secdev}
{\sc Huang, K., Sampson, J., and Jaeger, T.}
\newblock Assessing the impact of efficiently protecting ten million stack objects from memory errors comprehensively.
\newblock In {\em Proceedings of the 2023 IEEE Secure Development Conference (IEEE SecDev)\/} (Oct. 2023), IEEE.

\bibitem{huang19oakland}
{\sc Huang, Z., Lie, D., Tan, G., and Jaeger, T.}
\newblock Using safety properties to generate vulnerability patches.
\newblock In {\em 2019 IEEE Symposium on Security and Privacy (SP)\/} (San Francisco, CA, USA, 2019), IEEE, pp.~539--554.

\bibitem{intelmultithread}
{\sc Intel}.
\newblock Intel guide for developing multithreaded application.
\newblock \url{https://www.intel.com/content/dam/develop/external/us/en/documents/gdma-2-165938.pdf}.
\newblock 2011.

\bibitem{intelonetbb}
{\sc Intel}.
\newblock Intel oneapi threading building blocks.
\newblock \url{https://www.intel.com/content/www/us/en/developer/tools/oneapi/onetbb.html#gs.63k1wf}.
\newblock Accessed on Mar 7, 2024.

\bibitem{intelmpxperf}
{\sc Intel}.
\newblock Intel mpx explained - performance evaluation.
\newblock \url{https://intel-mpx.github.io/performance/}, 2018.
\newblock Accessed on May 23, 2023.

\bibitem{intel64bit}
{\sc Intel}.
\newblock Envisioning a simplified intel architecture.
\newblock \url{https://www.intel.com/content/www/us/en/developer/articles/technical/envisioning-future-simplified-architecture.html}, 2023.
\newblock Accessed on May 21, 2023.

\bibitem{bopc}
{\sc Ispoglou, K.~K., AlBassam, B., Jaeger, T., and Payer, M.}
\newblock Block oriented programming: Automating data-only attacks.
\newblock In {\em Proceedings of the 2018 ACM SIGSAC Conference on Computer and Communications Security\/} (New York, NY, USA, 2018), CCS '18, Association for Computing Machinery, p.~1868–1882.

\bibitem{jemalloc}
{\sc jemalloc}.
\newblock jemalloc — general purpose memory allocation functions.
\newblock \url{https://jemalloc.net/jemalloc.3.html}.
\newblock Accessed on Mar 7, 2024.

\bibitem{hextype}
{\sc Jeon, Y., Biswas, P., Carr, S., Lee, B., and Payer, M.}
\newblock Hextype: Efficient detection of type confusion errors for c++.
\newblock In {\em Proceedings of the 2017 ACM SIGSAC Conference on Computer and Communications Security\/} (New York, NY, USA, 2017), CCS '17, Association for Computing Machinery, p.~2373–2387.

\bibitem{AKP-wikipedia}
{\sc Karenina, A.}
\newblock Anna karenina principle, 2023.

\bibitem{dirtypipe}
{\sc Kellermann, M.}
\newblock {The Dirty Pipe Vulnerability}.
\newblock \url{https://dirtypipe.cm4all.com/}.
\newblock Accessed on May 13, 2023.

\bibitem{aos20micro}
{\sc Kim, Y., Lee, J., and Kim, H.}
\newblock Hardware-based always-on heap memory safety.
\newblock In {\em 2020 53rd Annual IEEE/ACM International Symposium on Microarchitecture (MICRO)\/} (Athens, Greece, 2020), IEEE, pp.~1153--1166.

\bibitem{kuznetsov14osdi}
{\sc Kuznetsov, V., Szekeres, L., Payer, M., Candea, G., Sekar, R., and Song, D.}
\newblock Code-pointer integrity.
\newblock In {\em Proceedings of the 11th USENIX Conference on Operating Systems Design and Implementation\/} (USA, 2014), OSDI'14, USENIX Association, p.~147–163.

\bibitem{threadcfg}
{\sc Landi, W., and Ryder, B.~G.}
\newblock A safe approximate algorithm for interprocedural pointer aliasing.
\newblock {\em SIGPLAN Not. 39}, 4 (apr 2004), 473–489.

\bibitem{Lattner2007MakingCP}
{\sc Lattner, C., Lenharth, A., and Adve, V.}
\newblock Making context-sensitive points-to analysis with heap cloning practical for the real world.
\newblock In {\em Proceedings of the 28th ACM SIGPLAN Conference on Programming Language Design and Implementation\/} (2007), PLDI '07, p.~278–289.

\bibitem{dangnull}
{\sc Lee, B., Song, C., Jang, Y., Wang, T., Kim, T., Lu, L., and Lee, W.}
\newblock {Preventing Use-after-free with Dangling Pointers Nullification}.
\newblock In {\em Proceedings of the 2015 Network and Distributed System Security Symposium (NDSS)\/} (San Diego, CA, USA, 2015), The Internet Society.

\bibitem{caver}
{\sc Lee, B., Song, C., Kim, T., and Lee, W.}
\newblock Type casting verification: Stopping an emerging attack vector.
\newblock In {\em Proceedings of the 24th USENIX Conference on Security Symposium\/} (USA, 2015), SEC'15, USENIX Association, p.~81–96.

\bibitem{lemay21micro}
{\sc LeMay, M., Rakshit, J., Deutsch, S., Durham, D.~M., Ghosh, S., Nori, A., Gaur, J., Weiler, A., Sultana, S., Grewal, K., and Subramoney, S.}
\newblock Cryptographic capability computing.
\newblock In {\em MICRO-54: 54th Annual IEEE/ACM International Symposium on Microarchitecture\/} (New York, NY, USA, 2021), MICRO '21, Association for Computing Machinery, p.~253–267.

\bibitem{hybridglobalalias}
{\sc Li, G., Zhang, H., Zhou, J., Shen, W., Sui, Y., and Qian, Z.}
\newblock A hybrid alias analysis and its application to global variable protection in the linux kernel.
\newblock In {\em 32nd USENIX Security Symposium (USENIX Security 23)\/} (Anaheim, CA, Aug. 2023), USENIX Association, pp.~4211--4228.

\bibitem{pacmem}
{\sc Li, Y., Tan, W., Lv, Z., Yang, S., Payer, M., Liu, Y., and Zhang, C.}
\newblock Pacmem: Enforcing spatial and temporal memory safety via arm pointer authentication.
\newblock In {\em Proceedings of the 2022 ACM SIGSAC Conference on Computer and Communications Security\/} (New York, NY, USA, 2022), CCS '22, Association for Computing Machinery, p.~1901–1915.

\bibitem{kleak}
{\sc Liang, Z., Zou, X., Song, C., and Qian, Z.}
\newblock K-leak: Towards automating the generation of multi-step infoleak exploits against the linux kernel.
\newblock In {\em 31st Annual Network and Distributed System Security Symposium, NDSS\/} (2024).

\bibitem{ptmalloc}
{\sc Library, G.~C.}
\newblock Glibc wiki - mallocinternals.
\newblock \url{https://sourceware.org/glibc/wiki/MallocInternals}.
\newblock Accessed on Mar 7, 2024.

\bibitem{pacitup19usenix}
{\sc Liljestrand, H., Nyman, T., Wang, K., Perez, C.~C., Ekberg, J.-E., and Asokan, N.}
\newblock Pac it up: Towards pointer integrity using arm pointer authentication.
\newblock In {\em Proceedings of the 28th USENIX Conference on Security Symposium\/} (USA, 2019), SEC'19, USENIX Association, p.~177–194.

\bibitem{autoslab}
{\sc Lin, Z.}
\newblock How autoslab changes the memory unsafety game.
\newblock \url{https://grsecurity.net/how_autoslab_changes_the_memory_unsafety_game}, 2022.
\newblock Accessed on May 25, 2023.

\bibitem{grebe}
{\sc Lin, Z., Chen, Y., Wu, Y., Mu, D., Yu, C., Xing, X., and Li, K.}
\newblock Grebe: Unveiling exploitation potential for linux kernel bugs.
\newblock In {\em 2022 IEEE Symposium on Security and Privacy (SP)\/} (2022), pp.~2078--2095.

\bibitem{dirtycred}
{\sc Lin, Z., Wu, Y., and Xing, X.}
\newblock Dirtycred: Escalating privilege in linux kernel.
\newblock In {\em Proceedings of the 2022 ACM SIGSAC Conference on Computer and Communications Security\/} (New York, NY, USA, 2022), CCS '22, ACM.

\bibitem{camp24usenix}
{\sc Lin, Z., Yu, Z., Guo, Z., Campanoni, S., Dinda, P., and Xing, X.}
\newblock {CAMP}: Compiler and allocator-based heap memory protection.
\newblock In {\em 33rd USENIX Security Symposium (USENIX Security 24)\/} (Philadelphia, PA, Aug. 2024).

\bibitem{dep}
{\sc Linux}.
\newblock Linux 2.6.7. nx (no execute) support for x86.
\newblock https://lkml.org/lkml/2004/6/2/228, 2004.

\bibitem{Liu17CCS}
{\sc Liu, S., Tan, G., and Jaeger, T.}
\newblock Ptrsplit: Supporting general pointers in automatic program partitioning.
\newblock In {\em Proceedings of the 2017 ACM SIGSAC Conference on Computer and Communications Security\/} (New York, NY, USA, 2017), CCS '17, Association for Computing Machinery, p.~2359–2371.

\bibitem{clangstaticanalyzer}
{\sc LLVM}.
\newblock Clang static analyzer.
\newblock https://clang-analyzer.llvm.org/, 2023.

\bibitem{ubSAN}
{\sc LLVM}.
\newblock Clang undefined behavior sanitizer.
\newblock \url{http://clang.llvm.org/docs/UsersManual.html}, 2023.
\newblock Accessed: 2023-05-02.

\bibitem{llvm-loop-canonicalization}
{Canonicaliza natural loops}.
\newblock LLVM documentation at \url{https://llvm.org/docs/Passes.html#loop-simplify-canonicalize-natural-loops}, 2020.

\bibitem{llvm-loop-simplify}
{Loop Simplify Form}.
\newblock LLVM documentation at \url{https://llvm.org/docs/LoopTerminology.html#loop-simplify-form}, 2020.

\bibitem{llvm_union}
{Mapping High Level Constructs to LLVM IR - Union}.
\newblock \url{https://mapping-high-level-constructs-to-llvm-ir.readthedocs.io/en/latest/basic-constructs/unions.html}.

\bibitem{microsoft-report}
{\sc Microsoft}.
\newblock Trends, challenges, and strategic shifts in the software vulnerability mitigation landscape.
\newblock \url{https://github.com/Microsoft/MSRC-Security-Research/blob/master/presentations/2019_02_BlueHatIL/2019_01%20-%20BlueHatIL%20-%20Trends%2C%20challenge%2C%20and%20shifts%20in%20software%20vulnerability%20mitigation.pdf}, 2019.
\newblock Accessed on May 28, 2023.

\bibitem{windowsheap}
{\sc Microsoft}.
\newblock {Customize exploit protection Article 09/28/2022 12 contributors }.
\newblock \url{https://learn.microsoft.com/en-us/microsoft-365/security/defender-endpoint/customize-exploit-protection?view=o365-worldwide}, 2022.
\newblock Accessed on May 13, 2023.

\bibitem{nescheck}
{\sc Midi, D., Payer, M., and Bertino, E.}
\newblock Memory safety for embedded devices with nescheck.
\newblock In {\em Proceedings of the 2017 ACM on Asia Conference on Computer and Communications Security\/} (New York, NY, USA, 2017), ASIA CCS '17, Association for Computing Machinery, p.~127–139.

\bibitem{safeinit}
{\sc Milburn, A., Bos, H., and Giuffrida, C.}
\newblock Safelnit: Comprehensive and practical mitigation of uninitialized read vulnerabilities.
\newblock In {\em Network and Distributed System Security Symposium\/} (2017).

\bibitem{tdi}
{\sc Milburn, A., Van Der~Kouwe, E., and Giuffrida, C.}
\newblock Mitigating information leakage vulnerabilities with type-based data isolation.
\newblock In {\em 2022 IEEE Symposium on Security and Privacy (SP)\/} (San Francisco, CA, USA, 2022), IEEE, pp.~1049--1065.

\bibitem{sql_slammer}
{\sc Moore, D., Paxson, V., Savage, S., Shannon, C., Staniford, S., and Weaver, N.}
\newblock {Inside the Slammer worm}.
\newblock {\em IEEE Security \& Privacy 1}, 4 (2003), 33--39.

\bibitem{softbound}
{\sc Nagarakatte, S., Zhao, J., Martin, M.~M., and Zdancewic, S.}
\newblock Softbound: Highly compatible and complete spatial memory safety for c.
\newblock In {\em Proceedings of the 30th ACM SIGPLAN Conference on Programming Language Design and Implementation\/} (New York, NY, USA, 2009), PLDI '09, Association for Computing Machinery, p.~245–258.

\bibitem{Nagarakatte2010CETSCE}
{\sc Nagarakatte, S., Zhao, J., Martin, M.~M., and Zdancewic, S.}
\newblock Cets: Compiler enforced temporal safety for c.
\newblock In {\em Proceedings of the 2010 International Symposium on Memory Management\/} (New York, NY, USA, 2010), ISMM '10, Association for Computing Machinery, p.~31–40.

\bibitem{cryptag}
{\sc Nasahl, P., Schilling, R., Werner, M., Hoogerbrugge, J., Medwed, M., and Mangard, S.}
\newblock Cryptag: Thwarting physical and logical memory vulnerabilities using cryptographically colored memory.
\newblock In {\em Proceedings of the 2021 ACM Asia Conference on Computer and Communications Security\/} (New York, NY, USA, 2021), ASIA CCS '21, Association for Computing Machinery, p.~200–212.

\bibitem{ccured}
{\sc Necula, G.~C., Condit, J., Harren, M., McPeak, S., and Weimer, W.}
\newblock Ccured: Type-safe retrofitting of legacy software.
\newblock {\em ACM Trans. Program. Lang. Syst. 27}, 3 (may 2005), 477–526.

\bibitem{ccured-toplas}
{\sc Necula, G.~C., McPeak, S., and Weimer, W.}
\newblock Ccured: Type-safe retrofitting of legacy code.
\newblock {\em SIGPLAN Not. 37}, 1 (jan 2002), 128–139.

\bibitem{neug2017dimva}
{\sc Neugschwandtner, M., Sorniotti, A., and Kurmus, A.}
\newblock Memory categorization: Separating attacker-controlled data.
\newblock In {\em Detection of Intrusions and Malware, and Vulnerability Assessment\/} (Cham, 2019), R.~Perdisci, C.~Maurice, G.~Giacinto, and M.~Almgren, Eds., Springer International Publishing, pp.~263--287.

\bibitem{nginxCommonPitfalls}
{\sc Nginx}.
\newblock Nginx development guide - common pitfalls.
\newblock \url{https://nginx.org/en/docs/dev/development_guide.html#common_pitfalls}.
\newblock Accessed: 2024-04-10.

\bibitem{dieharder}
{\sc Novark, G., and Berger, E.~D.}
\newblock Dieharder: Securing the heap.
\newblock In {\em Proceedings of the 17th ACM Conference on Computer and Communications Security\/} (New York, NY, USA, 2010), CCS '10, Association for Computing Machinery, p.~573–584.

\bibitem{nsa-report}
{\sc NSA-CSS}.
\newblock Nsa releases guidance on how to protect against software memory safety issues.
\newblock \url{https://www.nsa.gov/Press-Room/News-Highlights/Article/Article/3215760/nsa-releases-guidance-on-how-to-protect-against-software-memory-safety-issues/}, 2022.

\bibitem{mpxreport}
{\sc Oleksenko, O., Kuvaiskii, D., Bhatotia, P., Felber, P., and Fetzer, C.}
\newblock Intel mpx explained: A cross-layer analysis of the intel mpx system stack.
\newblock {\em ACM SIGMETRICS Performance Evaluation Review\/} (2019).

\bibitem{smashing}
{\sc One, A.}
\newblock Smashing the stack for fun and profit.
\newblock {\em Phrack magazine\/} (1996).

\bibitem{chatgpt}
{\sc OpenAI}.
\newblock {ChatGPT}.
\newblock \url{https://chat.openai.com/}.
\newblock Accessed on May 13, 2023.

\bibitem{palit21oakland}
{\sc Palit, T., Firose~Moon, J., Monrose, F., and Polychronakis, M.}
\newblock Dynpta: Combining static and dynamic analysis for practical selective data protection.
\newblock In {\em 2021 IEEE Symposium on Security and Privacy (SP)\/} (2021), IEEE.

\bibitem{proskurin20oakland}
{\sc Proskurin, S., Momeu, M., Ghavamnia, S., Kemerlis, V.~P., and Polychronakis, M.}
\newblock xmp: Selective memory protection for kernel and user space.
\newblock In {\em 2020 IEEE Symposium on Security and Privacy (SP)\/} (2020), IEEE, pp.~563--577.

\bibitem{posixrealloc}
{\sc ptmalloc}.
\newblock realloc(3) - linux man page.
\newblock \url{https://linux.die.net/man/3/realloc}.
\newblock Accessed on Mar 7, 2024.

\bibitem{gpt}
{\sc Radford, A., Wu, J., Child, R., Luan, D., Amodei, D., and Sutskever, I.}
\newblock Language models are unsupervised multitask learners.
\newblock {\em OpenAI Blog\/} (June 2019).

\bibitem{pacmam22isca}
{\sc Ravichandran, J., Na, W.~T., Lang, J., and Yan, M.}
\newblock Pacman: Attacking arm pointer authentication with speculative execution.
\newblock In {\em Proceedings of the 49th Annual International Symposium on Computer Architecture\/} (New York, NY, USA, 2022), ISCA '22, Association for Computing Machinery, p.~685–698.

\bibitem{heartbleed}
{\sc Riku, Antti, Matti, and Mehta, N.}
\newblock Heartbleed.
\newblock \url{http://heartbleed.com/}, 2014.

\bibitem{rop}
{\sc Roemer, R., Buchanan, E., Shacham, H., and Savage, S.}
\newblock Return-oriented programming: Systems, languages, and applications.
\newblock {\em ACM Trans. Inf. Syst. Secur. 15}, 1 (mar 2012).

\bibitem{kalloc_type}
{\sc Security, A.}
\newblock Towards the next generation of xnu memory safety: kalloc\_type.
\newblock \url{https://security.apple.com/blog/towards-the-next-generation-of-xnu-memory-safety/}, 2022.

\bibitem{morris-worm-seeley}
{\sc Seeley, D.}
\newblock {A Tour of the Worm}.
\newblock \url{https://www.cs.unc.edu/~jeffay/courses/nidsS05/attacks/seely-RTMworm-89.html.}

\bibitem{arm-mte}
{\sc Serebryany, K.}
\newblock Arm memory tagging extension and how it improves c/c++ memory safety.
\newblock {\em login Usenix Mag. 44\/} (2019).

\bibitem{asan}
{\sc Serebryany, K., Bruening, D., Potapenko, A., and Vyukov, D.}
\newblock Addresssanitizer: A fast address sanity checker.
\newblock In {\em Proceedings of the 2012 USENIX Conference on Annual Technical Conference\/} (USA, 2012), USENIX ATC'12, USENIX, p.~28.

\bibitem{chex86}
{\sc Sharifi, R., and Venkat, A.}
\newblock Chex86: Context-sensitive enforcement of memory safety via microcode-enabled capabilities.
\newblock In {\em Proceedings of the ACM/IEEE 47th Annual International Symposium on Computer Architecture\/} (2020), ISCA '20.

\bibitem{sharir-pnueli}
{\sc Sharir, M., and Pnueli, A.}
\newblock {Two approaches to interprocedural data flow analysis}.
\newblock \url{https://www.cse.psu.edu/~trj1/cse598-f11/docs/sharir_pnueli1.pdf}, 1981.
\newblock Accessed on May 13, 2023.

\bibitem{freeguard17ccs}
{\sc Silvestro, S., Liu, H., Crosser, C., Lin, Z., and Liu, T.}
\newblock Freeguard: A faster secure heap allocator.
\newblock In {\em Proceedings of the 2017 ACM SIGSAC Conference on Computer and Communications Security\/} (New York, NY, USA, 2017), CCS '17, Association for Computing Machinery, p.~2389–2403.

\bibitem{value-range-book}
{\sc Simon, A.}
\newblock Value-range analysis of c programs: Towards proving the absence of buffer overflow vulnerabilities, 2008.

\bibitem{memsafe10scam}
{\sc Simpson, M.~S., and Barua, R.~K.}
\newblock {MemSafe: Ensuring the Spatial and Temporal Memory Safety of C at Runtime}.
\newblock In {\em Proceedings of the 2010 10th IEEE Working Conference on Source Code Analysis and Manipulation (SCAM)\/} (USA, 2010), John Wiley \& Sons, Inc., p.~93–128.

\bibitem{practicalrest}
{\sc Sinha, K., and Sethumadhavan, S.}
\newblock Practical memory safety with rest.
\newblock In {\em Proceedings of the 45th Annual International Symposium on Computer Architecture\/} (2018), ISCA '18, IEEE Press, p.~600–611.

\bibitem{yannis18ecoop}
{\sc Smaragdakis, Y., and Kastrinis, G.}
\newblock {Defensive Points-To Analysis: Effective Soundness via Laziness}.
\newblock In {\em 32nd European Conference on Object-Oriented Programming (ECOOP 2018)\/} (Dagstuhl, Germany, 2018), T.~Millstein, Ed., vol.~109 of {\em Leibniz International Proceedings in Informatics (LIPIcs)}, Schloss Dagstuhl--Leibniz-Zentrum fuer Informatik, pp.~23:1--23:28.

\bibitem{snow2013just}
{\sc Snow, K.~Z., Monrose, F., Davi, L., Dmitrienko, A., Liebchen, C., and Sadeghi, A.-R.}
\newblock {Just-in-time Code Reuse: On the Effectiveness of Fine-grained Address Space Layout Randomization}.
\newblock In {\em Proceedings of the 34th IEEE Symposium on Security and Privacy (S\&P)\/} (2013).

\bibitem{sophos2021state}
{\sc Sophos}.
\newblock The state of ransomware 2021.
\newblock \url{https://www.sophos.com/en-us/medialibrary/pdfs/whitepaper/sophos-the-state-of-ransomware-2021-wp.pdf}, 2021.
\newblock Accessed on May 13, 2023.

\bibitem{heap-feng-shui}
{\sc Sotirov, A.}
\newblock Heap feng shui in javascript.
\newblock {\em Black Hat Europe\/} (2007).

\bibitem{stamatogiannakis2022asleep}
{\sc Stamatogiannakis, M., Bos, H., Giuffrida, C., Mavroudis, V., and Papadopoulos, S.}
\newblock Asleep at the keyboard? assessing the security of github copilot's code contributions.
\newblock In {\em 2022 IEEE Symposium on Security and Privacy (SP)\/} (San Francisco, CA, USA, 2022), IEEE, pp.~754--768.

\bibitem{driller}
{\sc Stephens, N., Grosen, J., Salls, C., Dutcher, A., Wang, R., Corbetta, J., Shoshitaishvili, Y., Kruegel, C., and Vigna, G.}
\newblock Driller: Augmenting fuzzing through selective symbolic execution.
\newblock In {\em Proceedings of the 23rd Annual Network and Distributed System Security Symposium\/} (2016).

\bibitem{svfmta}
{\sc Sui, Y., Di, P., and Xue, J.}
\newblock Sparse flow-sensitive pointer analysis for multithreaded programs.
\newblock In {\em Proceedings of the 2016 International Symposium on Code Generation and Optimization\/} (New York, NY, USA, 2016), CGO '16, Association for Computing Machinery, p.~160–170.

\bibitem{sui2016fse}
{\sc Sui, Y., and Xue, J.}
\newblock On-demand strong update analysis via value-flow refinement.
\newblock In {\em Proceedings of the 2016 24th ACM SIGSOFT International Symposium on Foundations of Software Engineering\/} (New York, NY, USA, 2016), FSE 2016, Association for Computing Machinery, p.~460–473.

\bibitem{svf}
{\sc Sui, Y., and Xue, J.}
\newblock Svf: Interprocedural static value-flow analysis in llvm.
\newblock In {\em Proceedings of the 25th International Conference on Compiler Construction\/} (New York, NY, USA, 2016), CC 2016, ACM, p.~265–266.

\bibitem{sui2018tse}
{\sc Sui, Y., and Xue, J.}
\newblock Value-flow-based demand-driven pointer analysis for c and c++.
\newblock {\em IEEE Transactions on Software Engineering 46}, 8 (2018), 812--835.

\bibitem{sui2012issta}
{\sc Sui, Y., Ye, D., and Xue, J.}
\newblock Static memory leak detection using full-sparse value-flow analysis.
\newblock In {\em Proceedings of the 2012 International Symposium on Software Testing and Analysis\/} (2012), ISSTA 2012.

\bibitem{sui2014tse}
{\sc Sui, Y., Ye, D., and Xue, J.}
\newblock Detecting memory leaks statically with full-sparse value-flow analysis.
\newblock {\em IEEE Transactions on Software Engineering (TSE) 40}, 2 (2014).

\bibitem{google-report}
{\sc Taylor, A., Whalley, A., Jansens, D., and Oskov, N.}
\newblock An update on memory safety in chrome.
\newblock \url{https://security.googleblog.com/2021/09/an-update-on-memory-safety-in-chrome.html}, 2021.
\newblock Accessed on May 28, 2023.

\bibitem{Douglas11SBLP}
{\sc Teixeira, D., and Pereira, F. M.~Q.}
\newblock {The Design and Implementation of a Non-Iterative Range Analysis Algorithm on a Production Compiler}.
\newblock In {\em The 15th Brazilian Symposium on Programming Languages (SBLP)\/} (Brazil, 2011), CBSoft.

\bibitem{whitehousememsafe}
{\sc {The White House}}.
\newblock Press release: Future software should be memory safe.
\newblock \url{https://www.whitehouse.gov/oncd/briefing-room/2024/02/26/press-release-technical-report/}.
\newblock FEBRUARY 26, 2024.

\bibitem{AnnaKarenina-Tolstoy}
{\sc Tolstoy, L.}
\newblock {\em Anna Karenina}.
\newblock Wordsworth Editions, Tsarist Russia, 1995.

\bibitem{trailofbitscgc}
{\sc Trailofbits}.
\newblock {DARPA Challenge Binaries on Linux, OS X, and Windows}.
\newblock \url{https://github.com/trailofbits/cb-multios}, 2017.

\bibitem{encryptmemorysafe}
{\sc Unterguggenberger, M., Schrammel, D., Lamster, L., Nasahl, P., and Mangard, S.}
\newblock Cryptographically enforced memory safety.
\newblock In {\em Proceedings of the 2023 ACM SIGSAC Conference on Computer and Communications Security\/} (New York, NY, USA, 2023), CCS '23, Association for Computing Machinery, p.~889–903.

\bibitem{tat}
{\sc van~der Kouwe, E., Kroes, T., Ouwehand, C., Bos, H., and Giuffrida, C.}
\newblock Type-after-type: Practical and complete type-safe memory reuse.
\newblock In {\em Proceedings of the 34th Annual Computer Security Applications Conference\/} (New York, NY, USA, 2018), ACSAC '18, Association for Computing Machinery, p.~17–27.

\bibitem{van2017dangsan}
{\sc van~der Kouwe, E., Nigade, V., and Giuffrida, C.}
\newblock Dangsan: Scalable use-after-free detection.
\newblock In {\em Proceedings of the Twelfth European Conference on Computer Systems\/} (New York, NY, USA, 2017), EuroSys '17, Association for Computing Machinery, p.~405–419.

\bibitem{cybersecurityventuresrdr}
{\sc Ventures, C.}
\newblock Cybersecurity ventures' ransomware damage report.
\newblock \url{https://cybersecurityventures.com/cybersecurity-500/}.
\newblock Accessed on May 13, 2023.

\bibitem{auto_fengshui}
{\sc Wang, Y., Zhang, C., Zhao, Z., Zhang, B., Gong, X., and Zou, W.}
\newblock {MAZE}: Towards automated heap feng shui.
\newblock In {\em 30th USENIX Security Symposium (USENIX Security 21)\/} (Aug. 2021), USENIX Association, pp.~1647--1664.

\bibitem{javaheapcloningpldi}
{\sc Whaley, J., and Lam, M.~S.}
\newblock Cloning-based context-sensitive pointer alias analysis using binary decision diagrams.
\newblock In {\em Proceedings of the ACM SIGPLAN 2004 Conference on Programming Language Design and Implementation\/} (New York, NY, USA, 2004), PLDI '04, Association for Computing Machinery, p.~131–144.

\bibitem{ffmalloc21usenix}
{\sc Wickman, B., Hu, H., Yun, I., Jang, D., Lim, J., Kashyap, S., and Kim, T.}
\newblock Preventing {Use-After-Free} attacks with fast forward allocation.
\newblock In {\em 30th USENIX Security Symposium (USENIX Security 21)\/} (Virtual Event, Aug. 2021).

\bibitem{kepler}
{\sc Wu, W., Chen, Y., Xing, X., and Zou, W.}
\newblock Kepler: facilitating control-flow hijacking primitive evaluation for linux kernel vulnerabilities.
\newblock In {\em Proceedings of the 28th USENIX Conference on Security Symposium\/} (USA, 2019), SEC'19, USENIX Association, p.~1187–1204.

\bibitem{fuze17usenix}
{\sc Wu, W., Chen, Y., Xu, J., Xing, X., Gong, X., and Zou, W.}
\newblock {FUZE}: Towards facilitating exploit generation for kernel use-after-free vulnerabilities.
\newblock In {\em 27th {USENIX} Security Symposium ({USENIX} Security 18)\/} (Baltimore, 2018), pp.~781--797.

\bibitem{xu2022systematic}
{\sc Xu, F.~F., Alon, U., Neubig, G., and Hellendoorn, V.~J.}
\newblock A systematic evaluation of large language models of code.
\newblock In {\em Proceedings of the 6th ACM SIGPLAN International Symposium on Machine Programming\/} (New York, NY, USA, 2022), MAPS 2022, Association for Computing Machinery, p.~1–10.

\bibitem{infatpointer}
{\sc Xu, S., Huang, W., and Lie, D.}
\newblock In-fat pointer: Hardware-assisted tagged-pointer spatial memory safety defense with subobject granularity protection.
\newblock In {\em Proceedings of the 26th ACM International Conference on Architectural Support for Programming Languages and Operating Systems\/} (2021), ASPLOS '21.

\bibitem{mluafdetect}
{\sc Yan, H., Sui, Y., Chen, S., and Xue, J.}
\newblock Machine-learning-guided typestate analysis for static use-after-free detection.
\newblock In {\em Proceedings of the 33rd Annual Computer Security Applications Conference\/} (New York, NY, USA, 2017), ACSAC '17, Association for Computing Machinery, p.~42–54.

\bibitem{cred}
{\sc Yan, H., Sui, Y., Chen, S., and Xue, J.}
\newblock Spatio-temporal context reduction: A pointer-analysis-based static approach for detecting use-after-free vulnerabilities.
\newblock In {\em Proceedings of the 40th International Conference on Software Engineering\/} (New York, NY, USA, 2018), ICSE '18, ACM, p.~327–337.

\bibitem{uafchecker}
{\sc Ye, J., Zhang, C., and Han, X.}
\newblock Poster: Uafchecker: Scalable static detection of use-after-free vulnerabilities.
\newblock In {\em Proceedings of the 2014 ACM SIGSAC Conference on Computer and Communications Security\/} (New York, NY, USA, 2014), CCS '14, Association for Computing Machinery, p.~1529–1531.

\bibitem{freesentry}
{\sc Younan, Y.}
\newblock {FreeSentry: Protecting Against Use-after-free Vulnerabilities Due to Dangling Pointers}.
\newblock In {\em 22nd Annual Network and Distributed System Security Symposium (NDSS)\/} (San Diego, CA, USA, 2015), Internet Society.

\bibitem{Zhai2020UBITectAP}
{\sc Zhai, Y., Hao, Y., Zhang, H., Wang, D., Song, C., Qian, Z., Lesani, M., Krishnamurthy, S.~V., and Yu, P.}
\newblock Ubitect: A precise and scalable method to detect use-before-initialization bugs in linux kernel.
\newblock In {\em Proceedings of the 28th ACM Joint Meeting on European Software Engineering Conference and Symposium on the Foundations of Software Engineering (ESEC/FSE)\/} (2020).

\bibitem{increlux}
{\sc Zhai, Y., Hao, Y., Zhang, Z., Chen, W., Li, G., Qian, Z., Song, C., Sridharan, M., Krishnamurthy, S.~V., Jaeger, T., and Yu, P.}
\newblock Progressive scrutiny: Incremental detection of ubi bugs in the linux kernel.
\newblock In {\em Proceedings 2022 Network and Distributed System Security Symposium\/} (San Diego, CA, USA, 2022).

\bibitem{tprunify}
{\sc Zhai, Y., Qian, Z., Song, C., Sridharan, M., Jaeger, T., Yu, P., and Krishnamurthy, S.~V.}
\newblock Don’t waste my efforts: Pruning redundant sanitizer checks of developer-implemented type checks.
\newblock In {\em 33rd USENIX Security Symposium (USENIX Security 24)\/} (Philadelphia, PA, USA, 2024), USENIX Association.

\bibitem{vtrust}
{\sc Zhang, C., Carr, S.~A., Li, T., Ding, Y., Song, C., Payer, M., and Song, D.}
\newblock Vtrust: Regaining trust on virtual calls.
\newblock In {\em Symposium on Network and Distributed System Security (NDSS)\/} (San Diego, CA, USA, 2016), The Internet Society.

\bibitem{asan--}
{\sc Zhang, Y., Pang, C., Portokalidis, G., Triandopoulos, N., and Xu, J.}
\newblock Debloating address sanitizer.
\newblock In {\em 31st USENIX Security Symposium (USENIX Security 22)\/} (Boston, MA, Aug. 2022), USENIX Association, pp.~4345--4363.

\bibitem{zhou2022fat}
{\sc Zhou, J., Criswell, J., and Hicks, M.}
\newblock Fat pointers for temporal memory safety of c.
\newblock {\em Proc. ACM Program. Lang. 7}, OOPSLA1 (apr 2023).

\bibitem{uafdetector}
{\sc Zhu, K., Lu, Y., and Huang, H.}
\newblock Scalable static detection of use-after-free vulnerabilities in binary code.
\newblock {\em IEEE Access 8\/} (2020), 78713--78725.

\bibitem{syzbridge}
{\sc Zou, X., Hao, Y., Zhang, Z., Pu, J., Chen, W., and Qian, Z.}
\newblock Syzbridge: Bridging the gap in exploitability assessment of linux kernel bugs in the linux ecosystem.
\newblock In {\em 31st Annual Network and Distributed System Security Symposium, NDSS\/} (2024).

\bibitem{syzscope}
{\sc Zou, X., Li, G., Chen, W., Zhang, H., and Qian, Z.}
\newblock {SyzScope}: Revealing {High-Risk} security impacts of {Fuzzer-Exposed} bugs in linux kernel.
\newblock In {\em 31st USENIX Security Symposium (USENIX Security 22)\/} (Aug. 2022), USENIX Association.

\end{thebibliography}

\appendix

\clearpage
\onecolumn

%kaiming: Leave the appendix as one column for now for good looking to distribute to others before camera-ready.
\section{Appendix}
\subsection{SPEC Benchmarks}
\label{append:spec}

We present the additional evaluation results for SPEC CPU2006 and SPEC CPU2017 Benchmarks in this section. Table~\ref{tab:appendspec} shows the per-benchmark stats on the objects that pass \tool's static safety validation for SPEC CPU2006 and SPEC CPU2017 benchmarks. For quantifying the time elapsed in \tool's static safety validation on SPEC benchmark, Table~\ref{tab:appendtime} shows the time consumed per benchmark in each of the validation phase for SPEC CPU2006 benchmarks.

\if 0
\begin{table*}[]
\aboverulesep=0ex 
\belowbottomsep=0.5ex
\belowrulesep=0ex
\abovetopsep=0.5ex
    \centering
        %\scalebox{0.85}{
        \resizebox{0.8\columnwidth}{!}{
	\begin{tabular}{l|r|r}
	    \toprule
		\ &\multicolumn{1}{c|}{\bf\em Total }&\multicolumn{1}{c}{\bf\em Uriah-Protected}\\ \midrule%&\multicolumn{1}{c|}{\bf\em Memory}\\
		
		{\bf\em 400.perlbench}&360,584,980&245,869,358 (68.1\%)\\%&502 MB\\
		
		{\bf\em 401.bzip2}&160&132 (82.5\%)\\%&122 MB\\
		
		{\bf\em 429.mcf}&5&5 (100\%)\\%&1,364 MB\\
		
		{\bf\em 445.gobmk}&654,582&401,256 (61.68\%)\\%&185 MB\\
		
		{\bf\em 456.hmmer}&2,464,253&1,720,456 (69.8\%)\\%&52 MB\\
    
		{\bf\em 458.sjeng}&12&9 (75.0\%)\\%&276 MB\\
            
		{\bf\em 462.libquantum}&175&125 (71.4\%)\\%&130 MB\\
            
		{\bf\em 464.h264ref}&168,025&118,628 (70.6\%)\\%&172 MB\\
            
		{\bf\em 470.lbm}&7&5 (71.4\%)\\%&420 MB\\
            
		{\bf\em 482.sphinx3}&13,857,545&7,921,685 (57.2\%)\\%&359 MB\\
            
		{\bf\em 471.omnetpp}&267,431,654&172,760,848 (64.6\%) \\
		
		{\bf\em 473.soplex}&236,481&161,280 (68.2\%)\\
         
		{\bf\em 433.milc}&6,582&5,360 (81.4\%)\\%&636 MB\\
            
		{\bf\em 483.astar}&4,852,696&3,697,754 (76.2\%)\\
            
		{\bf\em 444.namd}&1,375&1,039 (75.6\%)\\%&134 MB\\
		
             \midrule
            {\bf\em AVERAGE}&----&{\bf 73.0\%}\\
            \bottomrule
	\end{tabular}
 }
        \captionsetup{font=footnotesize}
	\caption{Percentage of memory allocations protected by \tool}
	\label{tab:Protection of Heap Allocations at Runtime Abstract}
\end{table*}
\fi

\begin{table}[]
\aboverulesep=0ex 
\belowbottomsep=0.5ex
\belowrulesep=0ex
\abovetopsep=0.5ex
\centering
\resizebox{.45\columnwidth}{!}{
\begin{tabular}{l|rrrrrrr}
\toprule
\multicolumn{1}{c|}{} &
  \multicolumn{1}{c}{\textbf{Foundation}} &
  %\multicolumn{1}{c}{\textbf{Porting}} &
  \multicolumn{1}{c}{\textbf{Static}} &
  \multicolumn{1}{c}{\textbf{SymExec}} &
  \multicolumn{1}{c}{\textbf{Total}} \\ \midrule
\textbf{perlbench}  &12,378.3s  &2,761.3s  &3,254.6s  &18,394.2s    \\
\textbf{bzip2}      &187.2s  &252.5s  &74.3s  &514.0s    \\
\textbf{mcf}        &13.2s  &8.6s  &12.5s  &34.3s     \\
\textbf{gobmk}      &3386.1s  &497.4s  &832.6s  &4716.1s     \\
\textbf{hmmer}      &733.2s  &165.3s  &171.6s  &1,070.1    \\
\textbf{sjeng}      &342.6s  &572.6s  &74.6s  &989.8    \\
\textbf{libquantum} &256.5s  &46.8s  &181.5s  &484.8s     \\
\textbf{h264ref}    &174.3s &15.7s  &33.4s  &223.4s    \\
\textbf{lbm}        &2,425.7s  &1,654.4s  &3,127.1s  &7,207.2s    \\
\textbf{sphinx3}    &694.2s  &455.1s  &323.5s  &1,472.8s    \\
\textbf{milc}       &4,233.6s  &2,176.8s  & 6,564.2s &12,974.6s    \\
\textbf{omnetpp}    &584.2s  &344.9s  &281.1s  &1,210.2s    \\
\textbf{soplex}     &382.5s  &1.2s  &47.0s  &430.7s     \\
\textbf{namd}       &3,121.4s  &1,962.6s  &899.4s  &5,983.4s    \\
\textbf{astar}      &858.2s  &439.7s  &914.3s  &2,212.2s   \\ \bottomrule
\end{tabular}
}\captionsetup{font=footnotesize}

	\caption{Time Elapsed in Each Phase of \tool's Static Safety Validation on SPEC CPU2006 Benchmarks}

	\label{tab:appendtime}
\end{table}

\begin{table*}[]
    \aboverulesep=0ex 
    \belowbottomsep=0.5ex
    \belowrulesep=0ex
    \abovetopsep=0.5ex
    \centering   
        \resizebox{0.9\textwidth}{!}{
	\begin{tabular}{l|r|rr|rrr|r|r}
	    \toprule
		 \multirow{2}{*} &\multirow{2}{*}{\bf\em Total}&\multirow{2}{*}{\bf\em VR-Spatial }&\multirow{2}{*}{\bf\em Uriah-Spatial}&\multirow{2}{*}{\bf\em CCured-Type}&\multirow{2}{*}{\bf\em CTCA-Type}&\multirow{2}{*}{\bf\em Uriah-Type\ }&\multicolumn{1}{c|}{\bf\em VR-Spatial+}&\multicolumn{1}{c}{\bf\em Uriah-Spatial+}\\
            
            &&&&&&&\multicolumn{1}{c|}{\bf\em CCured-Type}&\multicolumn{1}{c}{\bf\em Uriah-Type}\\
            \midrule
            {\bf\em perlbench}&319&186 (58.3\%)&241 (75.5\%)&206 (64.6\%)&258 (80.9\%)&271 (85.0\%)&154 (48.3\%)&230 (72.1\%)\\
	    
		{\bf\em bzip2}&5&5 (100\%)&5 (100\%)&2 (40.0\%)&4 (80.0\%)&5 (100\%)&2 (40.0\%)&4 (80.0\%)\\
		%\hline
		%{\bf\em gcc}&&&&&&&&&&&\\
		
		{\bf\em mcf}&4&4 (100\%)&4 (100\%)&0 (0.0\%)&4 (100\%)&4 (100\%)&0 (0.0\%)&4 (100\%)\\
		
		{\bf\em gobmk}&29&19 (65.5\%)&23 (79.3\%)&10 (34.5\%)&15 (51.7\%)&19 (65.5\%)&9 (31.0\%)&16 (55.2\%)\\
		
		{\bf\em hmmer}&350&238 (68.0\%)&282 (80.6\%)&73 (20.9\%)&215 (61.4\%)&256 (73.1\%)&65 (18.6\%)&240 (68.6\%)\\
		
		{\bf\em sjeng}&12&10 (83.3\%)&10 (83.3\%)&3 (25.0\%)&9 (75.0\%)&9 (75.0\%)&3 (25.0\%)&9 (75.0\%)\\
		
		{\bf\em libquantum}&19&13 (68.4\%)&15 (78.9\%)&7 (36.8\%)&16 (84.2\%)&16 (84.2\%)&5 (26.3\%)&14 (73.7\%)\\
		
		{\bf\em h264ref}&103&76 (73.8\%)&81 (78.6\%)&29 (28.2\%)&87 (84.5\%)&87 (84.5\%)&22 (21.4\%)&75 (72.8\%)\\
		
		{\bf\em lbm}&7&4 (57.1\%)&5 (71.4\%)&7 (100\%)&7 (100\%)&7 (100\%)&4 (57.1\%)&5 (71.4\%)\\
		
		{\bf\em sphinx3}&138&66 (47.8\%)& 78 (56.5\%)&59 (42.8\%)&113 (81.9\%)&120 (87.0\%)&43 (31.2\%)&70 (50.7\%)\\
		
		{\bf\em milc}&55&41 (74.5\%)&47 (85.5\%)&8 (14.5\%)&47 (85.5\%)&49 (89.1\%)&8 (14.5\%)&45 (81.8\%)\\
		
		{\bf\em omnetpp}&859&578 (67.3\%)&600 (69.8\%)&402 (46.8\%)&713 (83.0\%)&735 (85.6\%)&342 (39.8\%)&525 (61.2\%)\\
		
		{\bf\em soplex}&242&165 (68.2\%)&172 (71.1\%)&137 (56.6\%)&190 (78.5\%)&202 (83.5\%)&115 (47.5\%)&161 (66.5\%)\\
		
	    {\bf\em namd}&29&22 (75.9\%)&24 (82.8\%)&7 (24.1\%)&24 (82.8\%)&24 (82.8\%)&7 (24.1\%)&24 (82.8\%)\\
            
		{\bf\em astar}&48&28 (58.3\%)&39 (81.2\%)&15 (31.3\%)&36 (75.0\%)&38 (79.2\%)&11 (23.0\%)&34 (71.0\%)\\
		\midrule
		{\bf\em AVERAGE}&----&{\bf 71.1\%}&{\bf 79.6\%}&{\bf 37.7\%}&{\bf 80.3\%}&{\bf 85.0\%}&{\bf 29.8\%}&{\bf 72.2\%}\\
		\midrule\midrule
  
            {\bf\em perlbench\_s}&705&509 (72.2\%)&616 (87.4\%)&240 (34.0\%)&567 (80.4\%)&625 (88.7\%)&227 (32.2\%)&544 (77.2\%)\\

        {\bf\em gcc\_s}&5,970&4,365 (73.1\%)&4,847 (81.2\%)&2,127 (35.6\%)&4,978 (83.4\%)&5,259 (88.1\%)&1,998 (33.5\%)&4,736 (79.3\%)\\
        
        {\bf\em mcf\_s}&18&12 (66.7\%)&16 (88.9\%)&11 (61.1\%)&14 (77.8\%)&15 (83.3\%)&5 (27.8\%)&14 (77.8\%)\\

        {\bf\em xalancbmk\_s}&2,992&2,133 (71.3\%)&2,394 (80.0\%)&1,042 (34.8\%)&2,439 (81.5\%)&2,394 (80.0\%)&757 (25.3\%)&1,922 (64.2\%)\\
        
        {\bf\em deepsjeng\_s}&30&20 (66.7\%)&28 (93.3\%)&10 (33.3\%)&24 (80.0\%)&24 (80.0\%)&8 (26.7\%)&23 (76.7\%)\\

        {\bf\em x264\_s}&82&56 (68.3\%)&68 (82.9\%)&27 (32.9\%)&62 (75.6\%)&72 (87.8\%)&23 (28.0\%)&64 (78.0\%)\\

        {\bf\em lbm\_s}&5&3 (60.0\%)&4 (80.0\%)&4 (80.0\%)&4 (80.0\%)&4 (80.0\%)&1 (20.0\%)&4 (80.0\%)\\

        {\bf\em omnetpp\_s}&730&502 (68.8\%)&598 (81.9\%)&244 (33.4\%)&551 (75.5\%)&654 (89.6\%)&242 (33.2\%)&563 (77.1\%)\\

        {\bf\em imagick\_s}&98&68 (69.4\%)&77 (78.6\%)&33 (33.7\%)&76 (77.6\%)&83 (84.7\%)&31 (31.6\%)&75 (76.5\%)\\

        {\bf\em leela\_s}&50&34 (68.0\%)&40 (80.0\%)&28 (56.0\%)&40 (80.0\%)&40 (80.0\%)&14 (28.0\%)&34 (68.0\%)\\

        {\bf\em nab\_s}&104&72 (69.2\%)&88 (84.6\%)&34 (32.7\%)&79 (76.0\%)&83 (79.8\%)&26 (25.0\%)&81 (77.9\%)\\

        {\bf\em xz\_s}&751&549 (73.1\%)&618 (82.3\%)&252 (33.6\%)&602 (80.2\%)&636 (84.7\%)&199 (26.5\%)&559 (74.4\%)\\

        \midrule

        {\bf\em AVERAGE}&----&{\bf 68.9\%}&{\bf 83.4\%}&{\bf 41.8\%}&{\bf 79.0\%}&{\bf 83.9\%}&{\bf 28.2\%}&{\bf 75.6\%}\\

            \bottomrule
	\end{tabular}
}
        \captionsetup{font=footnotesize}
	\caption{Incremental Safety Improvement of \tool on SPEC CPU2006 and SPEC CPU2017 Benchmarks}
	\label{tab:appendspec}

\end{table*}

\end{document}